\documentclass[11pt,a4paper]{article}
\pdfoutput=1
\usepackage{graphicx}
\usepackage{afterpage}
\usepackage{amssymb}
\usepackage{amsmath}
\usepackage{dsfont}
\usepackage{multirow}
\usepackage{epstopdf}

\usepackage{url}
\usepackage{hyperref}

\newcommand{\be}{\begin{equation}}

\newcommand{\ee}{\end{equation}}
\newcommand{\bea}{\begin{eqnarray}}
\newcommand{\eea}{\end{eqnarray}}
\newcommand{\bi}{\begin{itemize}}
\newcommand{\ei}{\end{itemize}}
\newcommand{\ben}{\begin{enumerate}}
\newcommand{\een}{\end{enumerate}}
\newcommand{\la}{\left\langle}
\newcommand{\ra}{\right\rangle}
\newcommand{\lc}{\left[}
\newcommand{\rc}{\right]}
\newcommand{\lp}{\left(}
\newcommand{\rp}{\right)}

\def\frac#1#2{{{#1}\over {#2}}}
\def\gsim{\mathrel{\rlap{\lower4pt\hbox{\hskip1pt$\sim$}}
    \raise1pt\hbox{$>$}}}         
\def\lsim{\mathrel{\rlap{\lower4pt\hbox{\hskip1pt$\sim$}}
    \raise1pt\hbox{$<$}}}         

\def\beq{\begin{equation}}  
\def\eeq{\end{equation}}  

\def \n0{N_j^{(0)}}

\def\lapprox{\lower .7ex\hbox{$\;\stackrel{\textstyle <}{\sim}\;$}}
\def\gapprox{\lower .7ex\hbox{$\;\stackrel{\textstyle >}{\sim}\;$}}

\def\GeV{{\rm GeV}}

\usepackage{cite}

\begin{document}
\vspace{-2.0cm}
\begin{flushright}
CERN-PH-TH/2012-311\\
ICTP-SAIFR/2013-003 \\
\end{flushright}

\begin{center}
{\large\bf Scale-Invariant Resonance Tagging in Multijet Events \\
and New Physics  in Higgs Pair Production}
\vspace{0.6cm}

\renewcommand{\thefootnote}{\fnsymbol{footnote}}

Maxime Gouzevitch$^1$, Alexandra Oliveira$^{2,3}$,  Juan Rojo$^4$, \\
Rogerio Rosenfeld$^{2,4}$, Gavin P. Salam$^{4,5,}$\footnote{On leave from Department of Physics, Princeton University, Princeton, NJ 08544, USA.} and Veronica
Sanz$^{6,7}$ \\

\renewcommand{\thefootnote}{\arabic{footnote}}
\setcounter{footnote}{0}
\vspace{1.cm}
{\it
~$^1$ Universit\'e de Lyon, Universit\'e Claude Bernard Lyon 1, CNRS-IN2P3, Institut de Physique Nucl\'eaire de Lyon, Villeurbanne,
France\\
 ~$^2$ Instituto de F\' {\i}sica Te\'orica, Universidade Estadual Paulista and\\
ICTP South American Institute for Fundamental Research,\\ S\~ao Paulo, SP 01140-070, Brazil\\
~$^3$ Institut de Physique Theorique, CEA-Saclay,F-91191 Gif-sur-Yvette Cedex, France.\\
~$^4$ PH Department, TH Unit, CERN, CH-1211 Geneva 23, Switzerland \\
~$^5$  LPTHE, CNRS UMR 7589, UPMC Univ.\ Paris 6, Paris 75252, France\\
~$^6$ Department of Physics and Astronomy, York University, \\ Toronto, ON, Canada, M3J 1P3\\
~$^7$ Department of Physics and Astronomy, University of Sussex, \\ Brighton BN1 9QH, UK\\}
\end{center}

\vspace{0.5cm}

\begin{center}
{\bf \large Abstract:}
\end{center}
We study resonant pair production of heavy particles in fully hadronic
final states by means of jet substructure techniques.
We propose a new resonance tagging strategy that smoothly interpolates
between the highly boosted and fully resolved regimes, leading to
uniform signal efficiencies and background rejection
rates across a broad range of  masses.
 Our method makes it possible to efficiently replace independent experimental searches, based on different
final state topologies, with a single common analysis.
As a case study, we apply our technique to pair production of
 Higgs bosons decaying into $b\bar{b}$ pairs in generic
New Physics scenarios.
We adopt as benchmark models
radion and massive KK graviton 
production in warped
extra dimensions.
 We find that despite the
overwhelming QCD background, the $4b$ final state has enough sensitivity
to provide a complementary handle in searches for enhanced
Higgs pair production at the LHC.

\clearpage

\tableofcontents

\clearpage

\section{Introduction}
\label{sec-intro}

Jets are a ubiquitous component of the LHC program,
relevant for
precision Standard Model measurements, Higgs boson 
characterization and Beyond the Standard
Model searches ~\cite{Ellis:2007ib,Salam:2009jx}. 
In particular, searches for New Physics in multijet events
are an important element of the LHC physics program. 
 New resonances and contact interactions
have been searched for by ATLAS and CMS in final states with
two jets~\cite{Khachatryan:2010jd, CMS:2012yf, Chatrchyan:2013qha, Atlas:2012pu,Chatrchyan:2011ns,Aad:2011fq,Aad:2011aj,Collaboration:2010eza}, 
four jets~\cite{FourATLAS, FourCMS}, six jets~\cite{CMSthreejets, SixCDF,ATLAS:2012dp}, eight jets \cite{EightCMS} and 
up to ten jets for the semi-classical black holes
searches~\cite{Chatrchyan:2012taa, BlackHoles2012}. 

A  challenge in searches for new phenomena in multijet
final states is  the
prohibitively large QCD multijet 
background.
A range of techniques is then required in order
to identify  particular categories of
jets,  making it possible to reduce this background. 
Among those that have been validated and applied 
to searches, one can mention $b$--tagging~\cite{CMS:2012yf,atlasb1,atlasb2}, 
jet shapes for
quark/gluon and other flavour identification studies~\cite{qgid,Aad:2012ma} 
and jet substructure 
tools~\cite{VVboosted,Aad:2012meb,cms-subjets,ATLAS:2012am,cms-subjets2,atlassubjets,atlassubjets2}. 
Stringent constraints
on a variety of new physics models have been obtained this way,
with many more expected with the full 2011-2012 dataset.

From the kinematic point of view, 
the most common scenario is that of a heavy resonance $X$ produced
in the  s-channel which then decays back into 
a pair of quark or gluon jets. 
However, there 
is a large class of models where paired production of
resonances dominates, processes of the form
$pp\to X \to 2Y \to 4{~\rm partons}$, with $Y$ being
another massive particle.
The mediator $X$ of this production 
might be an
exotic particle from a new strongly coupled sector,
or a resonance from extra-dimensions, such as a massive graviton or a radion. 
The
$Y$ resonance could be either some BSM particle (sparticles in R-parity-violating supersymmetry,
colorons~\cite{coloron}, axigluons~\cite{Antunano:2007da}) or 
some SM particle ($W,Z$ or Higgs)
that subsequently decays into quarks and gluons.

These generic four parton processes
lead to very distinct final state signatures depending on the interplay
between the masses of the two intermediate resonances, $M_X$ and
$M_Y$.
 If the mass ratio is large, $M_X\gg M_Y$, the $Y$ resonances
 will be produced very boosted, and typically the decay products
of each of the two $Y$ resonances will be collimated into a single {\it fat}
jet. On the other hand, for $M_X\sim 2 M_Y$, the $Y$ resonances
will be produced nearly at rest, decaying into four well
separated jets. Existing searches assume either the highly
boosted or fully resolved regimes, and by doing so exclude
a potentially large region of the New Physics parameter space. 

It is the goal of this paper to design a jet reconstruction
and analysis strategy that
can be applied simultaneously to the boosted and resolved regimes. 
This will be achieved by merging the boosted-regime strategies, based
on jet substructure techniques, with a suitable  strategy for the resolved 
four-jet regime, based
on dijet mass pairings, together with a smooth interpolation between the 
two limits.
Such a strategy has the potential to make the experimental searches more
efficient and  allow a wider range of BSM models to be
probed within the same common analysis.

The approach that we will present here is fully general and model-independent,
assuming only that resonances are pair produced and then decay hadronically,
with no constraint on the absolute masses: indeed, at the parton level
the problem turns out to be scale independent, and the dynamics are
completely determined by the mass ratio $r_M\equiv M_X/2M_Y$. 
Of course
additional QCD radiation and confinement on the one hand and experimental
cuts on the other break this scale invariance, but we will see
that the general qualitative results are robust.

To provide a realistic application of our technique, we will examine
resonant  Higgs pair production, recently 
studied as a promising
probe of New Physics 
scenarios~\cite{Kribs:2012kz,Dolan:2012ac,Contino:2010mh}.
We will therefore derive model independent limits on BSM 
resonant Higgs pair production in the $4b$ final state.
We will then apply these bounds 
 in the context of warped extra
 dimensional
models, where
 Higgs pair production is mediated by either a spin zero
(radion) or spin two (massive Kaluza-Klein graviton) resonance.
We will show that a wide range of the parameter space
of the radion and massive KK graviton scenarios
 can be covered by present and future LHC data, and
that despite the
overwhelming QCD background, the $4b$ final state has enough sensitivity
to provide a useful handle in searches for enhanced
Higgs pair production at the LHC.

The outline of this paper is as follows. 
We begin in Sect.~\ref{sec:jetalg} by introducing
the general search strategy for pair produced resonances
that can be applied simultaneously to the boosted
and resolved regimes.
In Sect.~\ref{sec:models} we review the
theoretical models for resonant Higgs pair production 
in warped extra dimensions
scenarios. 
Then in
Sect.~\ref{sec:results} we apply the jet reconstruction strategy 
both to signal events and to the QCD multijet background, and
explore the potential for new physics searches in the $2H\to 4b$
channel.
In Sect.~\ref{sec:conclusions} we conclude and outline future
developments. 

\section{Scale-invariant resonance tagging}
\label{sec:jetalg}

Multijet signatures have long been recognized as an important
channel for Beyond the Standard Model searches at
hadron colliders~\cite{Chivukula:1991zk}. 
The main difficulty
in these channels is how to tame the overwhelming QCD multijet background.
Searches in multijet final states 
are commonly separated into boosted and resolved regimes.
An example of the former arises when light partons are produced from 
the decay of a heavy resonance.
Recently developed jet substructure techniques, reviewed for example
in~\cite{Abdesselam:2010pt,Altheimer:2012mn},
make it possible to substantially
improve the discrimination power in the boosted regime.
At the LHC the advent of jet substructure methods has made it
possible to study boosted production of the heavy Standard Model
particles, like $W$ and $Z$ bosons and top quarks, for $\sqrt{\hat{s}}$, the
centre-of-mass energy of the hard process, above 1 TeV.
Searches for new physics such as resonant production
of $VV$ or $t \bar{t}$~\cite{Chatrchyan:2012ku,Atlas:2012txa,Chatrchyan:2012cx,Aad:2012raa} or searches for boosted supersymmetric
particles and colored scalars~\cite{ATLAS:2012dp,FourATLAS} have also benefited from these
developments.

\begin{figure}[h]
\centering
\includegraphics[scale=0.33]{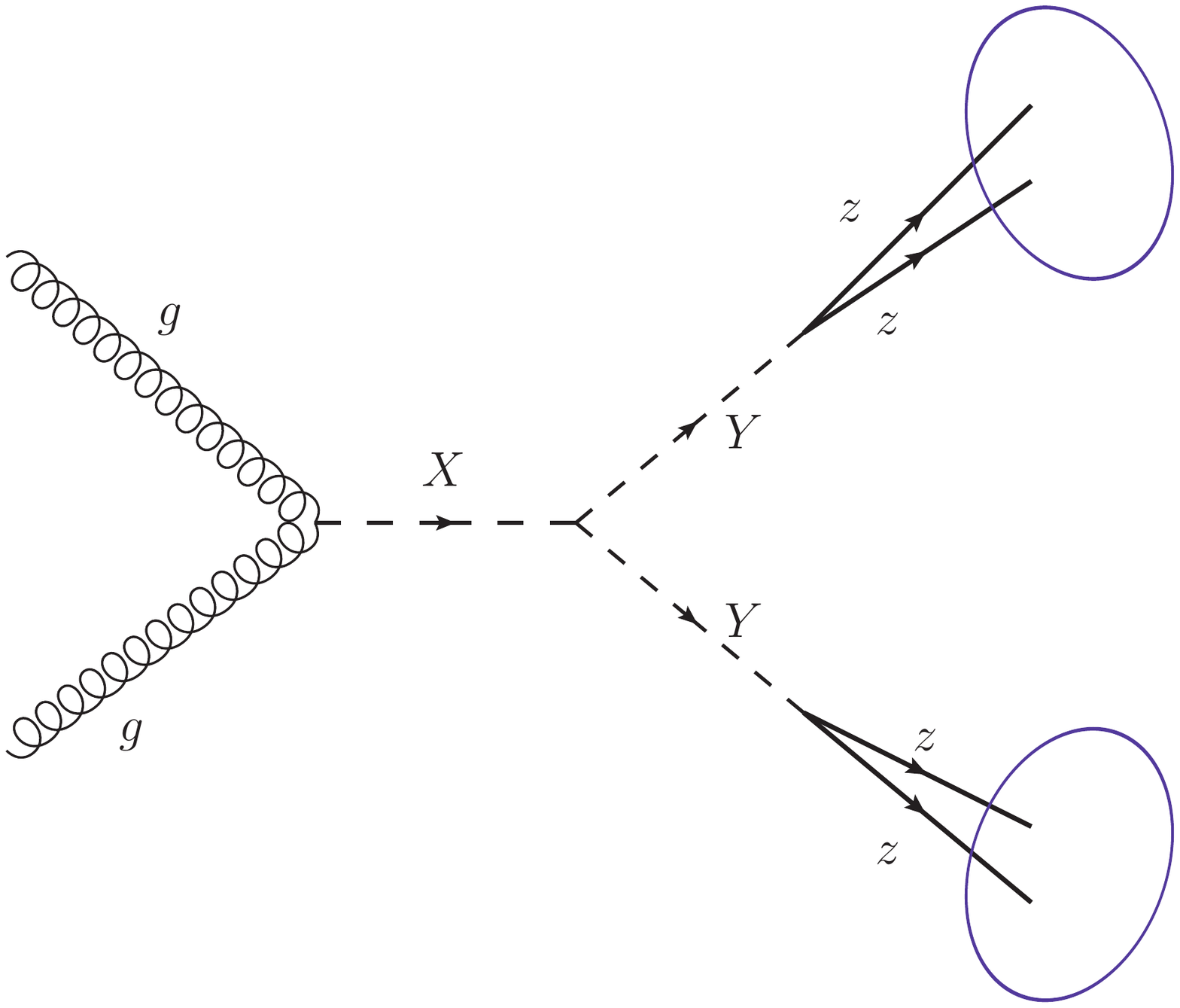}
\includegraphics[scale=0.35]{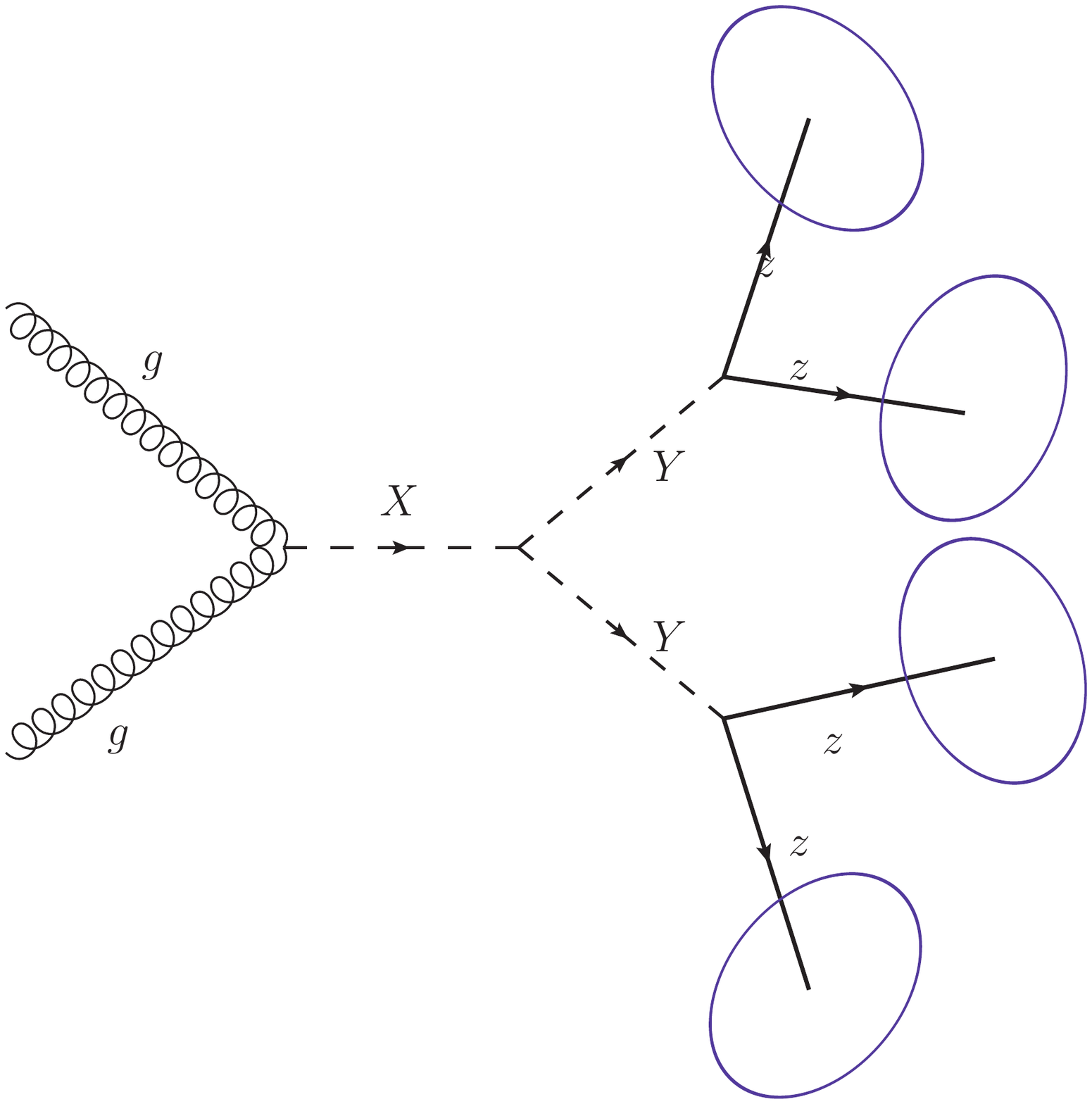}
\vspace{-25mm}
\caption{\small   Schematic diagrams
for the generic process $pp \to X \to 2 Y \to 4 z$ 
in the boosted (left plot) regime, corresponding
to large values of the mass ratio 
$r_M=M_X/2M_Y$, and in the resolved (right plot) regime,
corresponding to small values of $r_M$.
}
\label{fig:toy}
\end{figure}

In this section we introduce a general strategy for jet reconstruction
designed for searches of pair-produced resonances in fully hadronic
final states, which is simultaneously suitable for both the highly 
boosted and the fully resolved regimes and that smoothly 
interpolates between them.
The generic process we are interested in is the $s$-channel production of 
a resonance $X$ which then decays into a pair of
resonances $Y$,
 which in turn each decay into a pair of 
light Standard Model particles, labeled $z$,
\be
\label{eq:procbasic}
pp \to X \to 2 Y \to 4 z \, .
\ee
The ratio between the masses of the $X$ and $Y$ resonances
will determine the degree of boost of the $Y$ resonances
and consequently the angular distribution of their
decay products
$z$ that will be observed in the detector.
At parton level, neglecting the mass of the final state
particles $m_z$, the problem is scale invariant and is characterized
by a single dimensionless variable, denoted by
\be
\label{eq:rm}
r_{M} \equiv \frac{M_X}{2M_Y} \, ,
\ee
which is simply the boost factor
 from the $Y$ rest frame
to the $X$ rest frame.
In the  highly boosted regime, $r_M\to \infty$, while in 
 the fully resolved regime, where the intermediate resonances
$Y$ are produced at rest, $r_M=1$.
 Schematic diagrams
for the boosted and resolved topologies are shown in 
Fig.~\ref{fig:toy}.

If we assume that the heavy
resonance $X$ is produced at rest, so that the laboratory and center-of-mass reference frames coincide,
we can parametrize  
the four momenta of the 
$X\to YY$ decay with the  
convention that $P=\lp p_{T,x},p_{T,y},p_L,E\rp$. 
We then have
\bea
P_{X} &=& (0, 0, 0, M_{X}) \, , \\
\label{eq:4vect}
P_{Y_1} &=& \frac{M_X}{2}\lp \beta_Y \cdot \sin \theta^*_Y \cos \phi^*_Y ,  \beta_Y \cdot \sin \theta^*_Y \sin \phi^*_Y, \beta_Y \cdot \cos \theta^*_Y, 1\rp \nonumber \, , \\
P_{Y_2} &=& \frac{M_X}{2}\lp -\beta_Y \cdot \sin \theta^*_Y \cos \phi^*_Y, -\beta_Y \cdot \sin \theta^*_Y \sin \phi^*_Y , -\beta_Y \cdot \cos \theta^*_Y, 1\rp \nonumber \, ,
\eea
where $Y_1$ and $Y_2$ are the two decay products of the $X$ particle,
$\theta_Y^*$ is the angle of $Y_1$ with respect to the beam, and
$\phi^*_Y$ is the azimuthal angle.
The boost parameters from the laboratory frame to the rest frame of the $Y$ particles are given by
\bea
\beta_Y &=& |\vec{P}_Y|/E_{Y} = \sqrt{1-1/r_M^2} \, ,\\
\gamma_Y &=& 1/\sqrt{1-\beta^2_Y} = r_M \, .
\label{eq:boost}
\eea
As one can see, the boost of the $Y$ particles, $\beta_Y$, is independent of
the absolute masses of  $X$ and $Y$, and depends only on their ratio.
It is in this sense that  we can consider that
the problem at hand is scale invariant: $\gamma_Y$ does not
depend on any absolute mass scale.

Fig.~\ref{fig:toy} suggests that depending
on the value of the mass ratio $r_M$ the search
strategy should be 
different. For large $r_M$,
the resonances $Y$ will be
very boosted, and thus the angular distances of 
 their decay products will be small,
  while for low $r_M$ the four final
state particles will be well separated. 
Since we are mostly interested in the case in which
the final state particles $z$ are QCD partons, quarks
or gluons, we will end up either with two {\it fat}
jets (in the boosted regime), four well separated
jets (in the resolved regime) or one fat and two separated jets (in the
intermediate regime). 
Fat jets
are jets for which the substructure pattern
is unlikely to have arisen from QCD radiation.

Given that in general we do not
have information on the masses of the intermediate
resonances, we don't know {\it a priori} in which of the two
regimes we will find ourselves, and
it would be
beneficial to have a search strategy that
simultaneously explores all possibilities. 
It should exhibit
reasonably homogeneous efficiencies and background
mistag rates for any value the mass ratio $r_{M}$
within the physically allowed range.
Below, we will present such a combined strategy that
simultaneously explores the boosted and resolved regimes.

\label{sec:jettag}

In order to validate the performance of the strategy
that we will propose
we have generated events for the generic process
Eq.~(\ref{eq:procbasic})
using a toy Monte Carlo  simulation.
The heavier resonance $X$ is assumed
to be produced at rest in the laboratory frame, justified by the
fall-off at large masses of parton luminosities~\cite{Ball:2012wy},
and to decay into the two intermediate
resonances $Y$ with a homogeneous angular distribution, as if
it were a spin-zero particle. 
The massless
decay products of the $Y$ resonance decay are also assumed to
decay isotropically in the $Y$ rest frame.
In this toy simulation the possible widths of the intermediate resonances
are neglected, as well as the masses of the final
state particles $m_z$.

In view of the later applications to Higgs pair
production, we will set $M_Y=125$ GeV and vary $M_X$ 
in a wide range, although it should be clear that at parton
level the event classification will depend only on the
ratio $r_{M}$.
For each of the $M_X$ values in the range from 250 GeV (resolved
regime, $r_{M}=1$) to 5 TeV (highly boosted regime, $r_{M}=20$), we
have generated 50K toy MC events.

To study the performance
of the jet reconstruction strategy for a realistic collider
environment, the parton
level events from the toy Monte Carlo have been showered 
and hadronized
with
{\tt Pythia8}~\cite{pythia}, version 8.170. 
We have done this for
LHC centre of mass energies of 8 and 14 TeV, 
and we include also underlying event and multiple
interactions with the
default tune 4C of {\tt Pythia8}.
 Initial state radiation
has been modeled assuming that the resonance is produced
in the gluon-gluon channel\footnote{This is a good approximation
for the warped extra dimensions models that we will
introduce as benchmark scenarios in the next section.}.
Parton and hadron-level events were then clustered with the anti-$k_t$ jet 
algorithm~\cite{Cacciari:2008gp} with radius of $R=0.5$.
Such small radii ($R=0.5$ for CMS, $R=0.4$ for ATLAS) are used
in most experimental multijet analysis.

No additional cuts will be applied to the reconstructed jets
at the parton level, so as to avoid 
introducing any explicit breaking of scale invariance.
On the other hand, at hadron-level it becomes necessary to
introduce additional kinematic cuts, which
explicitly break scale invariance. 
In this section we will adopt the following set of basic 
kinematic cuts for jets in hadron-level events:
\be
p_{T,\rm jet}^{\rm min} \ge 25~\GeV, \qquad |y^{\rm jet}|\le 5\, , \qquad H_T\equiv \sum_{\rm jets} p_{T,{\rm jet}} \ge 100~\GeV \, . \label{eq:cuts}
\ee
In the $H_T$ variable the sum goes over
the four leading jets of the event above the $p_T^{\rm min}$ cut. 
These loose cuts have a very limited effect on the selection
efficiencies except at the smallest values of $r_M$.\footnote{
When presenting our final results in 
Sect.~\ref{sec:results} we will adopt more realistic cuts, in line with
those of
typical LHC searches.
}

\begin{figure}[t]
\centering
\includegraphics[scale=0.37]{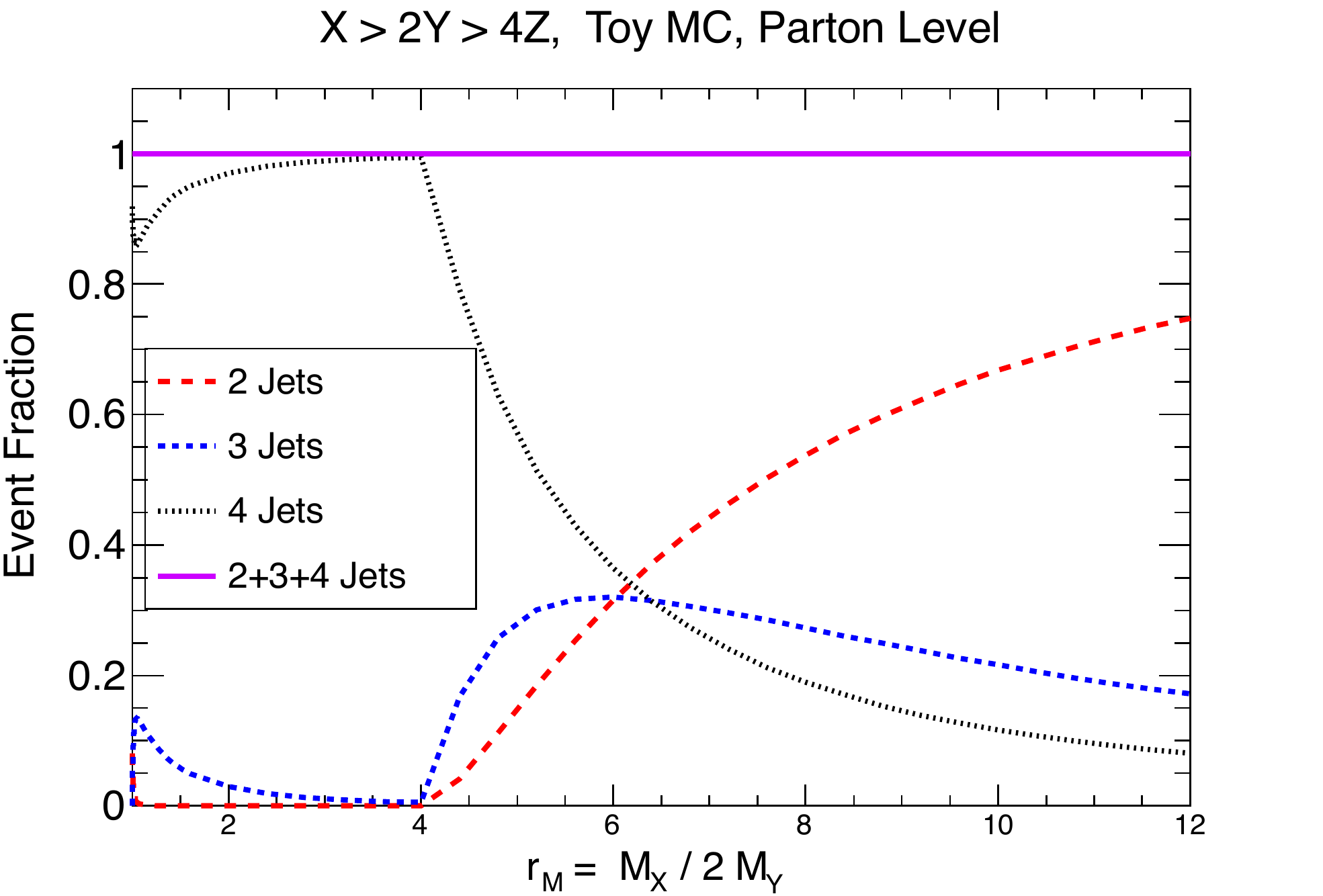}
\includegraphics[scale=0.37]{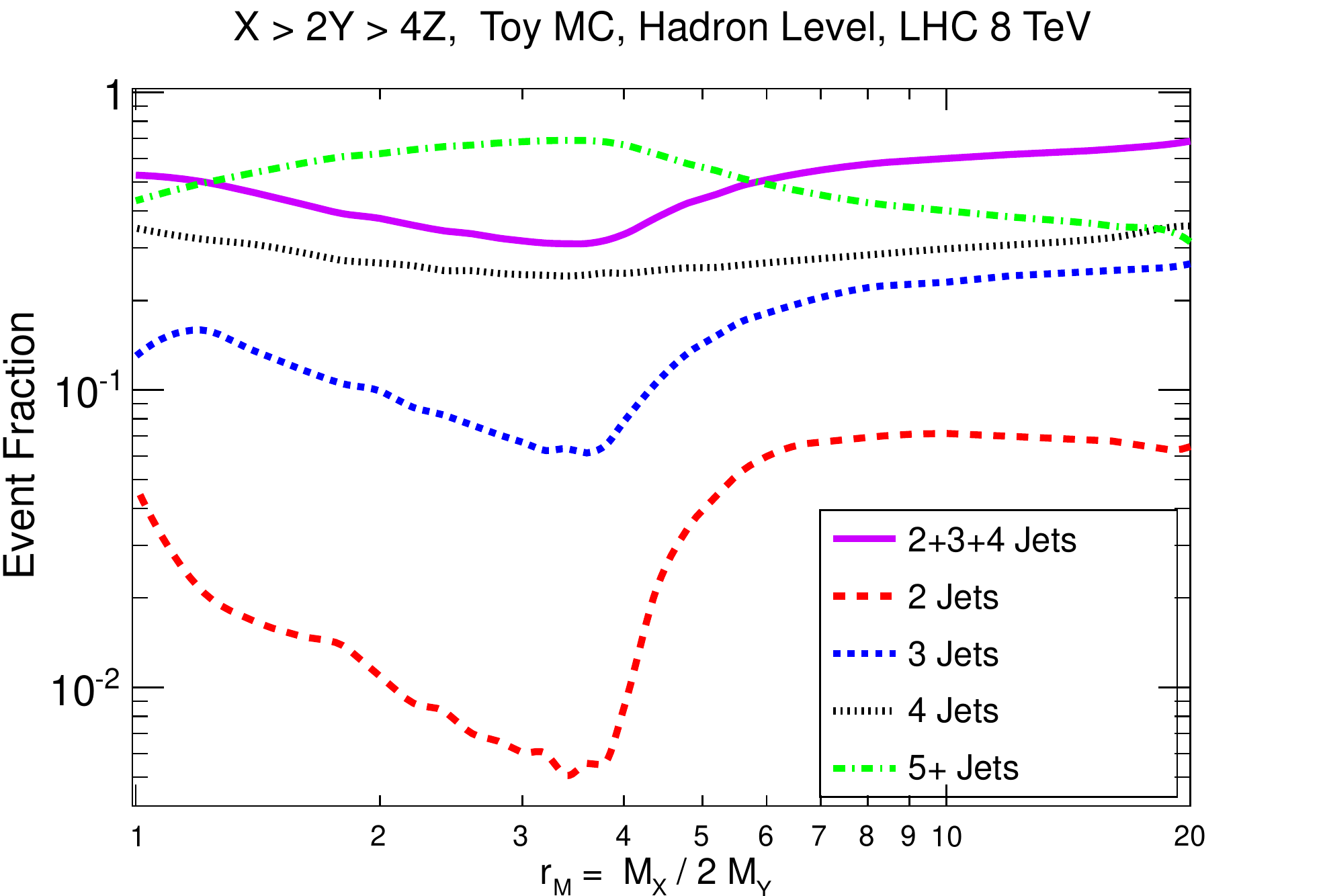}
\caption{\small  Left plot: the fraction of events with a given
number of reconstructed jets,
as a function of the resonance mass ratio
$r_{M}$ (Eq.~(\ref{eq:rm}), for parton-level toy Monte Carlo events.
No cuts have been applied to the final state particles. Right
plot: the same at hadron-level, with the basic
cuts Eq.~(\ref{eq:cuts}) applied. 
Note that at parton
level the only possible topologies are two-, three- and four-jet
events, so their sum is equal to the total number
of events.
A higher density of mass points has been used in the left-hand figure
than in the right-hand figure, and for $r_M \lesssim 1.5 $ only the
left-hand figure gives a faithful representation of the structures
that are present.  }
\label{fig:njets}
\end{figure}

In order to identify the three different regimes, boosted,
resolved, and mixed, 
useful information is provided by evaluating the fraction
of events with a given reconstructed jet topology.
We show the relative fractions of the different jet topologies
in  Fig.~\ref{fig:njets}
for both parton and hadron-level, as well as
the sum of events with two, three and four jets. 
At parton level, by construction, events can only have between two and
four jets, and we clearly see that four-jet events dominate at low
$r_M$, two-jet events in the boosted regime for large $r_M$, with the
three-jet case in between.\footnote{For $r_M=1$, symmetry
  considerations mean that the 3-jet rate is identically zero at
  parton level.}
At hadron-level both the shower and the kinematic cuts break scale
invariance, and now we can have events with fewer than two jets, at
very low $r_M$, due to the basic cuts Eq.~(\ref{eq:cuts}), and with
more than four jets due to parton radiation. 
Note though that
still between 30\% and 50\% of the events,
depending on the value of of $r_M$,  have between two and four jets.

Given that parton showering can significantly modify the number
of jets, an event classification based on the number of jets is not optimal
under realistic conditions. 
Instead, we use 
 an alternative classification, based
on the number of \emph{tagged} jets per event, that is, jets
that are found to have non-trivial substructure.
We will proceed as follows:
each of the two hardest anti-$k_t$ jets in the event is reclustered
using the Cambridge/Aachen algorithm~\cite{CA} with $R_{\rm sj}=1.3$
(where the subscript $sj$ means sub-jet),\footnote{ Any value
  substantially larger than the radius used for the anti-$k_t$ jets
  ($R=0.5$) would have been suitable. Our concrete choice facilitate
  the use of the analysis even with older versions of FastJet 
  (v2), which have a restriction $R<\pi/2$.} 
and processed
with the BDRS mass-drop tagger~\cite{Butterworth:2008iy}.
This tagger has two parameters $\mu$ and $y_{\rm cut}$. To determine
if a jet arises from a massive object, the last step
of the clustering for jet $j$ is undone, giving two subjets $j1$ and
$j2$, with $m_{j1} > m_{j2}$; 
if both are significantly lighter than the parent jet, $m_{j1} \le \mu
\cdot m_j$ and the splitting is not too asymmetric
\be
\label{eq:boosted-asymmetry-cut}
\frac{\min(p_{t,j1},p_{t,j2})^2}{m_{j}^2}\Delta R^2_{j1,j2} > y_{\rm cut} \, ,
\ee 
where $\Delta R_{j1,j2}$ is the angular separation between the two
subjets,
then $j$ is returned as the tagged
jet. 
Otherwise we replace $j$ with $j1$ and apply the unclustering to the
new $j$, repeating the procedure until we find a subjet for which the
mass-drop and asymmetry conditions are both satisfied.
If the procedure recurses to the point where it finds a
single-particle jet, then the jet is considered untagged.
We use the values $\mu = 0.67 $ and $y_{\rm cut} = 0.09$,
as in the original BDRS paper~\cite{Butterworth:2008iy}.

Our strategy is separated into two parts: the analysis chain,
which sets the flow of the event classification, and the 
quality requirements, which determine whether a given topology
is classified as a signal event or rejected as a background event.
We discuss these two parts in turn. 

\begin{figure}[t]
\centering
\includegraphics[width=\textwidth]{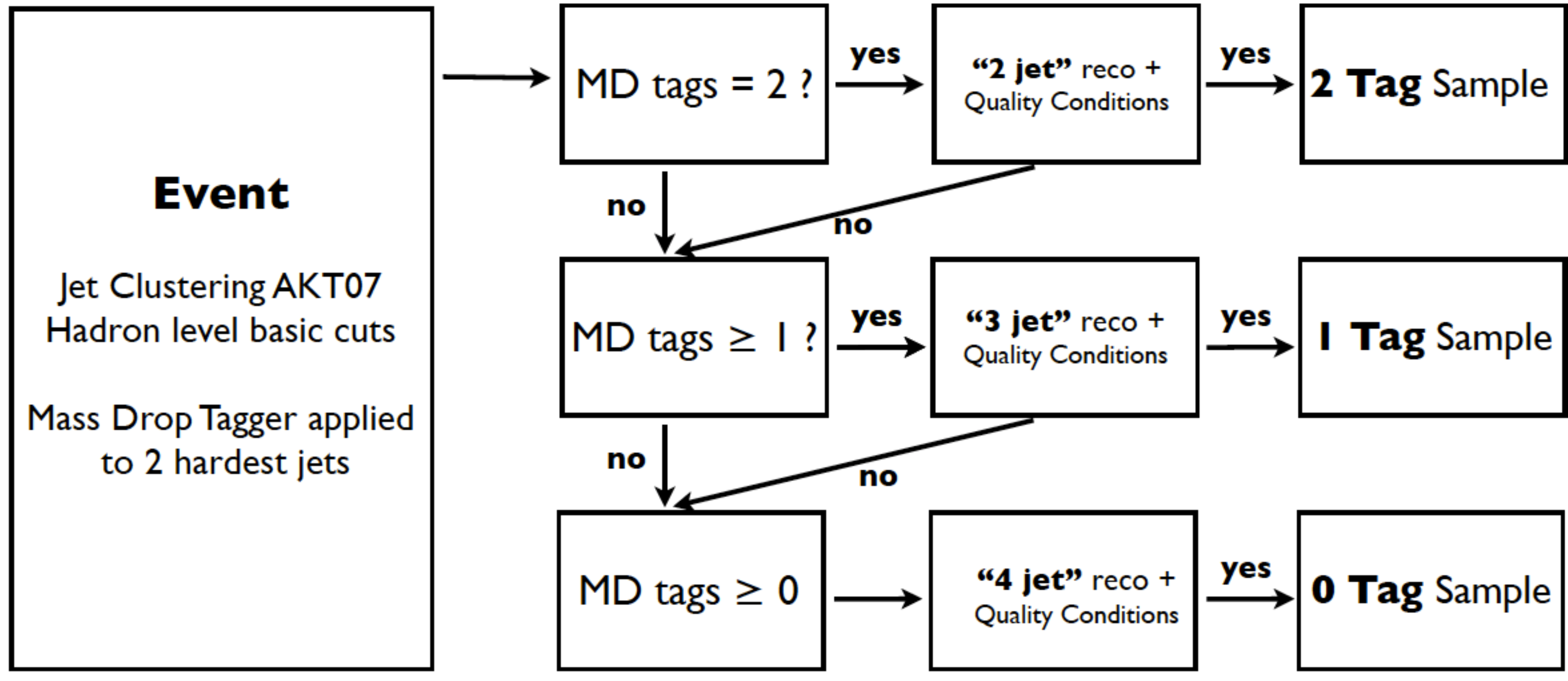}
\caption{\small Flow chart summarizing the basic structure of
the resonance-pair tagger algorithm. The quality conditions
are specified in the text. }
\label{fig:flowchart}
\end{figure}

{\bf Analysis chain}. 
We start by examining events
with at least two jets after basic cuts.
Summarized in the flow chart
of Fig.~\ref{fig:flowchart},
the analysis chain
depends on the number of mass-drop jet tags, that is, jets
that have been identified by the BDRS  tagger
to have an internal structure potentially not arising from
QCD radiation.

\begin{itemize}
\item If the two hardest jets in the event are mass-drop tagged, we
  examine if these two
  jets can be  identified as arising from the decay products of two
  boosted $Y$ resonances.
  This is established by verifying if the two $Y$ candidates satisfy
  the quality 
  conditions on their mass difference and angular separation listed
  below, in which case the event is assigned to the {\bf 2-tag}
  sample. 
\item If the event has a single mass-drop tag among the two
  hardest jets, or if the event had two mass-drop tags, but was not
  assigned to the  
  {\bf 2-tag} sample, then we examine whether the event can be
  classified as having an underlying three-jet topology,
  where the decays of one $Y$ resonance are collected into a single
  jet but not those of the other. 
Events
  with fewer than three jets after cuts are discarded.
  If there is a single mass-drop tag, the second $Y$ candidate is formed
  by adding the four-vectors of the other two hardest jets in the event.
  If there are two mass-drop tags but the event has been rejected in
  the {\bf 2-tag} category, we examine combinations whereby one of
  the tagged jets is taken to correspond to a first $Y$ candidate,
  while the other tagged jet is assumed to be a mistag and is combined
  with a third jet to make up the second $Y$ candidate.
  If
  the jet mass and angular quality
  requirements listed below are satisfied, the event is classified
  into the {\bf 1-tag} sample.
\item If no mass-drop tags are found in the event, or tags have been
  found but the event has failed to be assigned in either of the above
  categories, we examine the possibility of an underlying four-jet
  parton kinematics.
  Discarding events with fewer than four jets passing the basic cuts,
  we select the jet pairing such that the combination $ij$ and $kl$
of the five hardest jets in the event
  leads to jet masses $M_{ij}$ and $M_{kl}$ that minimizes the
  difference $|M_{ij}-M_{kl}|$, and use this pairing to reconstruct
  the two $Y$ candidates.\footnote{This choice, as compared to selecting only the four leading jets, improves the efficiency of the resolved configuration since often one gluon from large-angle initial-state radiation
 can have a larger $p_T$ than the four original $b$-jets. }  If these two candidates pass the mass and
  angular quality requirements given below, the event is classified as
  belonging to the {\bf 0-tag} sample.\footnote{Another approach
would be to determine if any pairings $ij$ and $kl$ of the
five hardest jets in the event satisfy the mass and angular quality requirements.
If there are multiple such pairings, then one uses the pairing that minimizes the
difference $|M_{ij}-M_{kl}|$ to reconstruct the two $Y$ candidates.
}
\end{itemize}

{\bf Quality requirements. } 
To identify the event as arising from the decay of the
$X$ resonance, Eq.~(\ref{eq:rm}), 
 additional mass and angular quality conditions are required,
which are essential to further suppress the QCD background.
Some of these requirements are designed so as to apply
similar
conditions to both the boosted and resolved topologies.

\begin{enumerate}
\item We require the masses of the two $Y$ candidates to be the same up to a
  given mass tolerance $f_m$, to account for experimental mass
  resolution, as well as mass smearing due to underlying event,
  hadronization and initial and final-state radiation:
  \be
  \Bigg| \frac{\lp m_{Y1}-m_{Y2}\rp}{\la m_{Y}\ra} \Bigg| \le f_m \, ,
  \ee
where $\la m_{Y}\ra$ is the average mass of the two reconstructed $Y$
resonances.
We assume in this work a fixed value\footnote{In realistic
  analysis, the typical mass resolution depends on the mass scale and
  jet kinematics. The details depend not just on the detector, but
  also on the experimental jet reconstruction techniques.} for the detector
mass resolution $f_m$ of
15\%~\cite{CMS:2012yf,Atlas:2012pu,Harris:2011bh}.
  This requirement cannot be made too stringent
  otherwise a large fraction of signal events would be 
discarded.
\item In the case in which the mass of the
  $Y$ resonance is known, the masses of the two $Y$ candidates
  must lie in a mass window around $M_Y$, where the width of the window
  is determined by the mass resolution of the detector.
  \be
  M_Y\lp 1-f_m\rp \le m_{Y1},m_{Y2} \le M_Y\lp 1+f_m\rp \ .
  \ee
  Since we will be considering Higgs pair-production, we will
  set $M_Y=M_H=125$ GeV in the following, though this requirement
  has a small impact in signal events, and is only relevant
  to suppress the QCD background. 
\item The separation in rapidity of the two $Y$ candidates
  must be smaller than some upper value, 
  \be
  \Delta y \equiv |y_{Y1}-y_{Y2}|\le \Delta y_{\rm max} \, ,
  \ee
  motivated by the fact that for a given mass of the $Y1, Y2$ system,
  background events, dominated by $t$-channel exchange, are enhanced
  in the forward region, while signal events, dominated by $s$-channel
  exchange, tend to be more central.
  We will take
  $\Delta y_{\rm max}=1.3$ in the following, a value optimized
  from the high mass dijet searches at the LHC~\cite{CMS:2012yf, Atlas:2012pu}.
\item Likewise, the separation in rapidity between the
two jets of a  $Y$ candidate in the resolved case, $y_{Yi,1}$
and $y_{Yi,2}$, with $i=1,2$,
  must also be smaller than some upper value, possibly different from before, 
  \be
  \Delta y \equiv |y_{Yi,1}-y_{Yi,2}|\le \Delta y_{\rm max}^{\rm res} \, ,
  \ee
  since for these kind of topologies, signal events will be produced
  closer in rapidity that QCD multijet production. 
We will take
  $\Delta y_{\rm max}^{\rm res}=1.5$ in the following, and we discuss below
the rationale for this choice.
\item To prevent excessively asymmetric configurations, whenever we have
  two resolved jets that correspond to a given $Y$-candidate,
  one with $p_T^{(1)}$ and the other with $p_T^{(2)}\le p_T^{(1)}$
  (in either the 1-tag sample or the 0-tag sample), we require
  \be
   p_T^{(2)} \ge y_{\rm cut } \cdot p_T^{(1)}  \, .
  \ee
  This cut plays a similar role as the asymmetry requirement in
  the BDRS mass-drop tagger, Eq.~(\ref{eq:boosted-asymmetry-cut}), but now in the case
  of resolved jets, and it helps reject events where a soft
  jet arises from final-state radiation (FSR).\footnote{
    To see the equivalence with the cut of
    Eq.~(\ref{eq:boosted-asymmetry-cut}), note that for reasonably
    small $\Delta R$, the mass of the $Y$ candidate is $m^2_j \simeq p_T^{(1)}
    p_T^{(2)} \Delta R_{12}^2$. Making use of the fact that
    $p_T^{(2)} < p_T^{(1)}$,
    Eq.~(\ref{eq:boosted-asymmetry-cut}) reduces to
    $p_T^{(2)}/p_T^{(1)} > y_\text{cut}$.
  }

\item With a similar motivation, for each two resolved
jets in a $Y$-candidate with mass $m_{Yi}$, with
$i=1,2$ we impose the
mass-drop condition on the masses of these two
resolved jets, $m_{Yi,1}$ and $m_{Yi,2}$, as follows
\be
{\rm max}\lp m_{Yi,1},m_{Yi,2}\rp \le \mu \cdot m_{Yi} \, , 
\ee
where $\mu$ is the same parameter as in the BDRS mass-drop
tagger.
Together with the asymmetry condition above, applying the mass
drop requirement also in the resolved jets ensure that the
same conditions hold for the three different possible
topologies, from the highly boosted to the fully resolved regimes.
Note, however, that for our default choice of $\mu=0.67$, the mass-drop
cut has only very limited impact on the final reconstruction efficiency.
\end{enumerate}
The values of the parameters used in our implementation
of the jet reconstruction strategy are summarized in 
Table~\ref{tab:jetparams}. 

\begin{table}[t]
\centering
\begin{tabular}{c|c|c|c|c|c}
\hline
\multicolumn{6}{c}{Jet Reconstruction}\\
\hline
$R$  &  $R_{\rm sj}$  &  $R_{\rm f}$  &  $n_{\rm filt}$  & $\mu$ & $y_{\rm cut}$ \\
0.5  &  1.3  &  0.3  &  3  & 0.67  & 0.09 \\
\hline
\multicolumn{6}{c}{ $\quad$}\\
\end{tabular}

\begin{tabular}{c|c|c}
\hline
\multicolumn{3}{c}{Basic cuts}\\
\hline
  $p_{T}^{\rm min}$  & $|y_{\rm max}|$ & $H_T^{\rm min}$ \\
  25 GeV & 5.0 & 100 GeV \\
\hline
\multicolumn{3}{c}{ $\quad$}\\
\end{tabular}

\begin{tabular}{c|c|c|c}
\hline
\multicolumn{4}{c}{Quality requirements}\\
\hline
$M_Y$     &  $\Delta y_{\rm max}$ & $\Delta y_{\rm max}^{\rm res}$  &  $f_m$ \\
125 GeV   &  1.3  & 1.5 &  0.15\\
\hline
\end{tabular}
\caption{\small Upper table: parameters that define the jet reconstruction
strategy, including mass-drop and filtering. Middle table: basic jet and
event
selection cuts. Lower table: parameters
of the quality requirements imposed on the tagged resonances.
See text for description.
\label{tab:jetparams}
}
\end{table}

In addition, in order to improve
on resolution, jet masses are
filtered~\cite{Butterworth:2008iy} as follows: the constituents
of each tagged jet are reclustered with a smaller radius $R_{\rm filt}={\rm min}\lp
\Delta R_{\rm sj,sj}/2,R_f\rp$, with $R_f=0.3$ and
 $\Delta R_{\rm sj,sj}$ the angular
distance between the two subjets after mass-drop in the boosted case.
Then only the three hardest subjets, $n_{\rm filt}=3$,
 are retained to account for at least one QCD emission.
The filtering procedure improves  mass resolution
of the reconstructed 
resonances~\cite{Butterworth:2008iy,Cacciari:2008gd} and makes
the procedure more resilient to soft radiation from the underlying
event and
 pile-up~\cite{Cacciari:2010te}.\footnote{In principle for the resolved
configuration, one could consider supplementing the analysis
chain with the inclusion of a large-angle radiation recovery
procedure, to improve on mass resolution, as 
advocated and used in~\cite{Cacciari:2008gd,Krohn:2009th,Chatrchyan:2011ns,Chatrchyan:2013qha,CMS:2012yf}. Such large-angle radiation recovery
procedure leads to the so called wide-jets in the CMS papers~\cite{Chatrchyan:2011ns,Chatrchyan:2013qha,CMS:2012yf}.  }

\begin{figure}[t]
\centering
\includegraphics[scale=0.37]{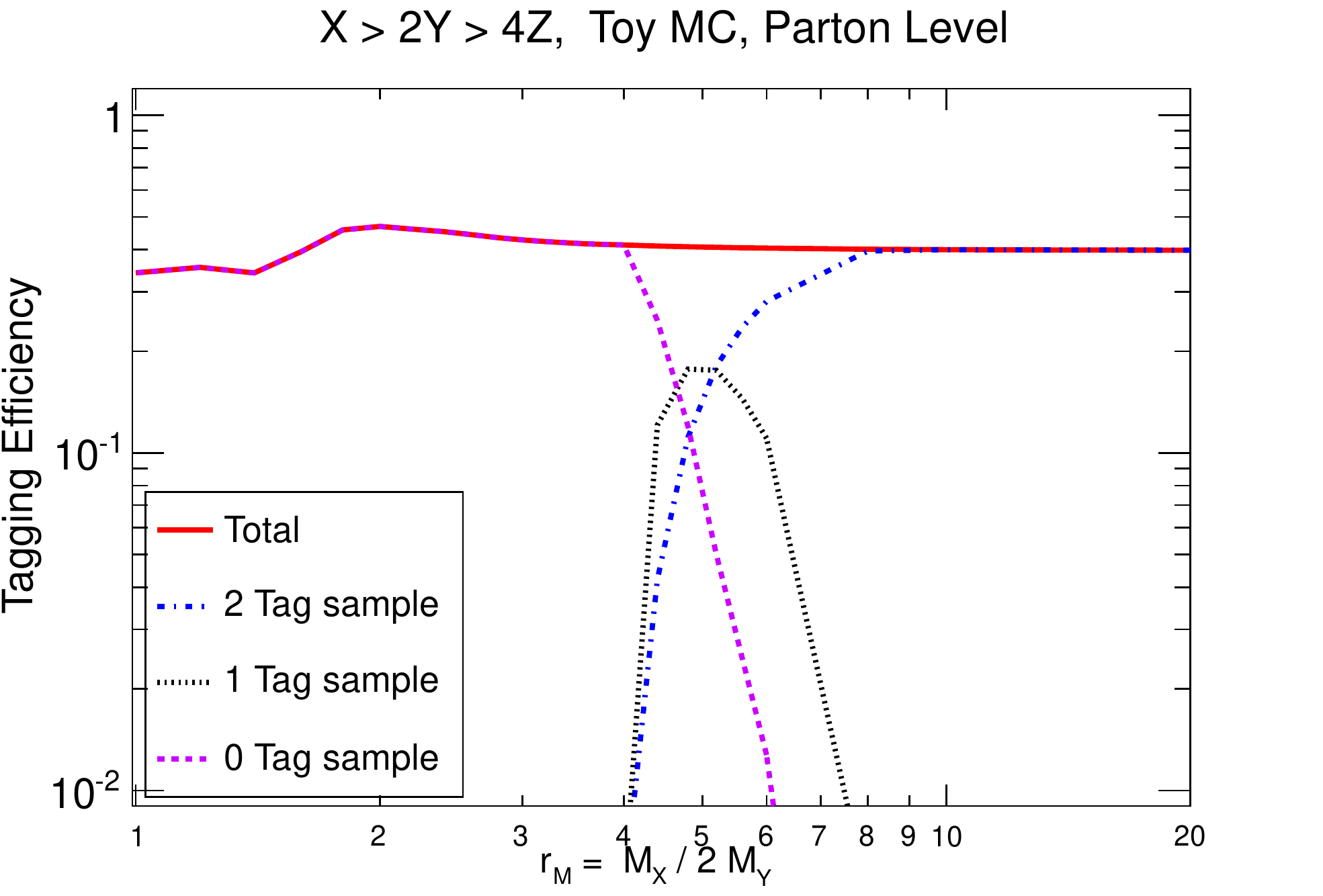}
\includegraphics[scale=0.37]{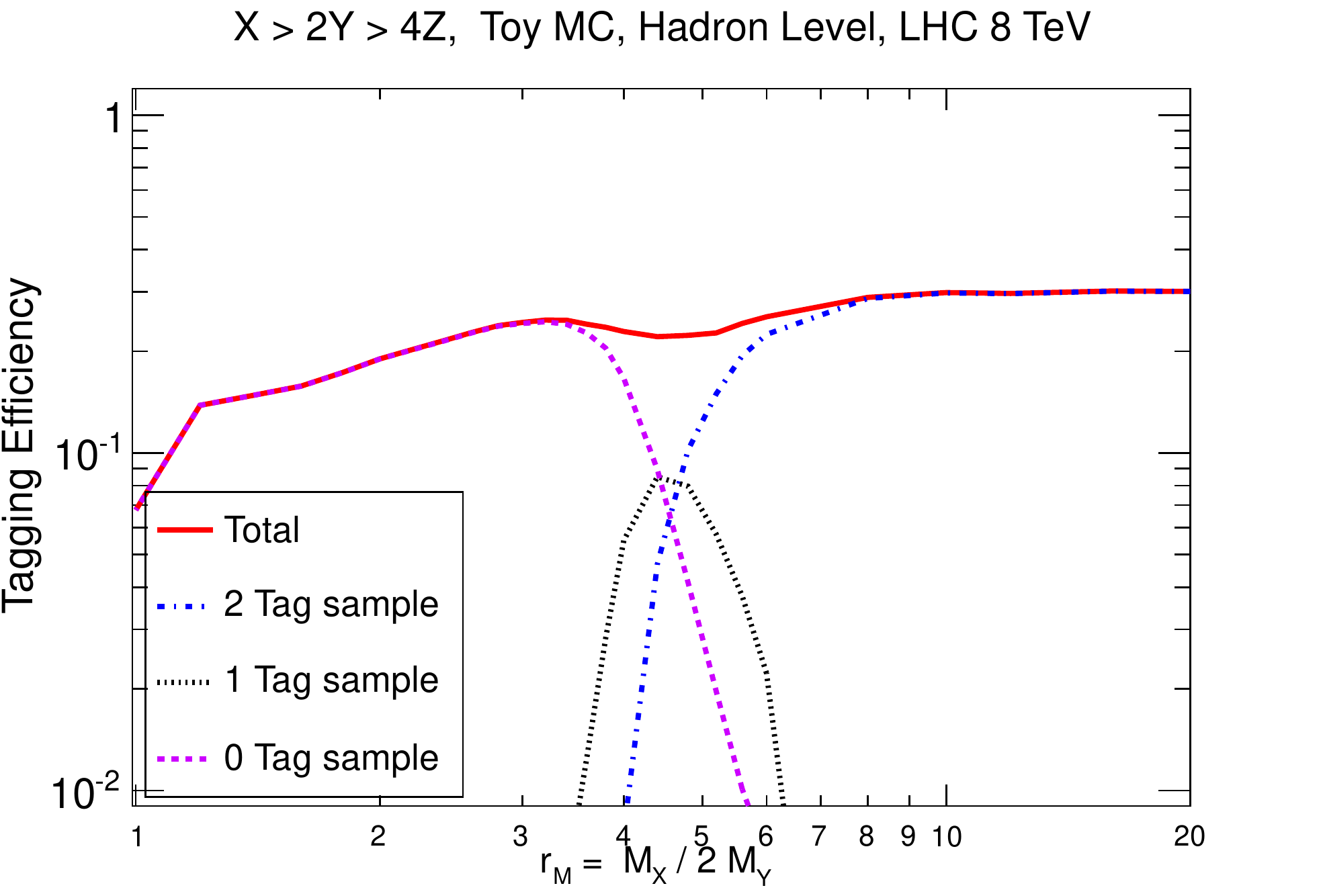}
\caption{\small  Left plot: The efficiency of the resonance
pair tagging algorithm as a function of resonances mass ratio
$r_{M}$ Eq.~(\ref{eq:rm}) for parton-level toy Monte Carlo events.
We show both the total efficiency and the break-up for the
2-tag, 1-tag and 0-tag samples.
No cuts have been applied to the final state particles.
Right plot: same for hadron-level events, which include
the basic jet selection cuts.}
\label{fig:efficiency-parton}
\end{figure}

This jet reconstruction strategy  has been implemented in 
 a code based on  {\tt FastJet3}~\cite{Cacciari:2011ma}, and we 
have processed the parton and hadron-level
toy Monte Carlo events through it.
We show in Fig.~\ref{fig:efficiency-parton} the efficiency of the resonance
pair tagging algorithm as a function of resonances mass ratio
$r_{M}$ for the parton and hadron-level toy Monte Carlo events.
We show both the total efficiency and the breakup of the efficiencies
corresponding to the 2-tag, 1-tag and 0-tag samples. 
The impact of the moderately loose selection cuts Eq.~(\ref{eq:cuts})
on the parton-level efficiencies
is negligible, and thus the differences between parton and hadron-level
arise from initial and final-state radiation.

At parton level, at low $r_M$, the 0-tag sample
dominates as expected from the resolved regime, while for
large $r_M$, the boosted regime, it is indeed the 2-tag sample
that dominates. 
The 1-tag sample is  important at intermediate
boosts. 
The combined efficiency is found to be rather flat in all
the mass range, between
30\% and 40\% for all mass values, showing that we are able to obtain
a reasonable tagging efficiency irrespectively of the degree of boost
of the $X$ resonance decay products. 
At hadron-level efficiencies
are somewhat 
lower due to additional parton radiation and underlying event at low masses, but still we obtain a reasonable tagging efficiency of between 20\% and
30\% in all the relevant range, approximately constant
for all topologies, except close to $r_M=1$.
 The 1-tag sample and the low $r_M$ 
0-tag sample are the ones most affected
by the transition from partons to hadrons.

Let us mention that the production threshold region close
to $r_M\sim 1$ is challenging from the jet reconstruction point of
view. First of all, there will be a substantial degree of
overlap between the decays products of the two Higgs bosons,
since the two are at rest, which leads to wrong mass pairings.
Second, it is quite frequent that large-angle initial-state
radiation (ISR) appears as
 additional jets, again confusing the pairing of the original jets.

\begin{figure}[t]
\centering
\includegraphics[scale=0.37]{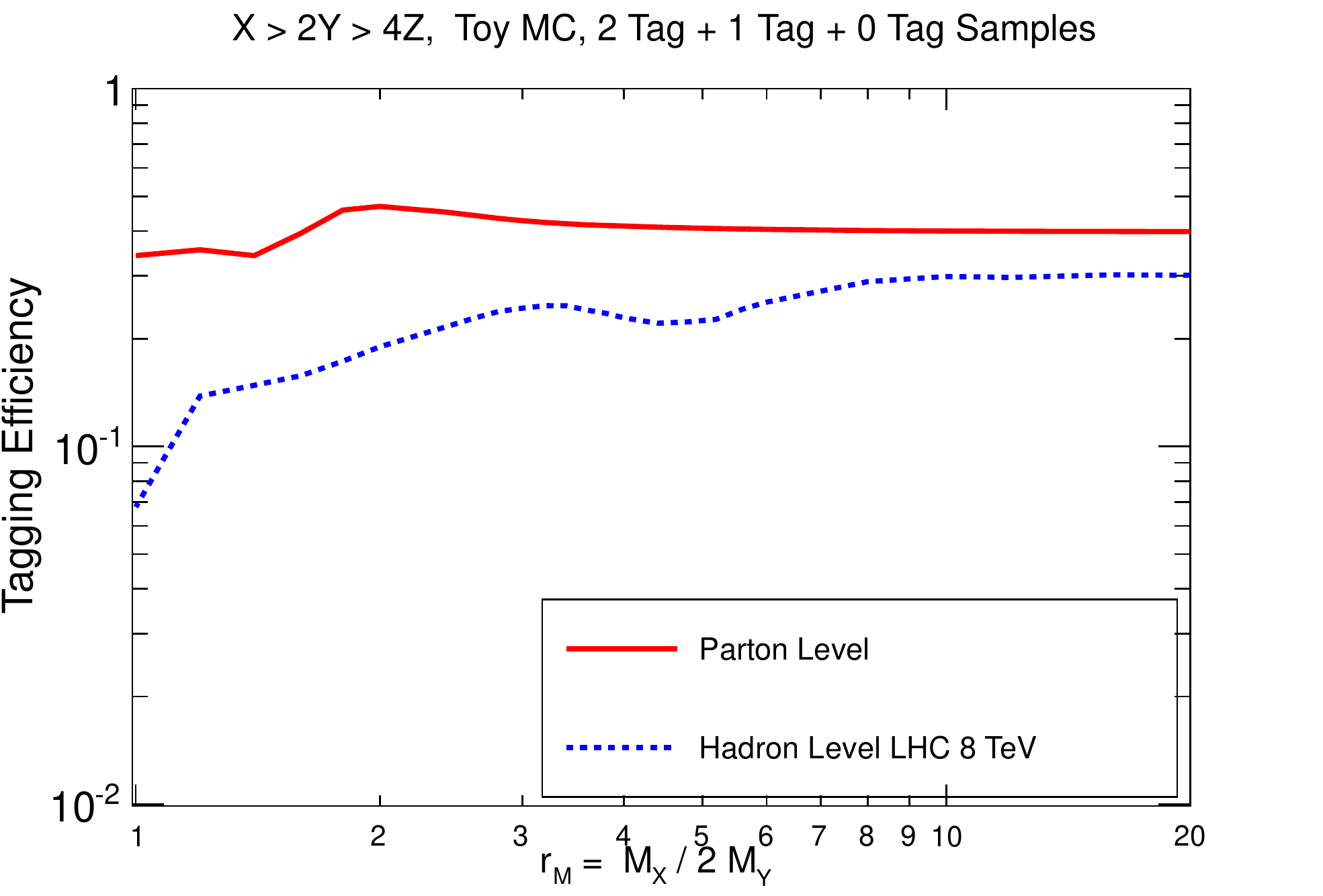}
\includegraphics[scale=0.37]{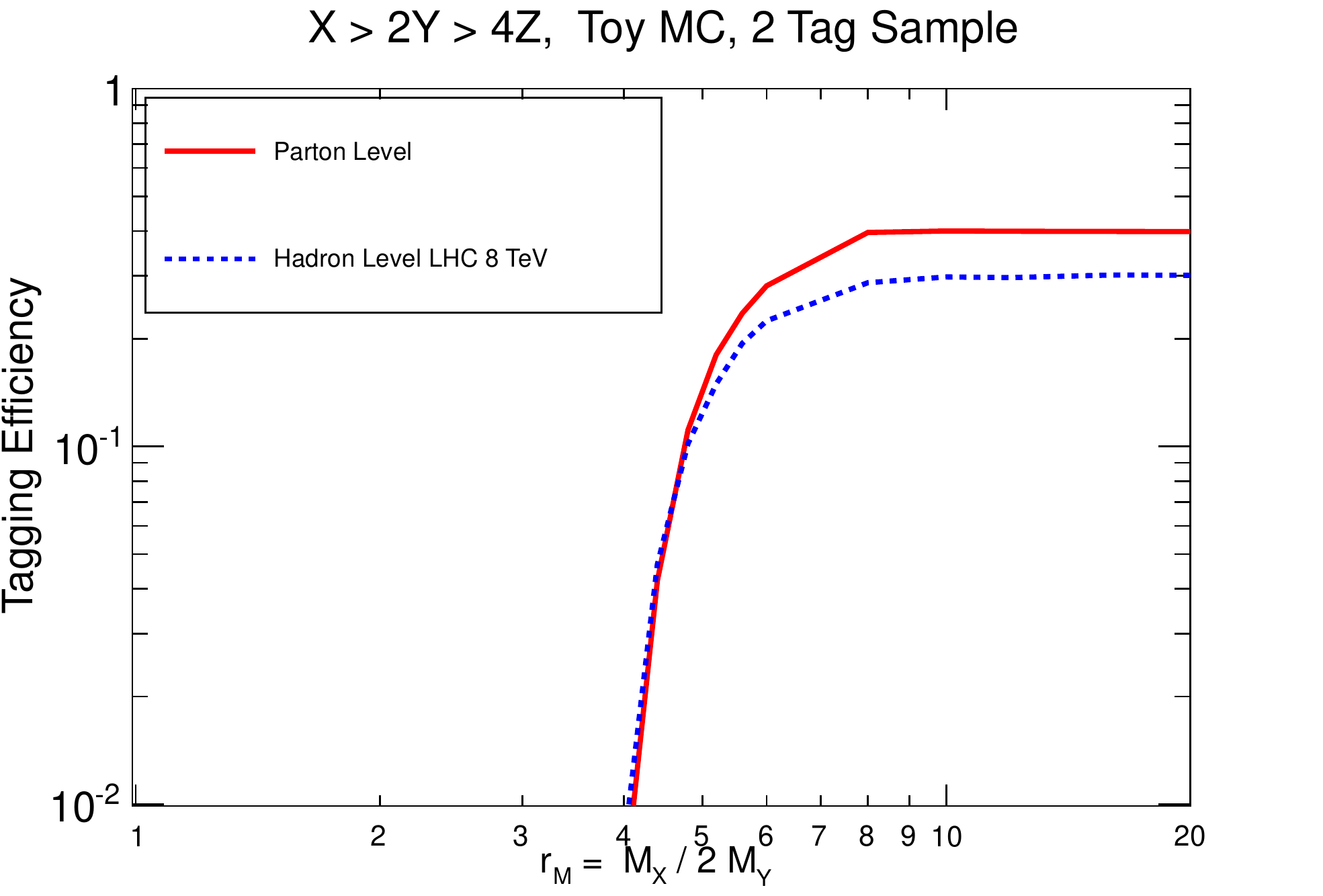}
\includegraphics[scale=0.37]{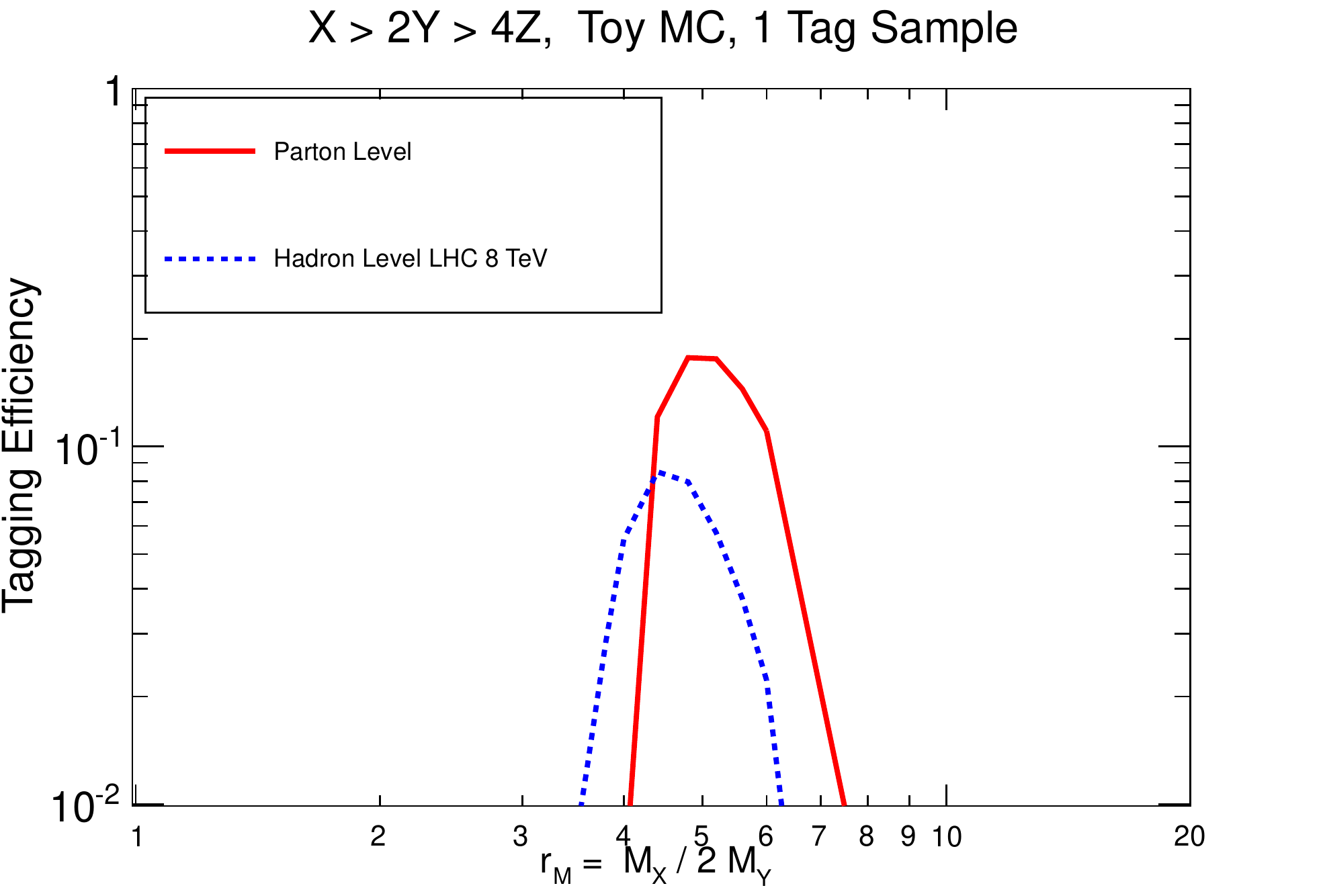}
\includegraphics[scale=0.37]{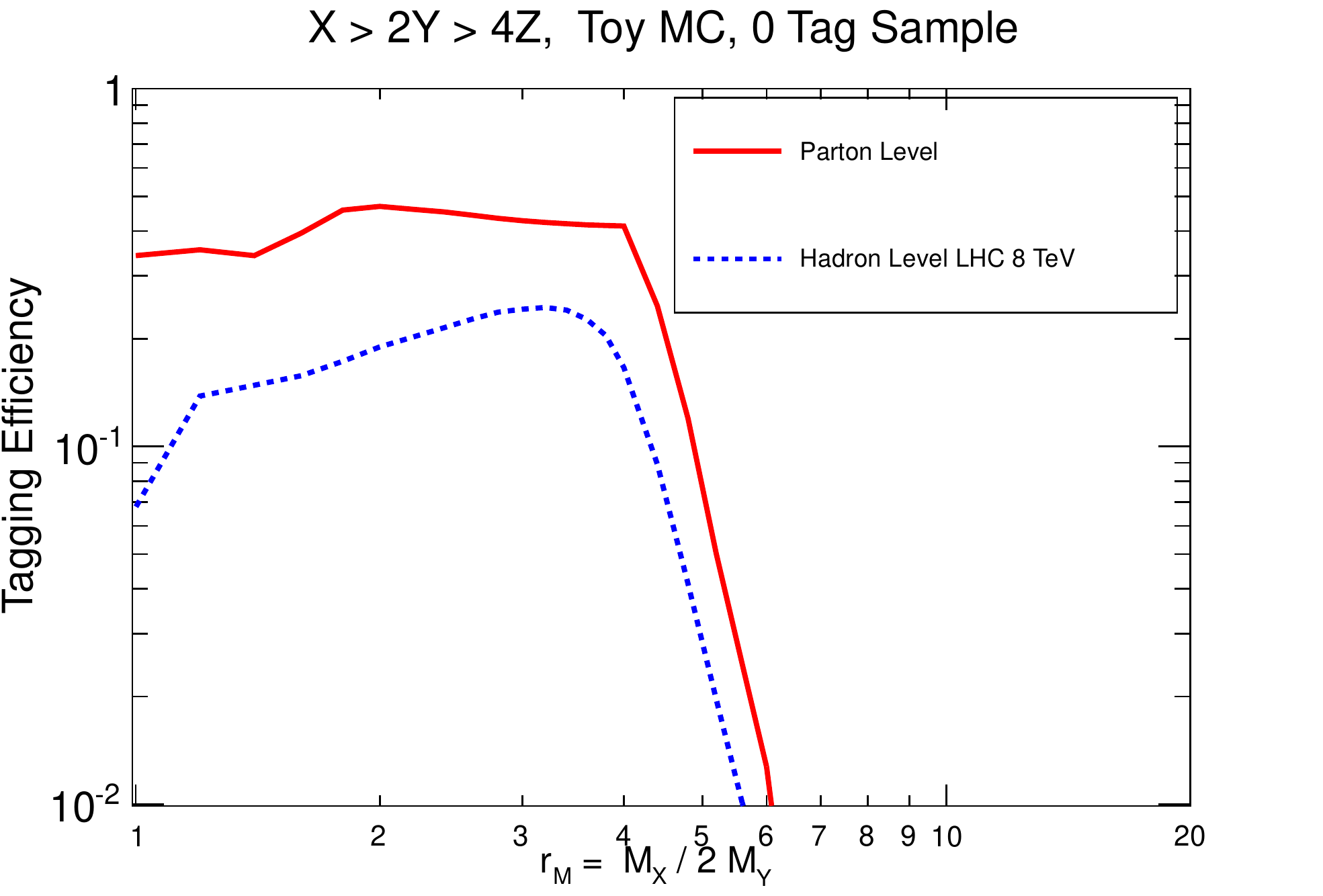}
\caption{\small  The efficiency of the resonance
pair tagging algorithm as a function of the resonance mass ratio
$r_{M}$, Eq.~(\ref{eq:rm}), for toy Monte Carlo events, comparing
parton-level and hadron-level results for LHC 8 TeV.
We show the total efficiency (upper left plot) and then the break-up for the
2-tag, 1-tag and 0-tag samples. The parton level results including
the basic selection cuts Eq.~(\ref{eq:cuts}) are very close
to the inclusive parton level case and thus not shown here.}
\label{fig:efficiency-parton+hadron}
\end{figure}

To help understand the efficiencies that we find,
let us recall that the asymmetry cut $y_{\rm cut}$ in the BDRS mass-drop tagger leads
to an upper bound on the efficiencies of signal events in the boosted
regime of approximately $\sim \lp 1-2y_{\rm cut}/(1+y_{\rm cut})\rp$
(this result is exact for the two-prong decay of a highly boosted scalar).
For the two tag sample at large $r_M$, we therefore expect
that at parton level the tagging efficiency is given by
\be 
\label{eq:2taglimit}
\epsilon_{\rm 2-tag}^{\rm lim} \equiv \epsilon_{\rm 2-tag}\lp r_M \gg 1 \rp=\lp 1-\frac{2y_{\rm cut}}{1+y_{\rm cut}}\rp^2 \cdot 
\frac{\exp(\Delta y_{\rm max})-1}{\exp(\Delta y_{\rm max})+1}  \sim 0.40 \, ,
\ee 
for our choice of parameters,
where
the last factor accounts for the contribution to the
total efficiency from the cut in $\Delta y$ between the two
$Y$--candidate jets.\footnote{The distribution of
$Y$ resonances is flat in $\cos \theta^*$, the decay angle of $Y$ in the
rest frame of the $X$ resonance, thus the cut
in $\Delta y_{\rm max}$ directly determines the maximum value 
of $\cos \theta^*_{\rm max}$ that will lead to resonance tagging.} 
This is exactly what is obtained in
Fig.~\ref{fig:efficiency-parton}. At hadron-level the efficiency
in the boosted regime is somewhat smaller due to the contamination
from initial-state radiation and the underlying event.

When we have a resolved $Y$ resonance candidate,
 the separation in rapidity between the
two jets must 
  be smaller than some upper value, $\Delta y_{\rm max}^{\rm res}$.
To determine the value of this cut, we note that in the small $R$ limit,
if such a cut is the only one applied to the final state,
the efficiency of the fully resolved case for $r_M \sim 1$ is given
by
\be
\epsilon_{\rm 0-tag}\lp r_M \sim 1 \rp = \lc \frac{\exp(\Delta y^{\rm res}_{\rm max})-1}{\exp(\Delta y^{\rm res}_{\rm max})+1}\rc^2 \, ,
\ee
so demanding that the efficiencies at low $r_M$ match the
asymptotic large $r_M$ value, Eq.~(\ref{eq:2taglimit}), 
we obtain
\be
\Delta y^{\rm res}_{\rm max} = \ln \lp \frac{1+\sqrt{\epsilon_{\rm 2-tag}^{\rm lim} }}{1-\sqrt{\epsilon_{\rm 2-tag}^{\rm lim} }} \rp \sim 1.5
\ee
for the default value of $y_{\rm cut}$ used in the mass-drop tagger algorithm.
With this choice, we can achieve at low $r_M$ the same efficiency
as at large $r_M$, at least in the parton-level case without
the basic kinematic cuts. 
Note that this cut ensures not just uniform signal efficiency, but it
also is useful for background rejection, especially in scenarios where
$M_Y/2$ is substantially larger than the jet $p_T^{\rm min}$ cut.

A final interesting comparison is that of the efficiencies
between parton and hadron-level, 
to gauge the robustness of our event classification based
on giving priority to the mass-drop tags.
This is useful
in order to understand the impact of parton showering and underlying event,
as well as of the basic kinematic cuts
in the tagging of the heavy resonances. Let us recall that the
only difference in the analysis chain between parton and hadron
level events are the basic cuts in Eq.~(\ref{eq:cuts}), and the
fact that in the 0-tag case we study the mass pairings of
the five leading jets.
Results are shown in Fig.~\ref{fig:efficiency-parton+hadron}. We show 
hadron-level results only for 8 TeV, those at 14 TeV are very similar.
The efficiency for the two tag sample is very similar
at parton and hadron-level, for all the values for $r_M$. 
The efficiencies for 0 and 1-tags sample have a similar shape but
a smaller magnitude, and the shape is somewhat shifted down
to lower $r_M$ values.
This small shift between the parton-level and hadron-level
efficiencies  is perhaps attributable to transverse boosts
induced by initial-state radiation.

\begin{figure}[t]
\centering
\includegraphics[scale=0.33]{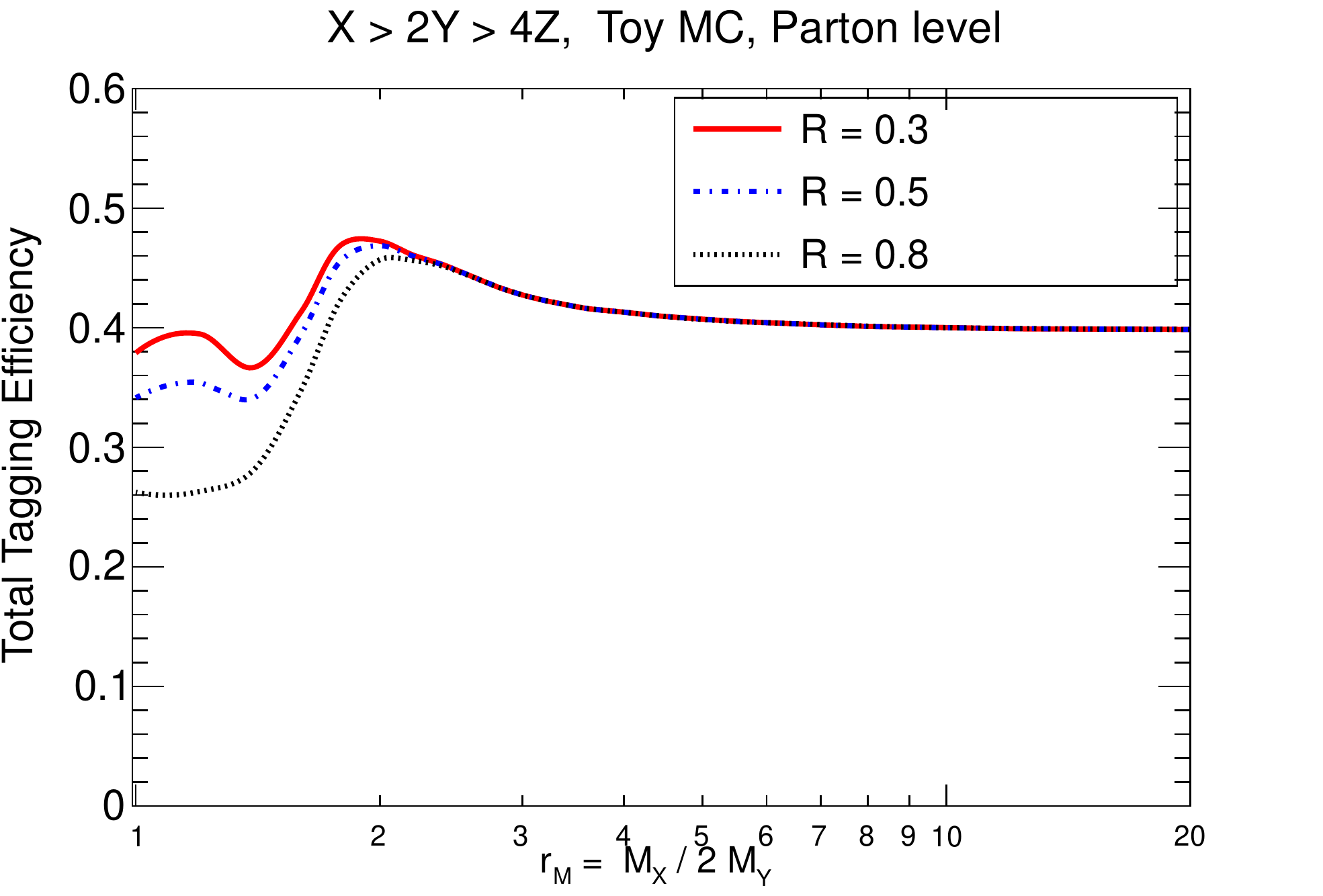}
\includegraphics[scale=0.33]{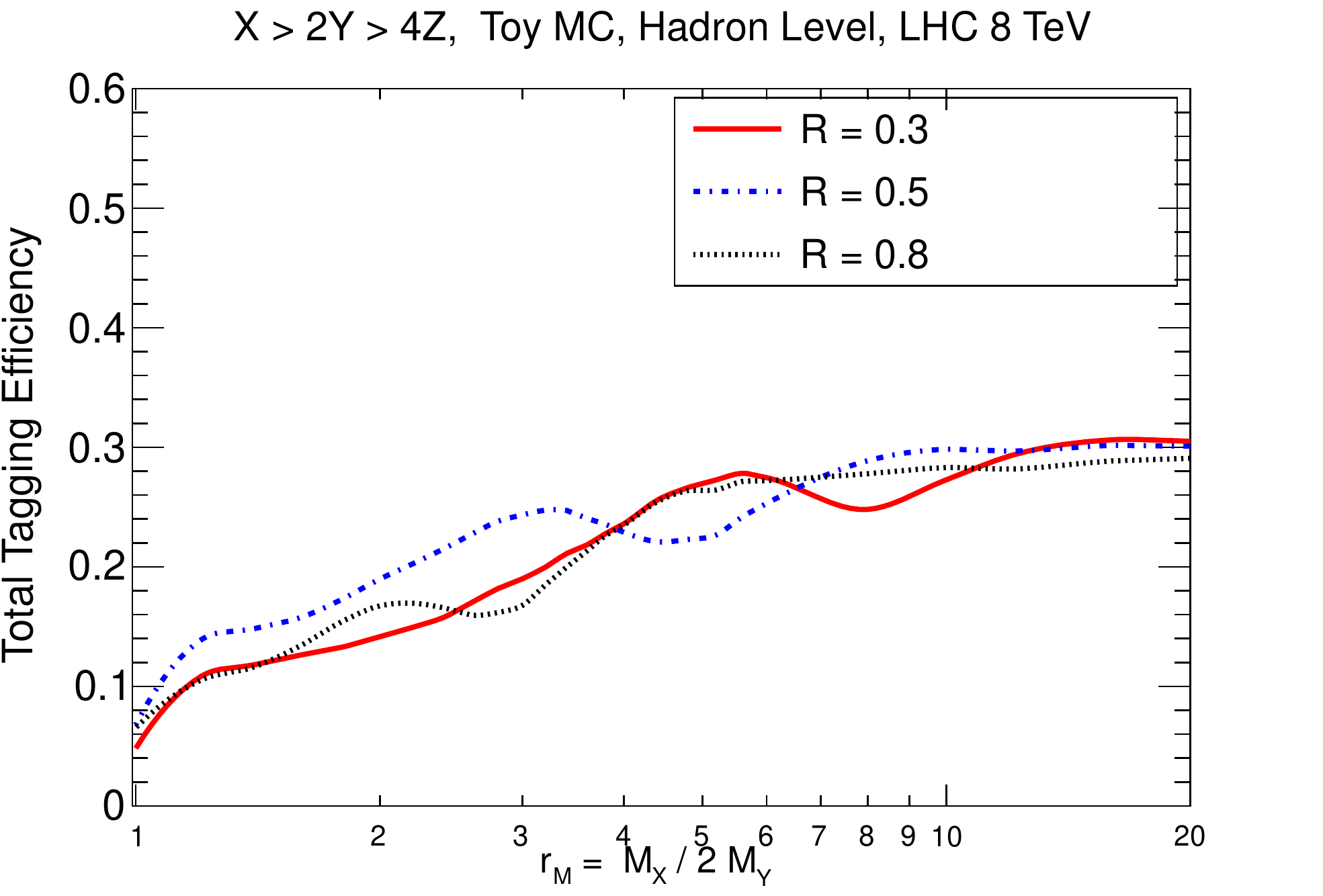}
\caption{\small  The total tagging efficiency for 
parton-level events (left plot) and for
hadron-level events (right plot)
 for different values of the jet radii $R$
as a function of the mass ratio $r_M$. The default
radius used in this paper is $R=0.5$. }
\label{fig:efficiency-hadron-Rdep}
\end{figure}

Since the jet reconstruction strategy that we advocate
is approximately scale invariant, one also expects the results
to be reasonably independent of the jet radius $R$ used
in the jet clustering: 
while the relative fraction of
2-tags, 1-tag and 0-tag events will of course vary with $R$,
their sum should be stable.
 Indeed, we show in 
Fig.~\ref{fig:efficiency-hadron-Rdep} the total efficiency
in parton and hadron-level events for three different radii, $R=0.3,0.5$
and 0.8. 
At parton level, results in the boosted regime are strictly
$R$--independent, as shown in Fig.~\ref{fig:efficiency-hadron-Rdep}.
Except at very low masses, parton-level results are
$R$--independent in all the mass range. 
For low  $r_M$, the degradation at parton level with
increasing $R$ arises in part because the likelihood  that
the decay products
from different $Y$ resonances end up in  a single jet is higher
for larger $R$.
Also
at hadron-level the total tagging efficiency is reasonably 
independent of
$R$.

To summarize,
 in this section we have presented our general strategy for a resonance
reconstruction analysis that can be applied simultaneously to the boosted
and the resolved regimes, with a smooth transition between the two
limits.
It is clear however that some of the details of the strategy can
be modified without affecting the general philosophy. 
One could study different ways of dealing with the four-jet events
rather than selecting the pairings which minimize the relative dijet masses,
like cuts in the angular distributions. 
It is also possible to extend the number of jets considered to build the
 resolved
$Y$-candidate in the 1-tag case up to the fourth or the fifth jet,
in analogy with the procedure used 
 for the
0-tag case.

These modifications could lead
to an overall improvement of the tagging efficiency, but the basic
strategy would be left unaffected.
Finally,
other substructure taggers could be used to classify events,
such as N-subjettiness~\cite{Thaler:2010tr} or
pruning~\cite{Ellis:2009su} among many others 
(see~\cite{QuirogaArias:2012nj} for a recent systematic comparison).
Note however that those taggers with an asymmetry cut, like 
mass--drop and pruning, are
special, because that cut can be linked with 3- and 4-jet analysis
parameters, as done in the present analysis.
In this respect, N-subjettiness is quite different, because it is cutting on
the radiation pattern in the jet.



\section{Resonant Higgs pair production in warped extra dimensional models}
\label{sec:models}

Now we discuss the benchmark models that we will consider for resonant
Higgs boson pair production. These models are based on the warped
extra dimensions scenario~\cite{RS1}, 
where  Higgs pair production is mediated
by either a Kaluza-Klein (KK) graviton or by a radion.
We will assume that the Higgs is the Standard Model boson~\cite{LHChiggs},
and consider its dominant decay into two pairs of $b\bar{b}$ quarks.
Higgs pair production in the Standard Model has
a small cross section~\cite{Hpair} (approximately  18 fb
at 14 TeV), but larger rates can be expected
in New Physics models~\cite{Contino:2010mh,NPdouble} like supersymmetry,
composite models, and warped extra dimensions.
With this motivation, in this section we review the theoretical expectations for resonant
Higgs pair production in the context of warped extra dimensional models, keeping in mind 
that the strategy proposed in this paper is equally valid for any other
Higgs pair production scenario.

Due to Bose symmetry, only resonances of spin zero and spin two can decay on-shell into a pair
of Higgs bosons. Both types are present in models with warped extra dimensions. They are referred
to as radion and KK-graviton, denoted by $\phi$ and $G$ respectively.
These models can naturally explain the large hierarchy between 
the Planck and electroweak scales by introducing a nontrivial geometry in the extra dimension.
The background metric for the case of a single extra dimension is given by
\begin{equation}
ds^2 =  e^{- 2 k y}  \eta_{\mu\nu} dx^\mu dx^\nu - dy^2,
\label{expky}
\end{equation}
where $y$ refers to the coordinate in the 5th dimension and $k$ is related to its curvature. 
The so-called ultraviolet (UV) and infrared (IR) branes are introduced at $y=0$ and $y=L$, respectively. Depending on the scenario, SM fields can be localized
in the IR brane or be allowed to explore the 5th dimension as well.
At each position in the extra-dimension $y_*$, the local cutoff is given by~\cite{Randall:2002tg}
\bea
\Lambda(y_*)=e^{-k y_*} \, \Lambda(y=0) \ .
\eea
If one assumes that the fundamental scale of the theory $\Lambda(y=0)$ 
is the reduced Planck mass $\overline{M}_{P} \sim 2.4 \times 10^{18}$ GeV, 
this scale would be locally shifted to lower values as one moves into the extra-dimension. Physics at the IR brane would have a cutoff at the TeV scale, if 
one requires a mild tuning of $k L ={\cal O}(50)$. This is the solution to the hierarchy problem in warped extra-dimensions, 
where Planck-scale physics appears as TeV physics via the warping in 
the local cutoff. 

As in any model with compact extra dimensions, one expects the existence of a tower of massive resonances for each
particle propagating in the extra-dimensions, called the Kaluza-Klein (KK) resonances. 
In particular, the graviton and the radion (and their associated KK towers) are described by the tensor and scalar quantum fluctuations of
the metric, introduced as an expansion in Eq.~(\ref{expky}) 
\bea
 g_{\mu\nu} = e^{-2 k y} \eta_{\mu\nu}\to e^{-2 (k y + F(x,y))} \, (\eta_{\mu \nu} + G_{\mu\nu} (x,y)) \ .
\eea
The fluctuation of the size of the extra dimension $y$ is described by the 4D scalar radion field, denoted here by $\phi$:
\bea
F(x,y) \propto  e^{2 k y} \, \phi(x) \ ,
\label{fr}
\eea
where $\phi(x)$ is the 4D wave function and $e^{2 k y}$ the localization
profile in the fifth dimension.
The fluctuations of 4D space-time are described by the graviton field $G_{\mu\nu} (x,y)$. 
The massless zero mode of this field corresponds to the usual graviton.
The first massive excitation, the lightest KK-graviton which will focus on, is
\bea
G_{\mu\nu}^{(1)} (x,y) \propto e^{2 k y} \, J_2 \left( e^{2 k y} \frac{m_G}{k} \right) \,  G_{\mu\nu}^{(1)} (x),
\label{fG}
\eea
where $J_2$ is a Bessel function and $m_G$ the graviton mass. 
The mass of the KK-graviton is related to $k/\overline{M}_{P}$ and
to the ultraviolet mass scale of the theory $\Lambda_G$ by
\bea
m_G =  \frac{k}{\overline{M}_{P}}  x_1 \Lambda_G \, ,
\eea
where $x_1 = 3.83$ is the first zero of the Bessel function $J_1$.
The three parameters, $k$, the UV mass scale  $\Lambda_G$ 
and the reduced Planck mass $\overline{M}_{P}$ are
related by $\Lambda_G = e^{-  k L} \overline{M}_{P}$. The mass scale
 $\Lambda_G$
is expected to take a value in the few TeV range.

\begin{figure}[t]
\centering
\includegraphics[scale=0.65]{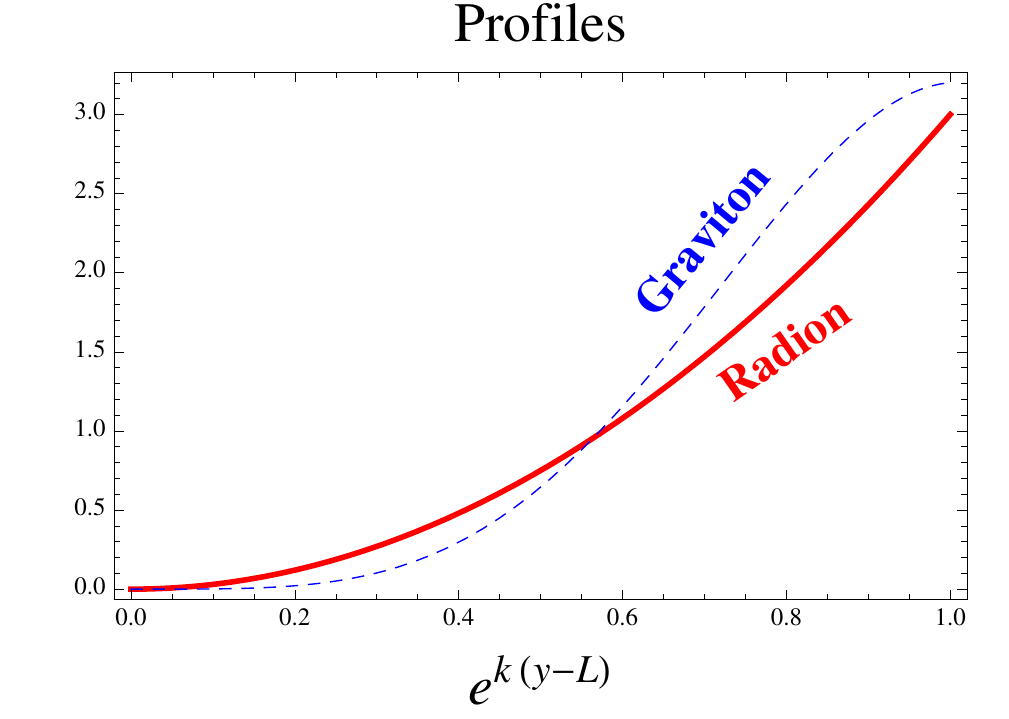}
\caption{\small  Relative localization of the graviton (blue-dashed) and radion (red-solid) in the extra-dimension. The profiles shown are the
$y$-dimension component of the wave functions Eqs.~\ref{fr} and~\ref{fG}. }
\label{localization}
\end{figure}

 We neglect the effect of localized kinetic terms, which would change the value of $x_1$~\cite{loc-kin}. 
We show the radion/graviton localization profiles in Fig.~\ref{localization}:
both fields are localized towards the infrared
brane, but the graviton localization is stronger. 

The radion and KK-graviton couplings to the SM particles are fixed by the action 
\bea
S = -\frac{1}{2} \int d^4 x d y \sqrt{g} \delta g_{MN} T^{MN} 
\eea
where $M,N$ are 5 D indexes and $T^{MN}$ is the 5D energy-momentum tensor involving all fields. 
After dimensional reduction, the effective coupling between the radion and 
KK-graviton lightest mode  and the SM is given by:
\begin{equation}
{\cal L} = -\frac{c_i}{\Lambda_G}   G^{\mu \nu (1)} \, T_{\mu \nu}^{i} - \frac{d_i}{\Lambda_\phi} \phi\, T^{\mu i}_{\mu}
\label{LG}
\end{equation}
where the $T^{i}_{\mu \nu}$ are the four-dimensional energy-momentum tensors of the Standard Model species $i=b, f, V, H, ...$,
and $V$ denotes a generic gauge boson. Here we are
neglecting corrections depending on the fermion localization parameters $c_L$ and $c_R$, that are small when the 
fermions are localized in IR brane such as the top quark and they are not large for the $b$ quark~\cite{chl}.
It is important to notice that the radion couples to the trace of the energy-momentum tensor, which vanishes at the classical level for massless gauge fields.
Note also that the radion scale is related to the KK-graviton scale by $\Lambda_\phi = \sqrt{6} \Lambda_G$ \cite{cgk}.

The coefficients $c_i$ and $d_i$ are  
proportional to the wavefunction overlap of the graviton/radion and the SM fields.
 For example, the Higgs is IR-localized, as its vacuum expectation value and mass are IR effects. 
The fact that the KK-graviton 
wave-function, Eq.~(\ref{fG}), is more peaked towards the IR brane than  the radion, Eq.~(\ref{fr}), as shown in Fig.~\ref{localization},
translates into a stronger coupling of the graviton to the Higgs, 
beyond what is expected from the
trivial rescaling of $\Lambda_\phi$ and $\Lambda_G$. 

In the original Randall-Sundrum model (RS1) all SM fields are localized in the IR brane, so all couplings are  $c_i \simeq {\cal O}(1)$. 
More realistic models, consistent with experimental constraints,
 must have SM fields in the bulk, 
leading to different values of the couplings $c_i$. 
A well motivated configuration, which we will refer to as {\it bulk } RS~\cite{BulkRS}, predicts that the SM fields communicating to the EWSB sector are peaked towards the IR brane. 
This is the case of the Higgs and longitudinal $W$ and $Z$ bosons, and possibly the top quarks. Light fermions would be localized near the UV brane, 
whereas massless gauge bosons are de-localized.
 The graviton and radion would then couple preferentially to IR localized fields, namely $h$, $W_L$, $Z_L$ and possibly $t$ as well. The coupling to $\gamma$ and $g$ is suppressed by a  volume factor $\simeq 1/k L$, and the coupling to light fermions (including the quarks in the proton) would be extremely suppressed. 
In summary, the two scenarios we are going to consider, RS1 and Bulk RS are defined by the following hierarchy of couplings of the KK graviton to SM particles:

\vspace{.2cm}

\fbox{%
\vspace{1cm}

\begin{minipage}{5.5 in}
\begin{center}
\bea
\textrm{ \bf RS1 scenario: } & & c_H = \textrm{ all the other } c_i \simeq {\cal O}(1)  \\
\textrm{ \bf Bulk RS scenario: } & & c_H \simeq c_{Z,W, t} \simeq {\cal O}(1) \simeq (k L) \,  c_{\gamma,g} \gg c_{u,d,\ell \ldots}
\eea
\end{center}
\vspace{0.cm}
\end{minipage}

}
\vspace{.2cm}

\begin{figure}[t]
\centering
\includegraphics[scale=0.31]{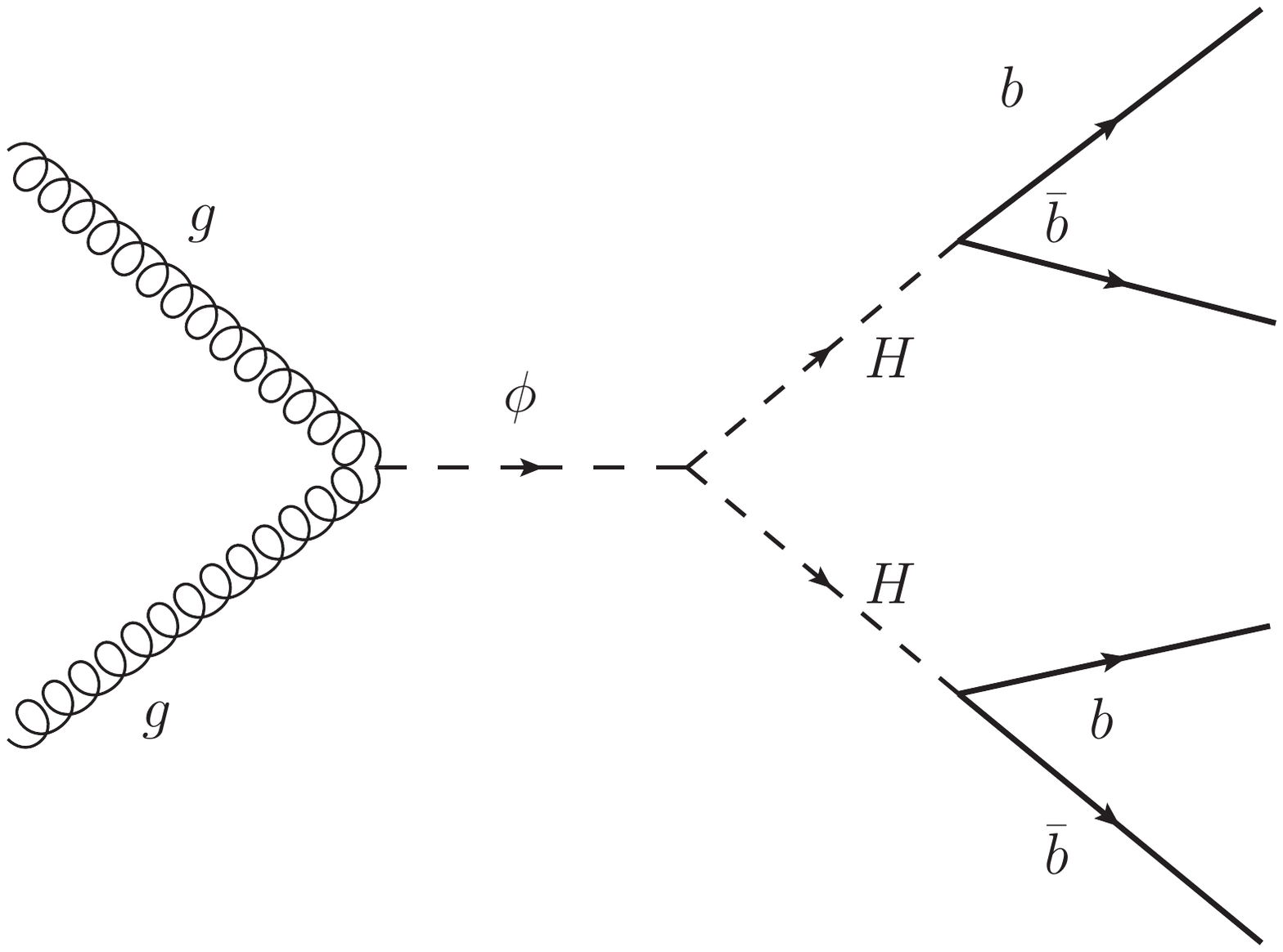}
\includegraphics[scale=0.31]{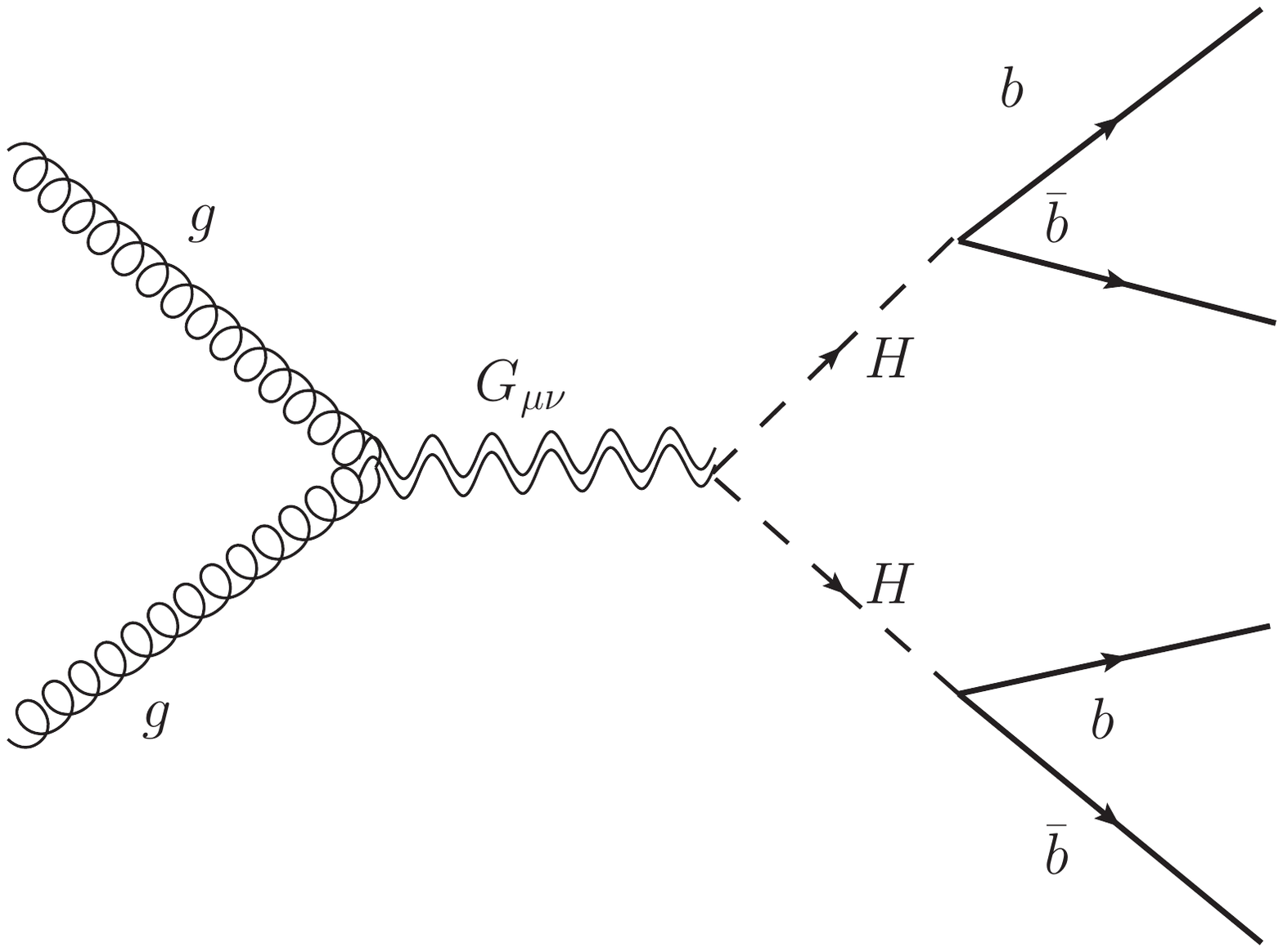}
\vspace{-40mm}
\caption{\small  Feynman diagrams
for Higgs pair production in warped extra
dimensions models mediated
by a radion $\phi$ (left plot) and a KK-graviton $G_{\mu\nu}$ (right plot).
The Higgs bosons then decay into a pair of $b$ quarks.
}
\label{fig:feynm}
\end{figure}

As an example, the stress-energy tensor for the Higgs field is given by
\bea
T_{\mu \nu}^{H} = \partial_\mu H \partial_\nu H - \frac{1}{2} g_{\mu \nu}  \left( \partial_\alpha H \partial^\alpha H - m_H^2 H^2 \right) 
\eea 
which results in the following  couplings,
\bea
{\cal L} \supset   -\frac{c_H}{\Lambda_G} G_{\mu\nu}^{(1)} \,  \partial^{\mu} H \partial^{\nu} H  +  \frac{d_H}{\Lambda_\phi} \phi \left( \partial_\alpha H \partial^\alpha H  - 2  m_H^2 H^2 \right) 
\eea
We do not consider a non-minimal Higgs coupling
to gravity, which would require the use of an ``improved"  energy-momentum tensor~\cite{ccj} and would lead to a Higgs-radion mixing~\cite{dggt,Grzadkowski:2012ng}, since mixing is not relevant in our case where we require
$M_{\phi}\ge 2M_{H}$. 
The Feynman diagrams relevant for Higgs pair production mediated
by a radion $\phi$ and a KK-graviton $G_{\mu\nu}^{(1)}$ are schematically shown
in Fig.~\ref{fig:feynm}.

In the remainder of this section we will discuss the production rates
of the radion and graviton in the RS1 and Bulk-RS scenarios.
 It
is beyond the scope of this paper to review the experimental
constraints on the parameter space of these models.
A discussion of the implications of recent measurements for limits on
extra dimension models can be found in~\cite{CMSGG, VVboosted, CMSdilep, ATLASdilep}. 
Note than when experimental limits  arise from decays to photons, leptons and four-fermion operators involving light fermions, they 
can be interpreted only in the context of 
 RS1 but not in the bulk RS scenario.
Bounds on the radion mass as a function of the cutoff scale $\Lambda_\phi$
have been compiled in Refs.~\cite{barger,sr}.

\subsection{Production rates at the LHC}

The production rates of the radion and the graviton at hadron
colliders will depend
on the respective couplings to the light quarks and gluons in
the incoming protons.
In RS1, the graviton couples to light quarks and gluons with the same coefficient $\mathcal{O}(1)$, whereas in the Bulk-RS scenario the couplings to light quarks are very suppressed.
The coupling of the graviton to gluons in the bulk RS model is given by
\bea
c_{g} = \frac{2 (1-J_0 (x_1))}{k L x_1^2 |J_2(x_1)|} \simeq 0.02
\eea
whereas in RS1 $c_g=1$.

The coupling of the radion $\phi$ to gluons (and to photons) vanishes at
tree level due to classical scale invariance. 
At 1-loop level it arises due to the trace anomaly, which is related to the beta function, and the top quark triangle diagram.
We denote by $\kappa^{\phi}_g$ the coupling of the radion to gluons defined by
\begin{equation}
\kappa^{\phi}_g \frac{\phi}{\Lambda_\phi} G_{\mu \nu}^a G^{a \mu \nu},
\end{equation}
where $\kappa^{\phi}_g$ is given by~\cite{chl}:
\begin{equation}
\kappa^{\phi}_g = - \frac{\alpha_s b_3}{8\pi} -\frac{1}{4 kL},
\label{anom}
\end{equation}
where we have neglected the top loop contributions. The coefficient of the QCD $\beta$ function is $b_3 = 8$.
The RS1 case corresponds to neglecting the volume suppressed term. 
Note that as compared to the graviton, for the radion production 
cross section there is less model flexibility in that the
coupling to the gluons is fixed independently of the localization 
of the SM fields.

In the following we will assume that gluon fusion is the dominant process for both radion and KK-graviton production at the LHC.
 This is certainly true for the radion, and also for the graviton in the bulk RS scenario, where the couplings
to the light quarks in the proton are very suppressed.
In the narrow width approximation, the production cross section via gluon fusion of a particle $X$ with mass $M$ is given by
\begin{equation}
\sigma_{pp \rightarrow X } (M,s)=  \int_{\tau}^1  L_{gg}(\hat{\tau}) \hat{\sigma}(gg \rightarrow X) (\hat{\tau} s)   d  \hat{\tau},
\end{equation}
where here $\tau=M^2/s$ and $L_{gg}$ is the gluon luminosity function.

\begin{figure}[t]
\centering
\includegraphics[scale=0.73]{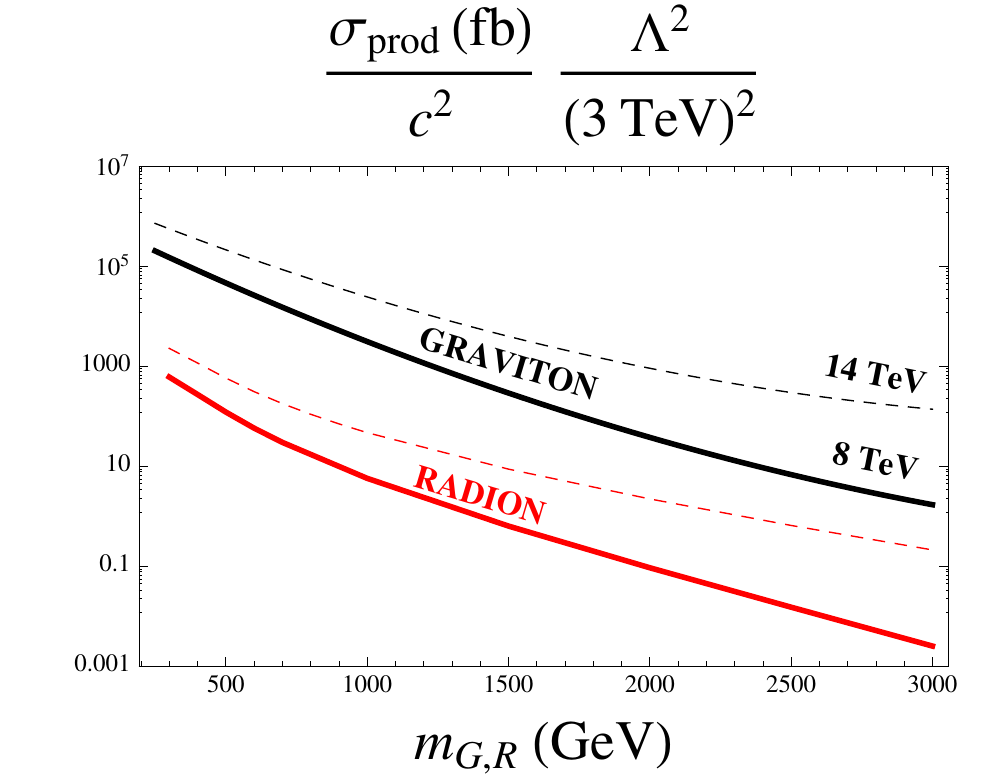}
\includegraphics[scale=0.73]{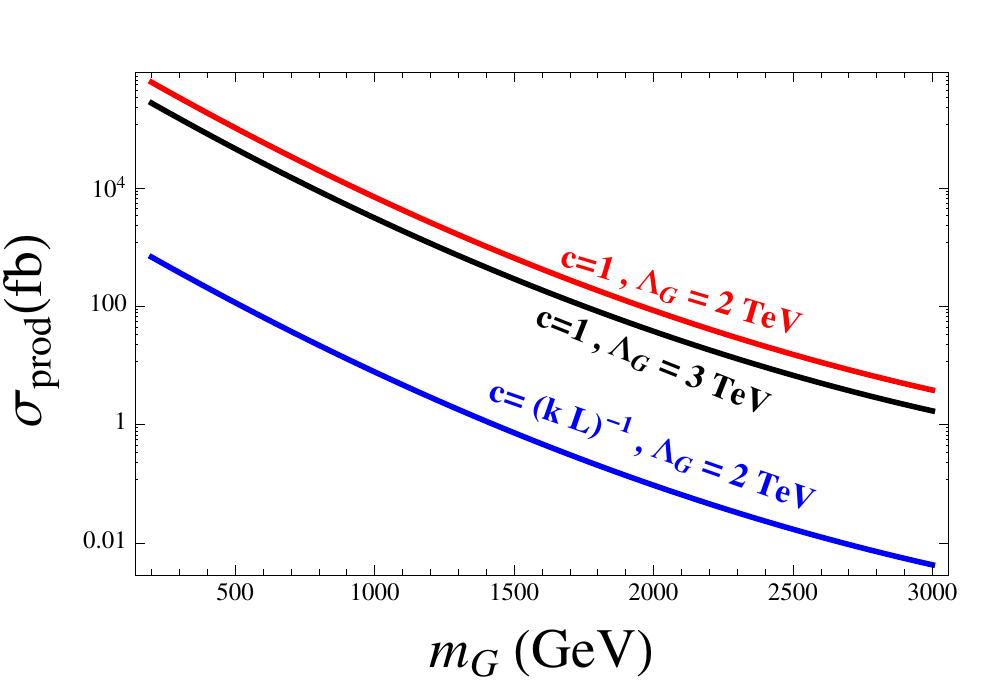}
\caption{\small (Left plot) Production cross sections as a function of the graviton and radion masses ($m_{G,R}$), where the trivial dependence on the scale $\Lambda=\Lambda_{G,R}$ and the coupling to gluons ($c=c_g$, 1) is factored out. Solid (dashed ) lines correspond to 8 (14) TeV. (Right plot) Graviton cross sections (at 8 TeV ), for specific choices of $c_g$=1, $1/kL$ corresponding to RS1 and bulk RS.  }
\label{xsec}
\end{figure}

We computed the production cross section of the processes $p p \to G$, $\phi$ at leading order (LO)  using {\tt Madgraph5}~\cite{Mad}. The results are shown in 
Fig.~\ref{xsec}.  We plot the production cross section, where we have factored out the trivial dependence on the coupling to gluons and the scale of dimension-five operators, $\Lambda_G$ and
we also show the cross section for specific values of those parameters.

Notice that the KK-graviton production cross section is larger than the corresponding radion cross section due to the fact that the radion coupling to gluons
is loop induced, whereas the KK-graviton has tree-level couplings to gluons.
 Also, the KK-graviton has five degrees of freedom, compared to the single degree of freedom of the radion. 

\subsection{Graviton and radion decays}

In RS1, with all the SM fields localized on the IR brane, a heavy graviton would decay democratically to all degrees of freedom. In the bulk RS, the Higgs and fields associated with EWSB are still IR localized, and 
using the equivalence theorem, one can show that~\cite{Agashe:2007zd}
\bea
\Gamma(G \to H H ) = \Gamma(G \to Z_L Z_L ) =  \Gamma(G \to W^+_L W^{-}_L )/2 = \frac{1}{960 \, \pi} \, \frac{m_G^3}{\Lambda_G^2}
\eea

In bulk RS, the width to gluons and photons is suppressed by the {\it effective} volume $k L$,
\bea
\Gamma (G\to gg ) = 8 \Gamma (G\to \gamma\gamma) \simeq 8 \frac{ \Gamma(G\to H H )}{k L} \simeq 10^{-1} \Gamma(G\to H H ) \ .
\eea

The graviton would also couple to fermions localized near the IR brane. In many models, third generation quarks are pushed towards the IR brane via a localization parameter $\nu$, which is a ratio of a 5D mass term, $M_{f}$, and the curvature, $\nu= M_{f}/k$.  
The effect of $\nu$ is as follows: for $\nu=1/2$, the {\it conformal value}, the fermion zero mode is delocalized in the extra dimension, as the profile is flat and does not prefer a particular location inside the extra-dimension. For $\nu>1/2$, the fermion zero-mode will be localized towards the IR brane, whereas for  $\nu<1/2$, the localization is near the UV brane.

The width to tops is given by 
\bea
\Gamma (G\to t \bar{t})  = \frac{1}{240 \, \pi} \, f(\nu_t)^2  \, \frac{m_G^3}{\Lambda_G^2}
\eea
for $m_G \gg 2 m_t$. 
We have defined
\bea
f(\nu_t) =\frac{3}{2} \,  \frac{1+2 \nu_t}{1-e^{-k L (1+2 \nu_t)}} \, \int_0^1 d y y^{2+2 \nu_t} \frac{J_2(3.83 y)}{J_2(3.83)} \ .
\eea

The branching ratio of the graviton to the Higgs depends on the top localization as
\bea
BR(G\to H H) (\nu_t) \approx \frac{1}{4} \,  \frac{1}{1+f(\nu_t)^2}
\eea

\begin{figure}[h!]
\centering
\includegraphics[scale=0.95]{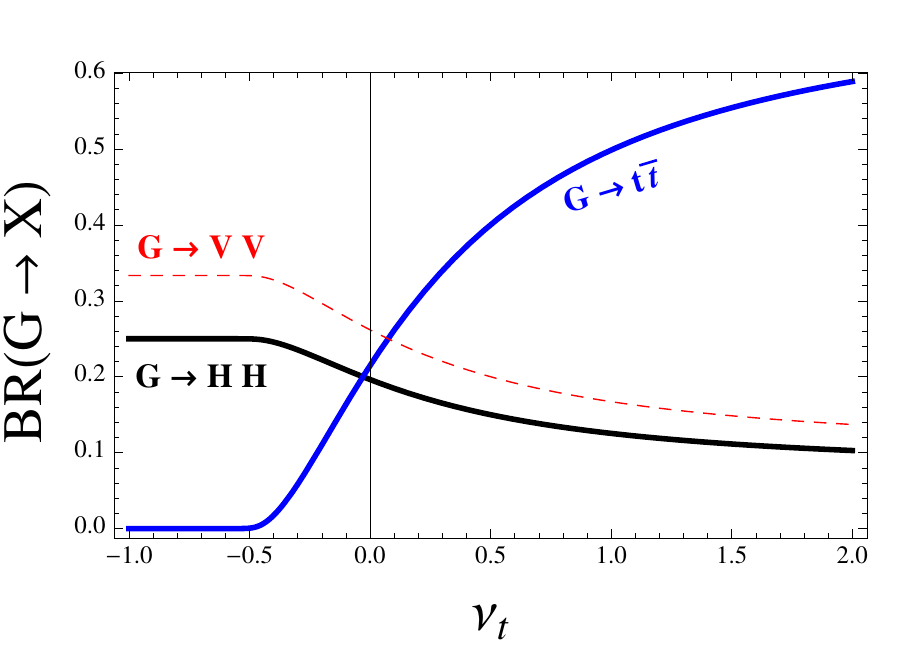}
\caption{\small Branching ratios of the graviton to the Higgs (black), tops (blue) and vector bosons (red-dashed), as a function of the top localization parameter $\nu_t$.}
\label{BRg}
\end{figure}

In Fig.~\ref{BRg} we show that the maximal branching ratio to a Higgs boson pair is 25\% (when the Higgs is 1/4 of the IR degrees of freedom), and quickly decreases as one increases the branching ratio to top quarks (increasing the value of $\nu_t$, and therefore the localization towards the IR brane).  

The dominant decay modes of the radion are into pairs of massive gauge bosons, Higgs bosons and top quarks.
Since the couplings are determined by the masses of the final state particles, and these masses arise from the TeV localized
Higgs boson, the RS1 and bulk RS couplings are the same at leading order. The corresponding widths (for large $m_\phi$)
are: 
\bea
\Gamma(\phi \to H H ) = \Gamma(\phi \to Z Z) = \Gamma(\phi \to W W ) /2 = \frac{1}{32 \pi} \frac{m_{\phi}^3}{\Lambda_{\phi}^2} 
\eea
and
\bea
\Gamma(\phi \to \bar{t} t ) = \frac{3}{8 \pi} \left(\frac{m_t}{m_\phi} \right) ^2\frac{m_{\phi}^3}{\Lambda_{\phi}^2}. 
\eea
Hence, for large radion masses the branching fraction to a pair of Higgs bosons is approximately
$25\%$,  independent of $\Lambda_\phi$, since the contribution from decays
to top quarks can be neglected.
 We note that for the smaller
$m_{\phi}$ values that are relevant for phenomenology the decay into
top quarks should in principle be taken into account, but in this work
for simplicity we will assume that $BR\lp \phi \to HH\rp=25\%$ independent
of the radion mass.

\subsection{Composite duals and model dependence}

So far, we have described Higgs pair production via gluon fusion into a radion or KK-graviton in warped extra-dimensions. In this context, the graviton cross sections are larger than the radion by at least an order of magnitude, and there is little room for changing this hierarchy. 

To test how robust  this prediction is, we would like to approach this model building in extra-dimensions from the point of view of holography. 
In this approach, models in warped extra dimensions are an {\it analogue computer} for strong interactions.
 This duality between 4D strongly-coupled theories and 5D weakly-coupled theories with gravity was inspired by the AdS/CFT correspondence, but took hold on a more qualitative basis~\cite{Lisa-Porratti} and has been used to build models of QCD~\cite{AdSQCD}, technicolor~\cite{HTC}, composite Higgs~\cite{composite-Higgs}, and even condensed matter systems~\cite{condensed-matter}, with some success. 
In this context, the KK resonances, a consequence of compactification, are the holograms of  massive resonances due to confinement. 

The KK graviton is therefore the dual of a spin-two bound state in a strongly coupled theory, very much like the $f_2$ of QCD~\cite{Ami}.
 One could then wonder how different the coupling structure of the
 $f_2$-like resonance would be with respect to the KK-graviton. As was shown in Ref.~\cite{Gimpostor}, the couplings we propose in Eq.~(\ref{LG}) saturate the possibilities, once Lorentz, gauge and CP invariance are assumed. No other structures are allowed up to dimension-six operators.
 Hence, our KK graviton analysis  can be directly generalized to strongly coupled sectors with spin-two resonances. Moreover, if the new strongly coupled sector participates in the electroweak symmetry breaking mechanism, a sizable coupling to Higgses would be expected. 

The dual of the radion would be the dilaton, the Goldstone boson of scale invariance.\footnote{It is unclear whether in QCD one would have such a creature, but some proposals are the $f(975)$ resonance~\cite{dilatonf975} of the $\sigma$ particle~\cite{sigma-dilaton}.} 
The dilaton couplings at tree level would be perfectly mimicked by the radion couplings, as the dilaton will couple to the trace of the stress tensor. 
This can be shown by writing down an effective theory where the dilaton is spurion of the scale symmetry~\cite{witek}. Within this analysis, the loop contributions to the dilaton to massless gauge bosons will follow the same structure as the anomalies written in Eqs.~(\ref{anom}). Therefore, our analysis of the radion couplings is also applicable to a dilaton in a composite sector. 

In summary, the structure of couplings we describe for the KK-gravitons and radion would be the same for the bound state duals. The main difference between the analysis in extra-dimensions and composite theories is the strength of the coupling.
 For example, one could imagine a composite theory where the spin-two resonance is made up of colorless techni-quarks, hence there would be no tree-level coupling to gluons, whereas the dilaton couplings would be determined by the scale symmetry. In this case, one would expect a larger production of dilaton than spin-two resonances.

\section{New Physics searches in the $HH \rightarrow 4b$ final state}
\label{sec:results}

In this section we apply the general resonance tagging strategy
presented in Sect.~\ref{sec:jetalg} to a particular scenario,
 namely the resonant Higgs boson pair production
pair which then decays into four $b$-quarks.
The results presented here are model independent and 
can be applied 
to any generic BSM model with enhanced  
Higgs pair production~\cite{Contino:2010mh,Dolan:2012ac,NPdouble},
though we  will provide an explicit interpretation of exclusion
limits in terms of the radion and graviton couplings
in the warped extra dimension models of Sect.~\ref{sec:models}.

First of all, we discuss the Monte Carlo event
generation for the signal and background events
with  {\tt MadGraph} and {\tt Pythia} and 
evaluate
 the tagging efficiency as a function
of $r_M$, to compare with the approximate
kinematics of the toy MC used in Sect.~\ref{sec:jetalg}. 
We recall that the main differences between the toy MC and
{\tt MadGraph} are that the latter includes the rapidity
distribution for the $X$ resonance and the correct
angular distributions of a spin-two particle in the case of the graviton.
Then we present the different assumptions
that underlie our
implementation
of $b$-tagging. 
A discussion of the background rejection capabilities of the 
tagging algorithm
follows, where we show that the combination of the resonance
tagging 
and $b$-tagging reduces the QCD multijet
background by several orders of magnitude. 
In the last
part of the section we present the implications in terms
of model independent searches in the $HH\to 4b$ final state,
and interpret these results in terms of exclusion ranges
in the parameter space of warped extra dimension models.\footnote{The feasibility of 
the $4b$ final state to probe
BSM resonant pair-production with jet substructure was also investigated 
in Ref.~\cite{Bai:2011mr} in the context of composite octet searches.}

\subsection{Monte Carlo signal event generation}

Our benchmark model
is $s$--channel Higgs boson pair production 
mediated by a radion or a massive Kaluza-Klein 
graviton 
resonance in scenarios
with warped extra dimensions. 
We have implemented these scenarios in the {\tt Madgraph5} Monte Carlo
program~\cite{Mad}.\footnote{Our results have been generated at
  leading order only; NLO corrections for resonant double-Higgs boson
  production have been calculated in the context of the minimal
  supersymmetric standard model, in the heavy top-mass limit, in
  Ref.~\cite{Dawson:1998py} and were found to be substantial, giving a
  $K$-factor of order $2$.
  Similar corrections are probably relevant to our radion case.
  However, given that other aspects of our study are probably not
  under control beyond a factor of two, e.g.\ the $b$-tagging
  assumptions for the background, we will conservatively not include
  the NLO signal enhancement.}
While the main motivation to study both radion and graviton simultaneously
is to cover a wider range of the model parameter
space, a useful by-product is 
to validate the jet finding strategy for two different 
angular distributions of the decay products. 
Indeed, from the
kinematic point of view the radion and graviton cases are identical
(for equal masses)
except for the different angular decay distributions of  spin-zero
and  spin-two particles.
Note that in the radion case, since the radion is an scalar, the
 kinematics and angular distribution will be very close
to those of the toy Monte Carlo  of Sec.~\ref{sec:jetalg} used
to validate the resonance tagging algorithm, with the only difference
arising from the rapidity distributions of the radion.

We have followed Ref.~\cite{Csaki:2007ns} 
to model the radion couplings to the  Higgs boson
 and to gluons using the {\tt FeynRules} framework~\cite{Christensen:2008py}.
 The implementation of the model 
has been based on the default {\tt MadGraph5} model with
 effective theory coupling of the Higgs to gauge bosons. 
In addition to the SM parameters, in the radion model we have
four additional parameters: the radion mass, $M_{\phi}$,
the ultraviolet mass scale of the theory, $\Lambda_{\phi}$,  
the radion-Higgs mixing parameter, $\xi$, 
and the compactification scale, $kL$. 
These parameters take the value
 $\Lambda_{\phi} = 3$ TeV and $kL = 35$, supplemented
by the no mixing condition that reads $\xi = 0$. The absence of
mixing is justified by the fact that the radion masses considered will
be always much larger than the Higgs mass. Any modification of 
 $\Lambda_{\phi}$ translates into a trivial rescaling of the total
rates.

To simulate the graviton production we 
have used the standard Randall--Sundrum model as implemented in {\tt MadGraph5}.
Here the relevant additional parameters are only the graviton mass
$M_G$ and the ultraviolet mass scale of the theory, $\Lambda_{G}$,
 chosen to be $\Lambda_{G} = 3$ TeV.
As pointed out on the Sect.~\ref{sec:models}, the
 mass scales $\Lambda_{\phi}$
and $\Lambda_G$ of 
the radion and the graviton are
 theoretically related. 
However, from the practical point of view we select
independently the parameters of the two models in order not to
impose additional constraints on the search ranges. 
For both  radion and graviton event generation the
narrow width approximation has been assumed.

We have generated events for radion and graviton production
for a range of masses between 250~GeV and 3~TeV. 
Higher  masses lead to too small cross sections to be of any phenomenological interest.
As in the case of the toy Monte Carlo events, {\tt Madgraph5}  parton level events were showered and hadronized using {\tt Pythia8} with the same settings
for underlying event and multiple interactions.

We have already discussed in Sect.~\ref{sec:jettag} the tagging
efficiency of the algorithm for the toy Monte Carlo kinematics,
both at parton and at hadron level. However, the
basic selection cuts did not match those 
of a realistic experimental analysis. 
We will use the following
selection cuts instead in the rest of this paper:
\be
p_{T}^{\rm min} \ge 50~\GeV, \qquad |\eta^{\rm jet}|\le 2.5\, , \qquad H_T\equiv \sum_{\rm jets} p_T^{\rm jet} \ge 300~\GeV \label{eq:cuts_final} \, .
\ee
These cuts are inspired by  
typical trigger and angular acceptances of the LHC
 experiments~\cite{FourATLAS, FourCMS}.

\begin{figure}[h!]
\centering
\includegraphics[scale=0.35]{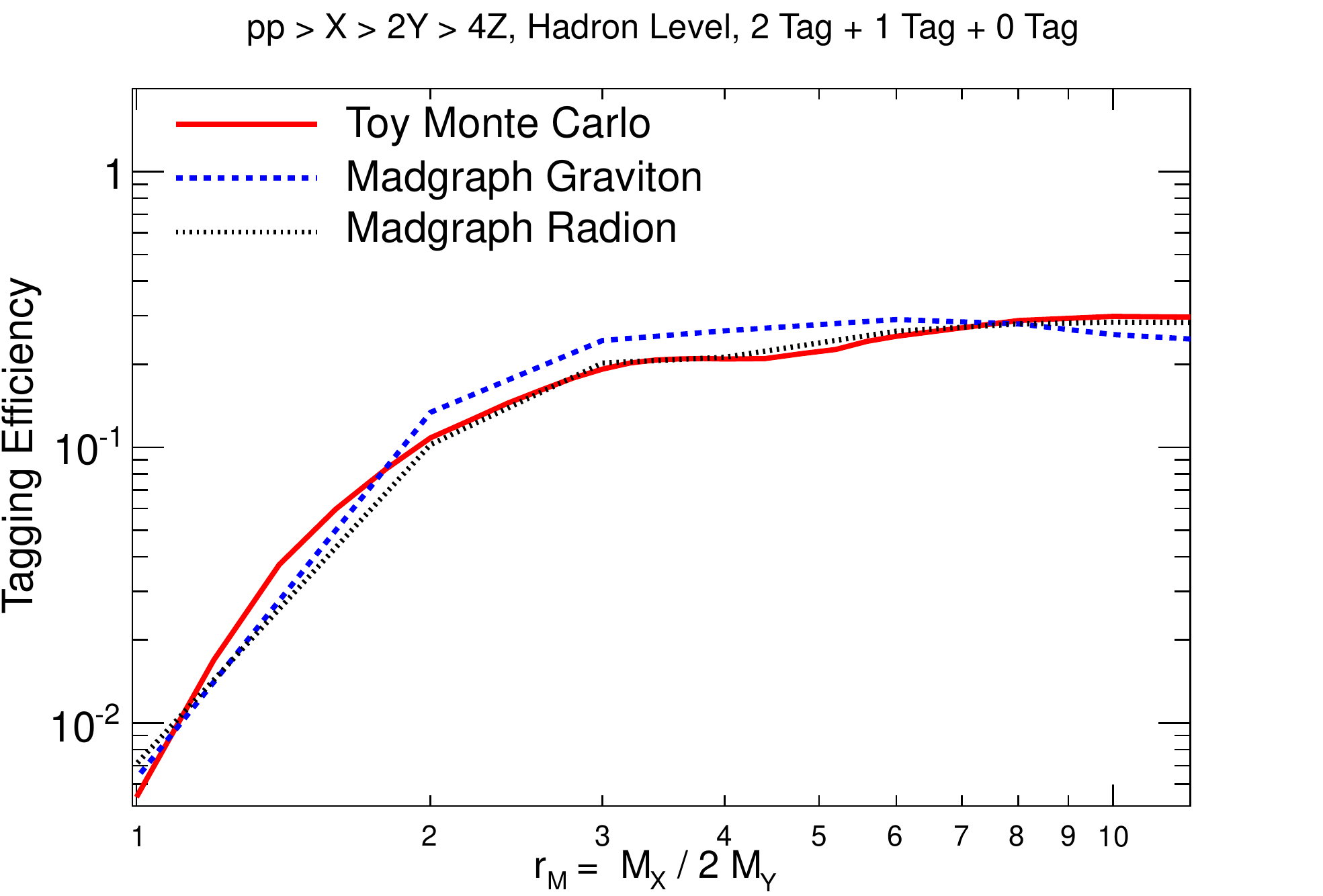}
\includegraphics[scale=0.35]{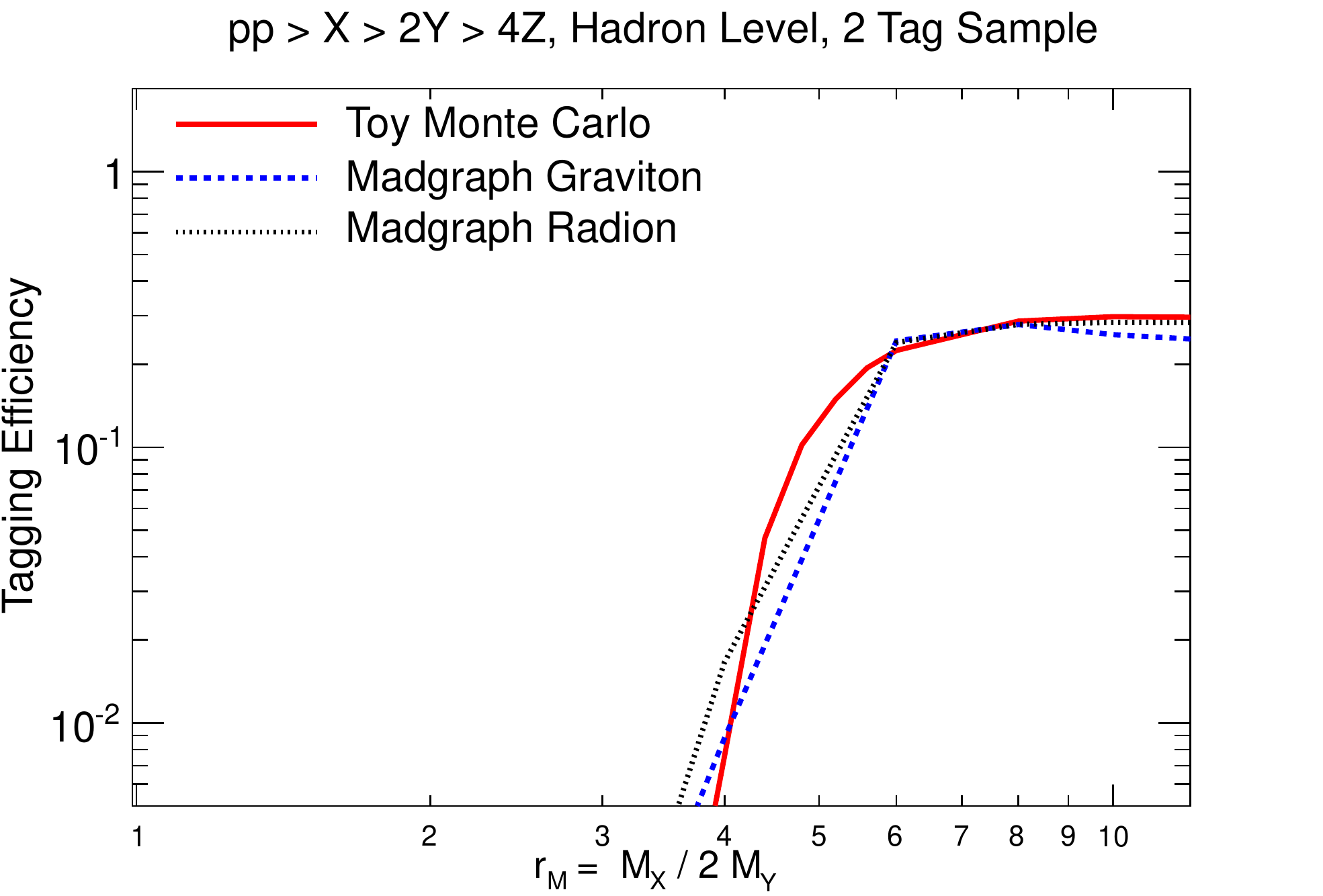}
\includegraphics[scale=0.35]{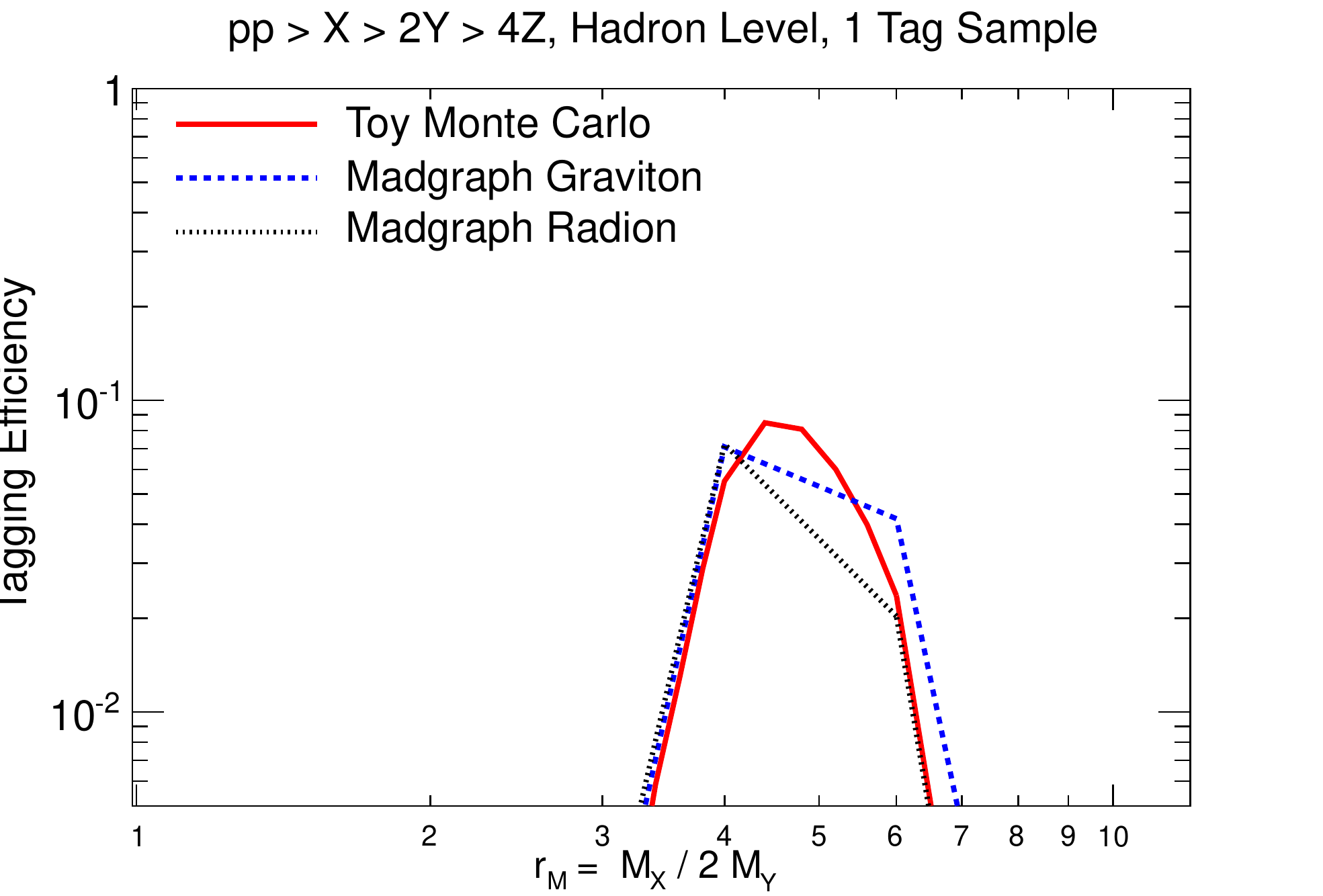}
\includegraphics[scale=0.35]{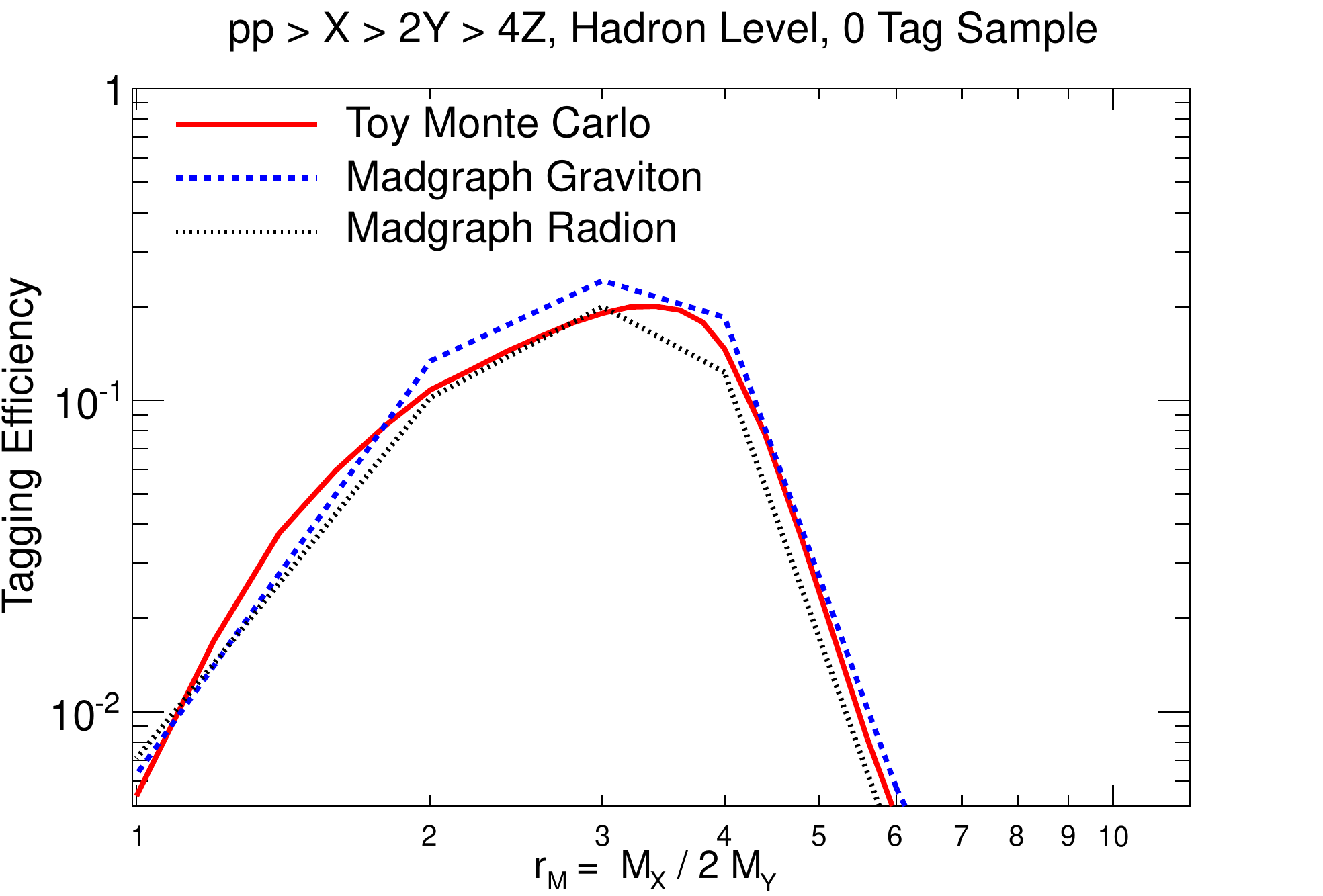}
\caption{\small  Comparison between the hadron-level tagging
efficiencies, at the 8 TeV LHC, for the toy Monte Carlo events and the radion
and graviton {\tt MadGraph5} events. We show the total efficiencies
and the breakup in different tagged samples, as a function
of the boost factor $r_M$. }
\label{fig:efficiency-madgraph}
\end{figure}

We show in Fig.~\ref{fig:efficiency-madgraph} the comparison
between the hadron-level tagging efficiencies at LHC 8 TeV
between the toy Monte Carlo events and the {\tt MadGraph}
radion and graviton events, as a function of the mass
ratio $r_M$. 
As we can see, the toy MC results agree well
with the radion events, which is a non trivial cross-check
that event generation is under control. 
Also, the efficiencies for the radion and graviton are very similar,
showing that the spin-zero vs. spin-two angular distributions do not
lead to any large differences at the level of the reconstruction.%
\footnote{At first sight this may appear to be surprising, given that
  the radion and graviton angular decay distributions are
  substantially different. However our $\Delta y_\text{max} = 1.3$ cut
  is sufficiently large that the integral of the $\Delta y$
  distribution up to $\Delta y_\text{max}$ is not too dissimilar in
  the two cases.}
Note that 
we have generated fewer mass points with {\tt MadGraph5} than
with the toy MC, hence the somewhat less smooth distributions
in the former case.

One significant difference between Fig.~\ref{fig:efficiency-madgraph}
and the results of Sec.~\ref{sec:jetalg}, is the much lower efficiency
in the low $r_M$ region.
It is a consequence of the larger $H_T$ cut in
Eq.~(\ref{eq:cuts_final}) than in Eq.~(\ref{eq:cuts}), which severely
reduces the fraction of tagged events when $m_X \lesssim 300$~GeV.
Insofar as the $H_T$ cut is present mainly to limit trigger bandwidth,
one could also imagine lowering it and then controlling bandwidth by means
of trigger-level $b$-tagging.

\subsection{B tagging}

The final state that we are interested in includes  four $b$-quarks from
the decays of the two Higgs bosons.
Therefore,
$b$-tagging will be an important ingredient to improve
the signal over background ratio.
We have adopted in this study a $b$-tagging scenario that we expect to
be realistic (possibly conservative), inspired by the ATLAS and CMS
capabilities~\cite{Khachatryan:2011wq,CMS:2012hd,atlasb1,atlasb2}.
The probability of tagging a $b$-quark is taken to be
$f_b=0.75$, the mistag probabilities of $c$-quarks, $f_c=0.10$, and of
light quarks and gluons, $f_l=0.03$.
We apply the $b$--tagging conditions
on the parton level events after showering but
before hadronization, that is, we tag $b$ quarks
rather than $B$ hadrons.
We will require one $b$-tag per Higgs candidate.
In detail it is implemented as follows:
\begin{itemize}
\item Determine the number of $b$-quarks within each of the two
Higgs candidates' jets. Such candidate jets can be a single
anti-$k_T$ jet with radius $R$ (in the boosted regime) or
a jet composed by the sum of two different anti-$k_T$ jets
(in the resolved limit).
\item A Higgs candidate jet 
is considered to be $b$-tagged if it contains at least
one $b$ quark with $p_{T,b} \ge p_{T,b}^{\rm min}=10$ GeV. The
$b$-tag efficiency is denoted by $f_b$.
\item A Higgs candidate jet which does not fulfill the previous condition,
but contains at least one $c$ quark with
 $p_{T,c} \ge p_{T,b}^{\rm min}$, will
be $b$-tagged with a mistag probability $f_c$.
\item A Higgs candidate jet which contains only light quarks
and gluons will be $b$-tagged with a mistag probability 
$f_l$.\footnote{
  Strictly speaking, since there are two jets in a ``resolved'' Higgs
  candidate, the mistag probability is closer to $2f_l$. 
  However, at the level of factors of two, our $b$-tagging estimates
  for the backgrounds are probably not accurate.
  On one hand, for example, ATLAS~\cite{atlasb1,atlasb2} obtains
  somewhat better light-jet rejection than the $f_l=0.03$ that we
  use.
  On the other hand, light-jet rejection factors will anyway depend on
  $p_t$ of the jets and potentially also their proximity to other jets.
}
\item $b$-tagged events are those
 for which the two Higgs candidates' jets have
been both $b$-tagged.
\end{itemize}
Therefore, events will be given different weights according
to the number of $b$ and $c$ quarks present in each
of the two Higgs candidate jets.
For instance, if the two Higgs candidates' jets
each contain at least one $b$ quark,
the event is assigned a weight $f_b^2$=0.56. 
This is the same probability for signal events and
for QCD background events where two $b$ quarks end up each
in a Higgs candidate jet.

We could also have considered a more optimistic scenario for the
$b$-tagging, in which each Higgs candidate is required to have two
$b$-tags. 
In particular CMS has demonstrated the ability to tag pairs of
$B$-hadrons even for angular separations $\Delta R_{b\bar b} <
0.4$~\cite{Khachatryan:2011wq}, which suggests that this scenario
could be viable also in the highly boosted limit where the $B$-hadrons
are within a single anti-$k_T$ jet.

\begin{figure}[t]
\centering
\includegraphics[scale=0.40]{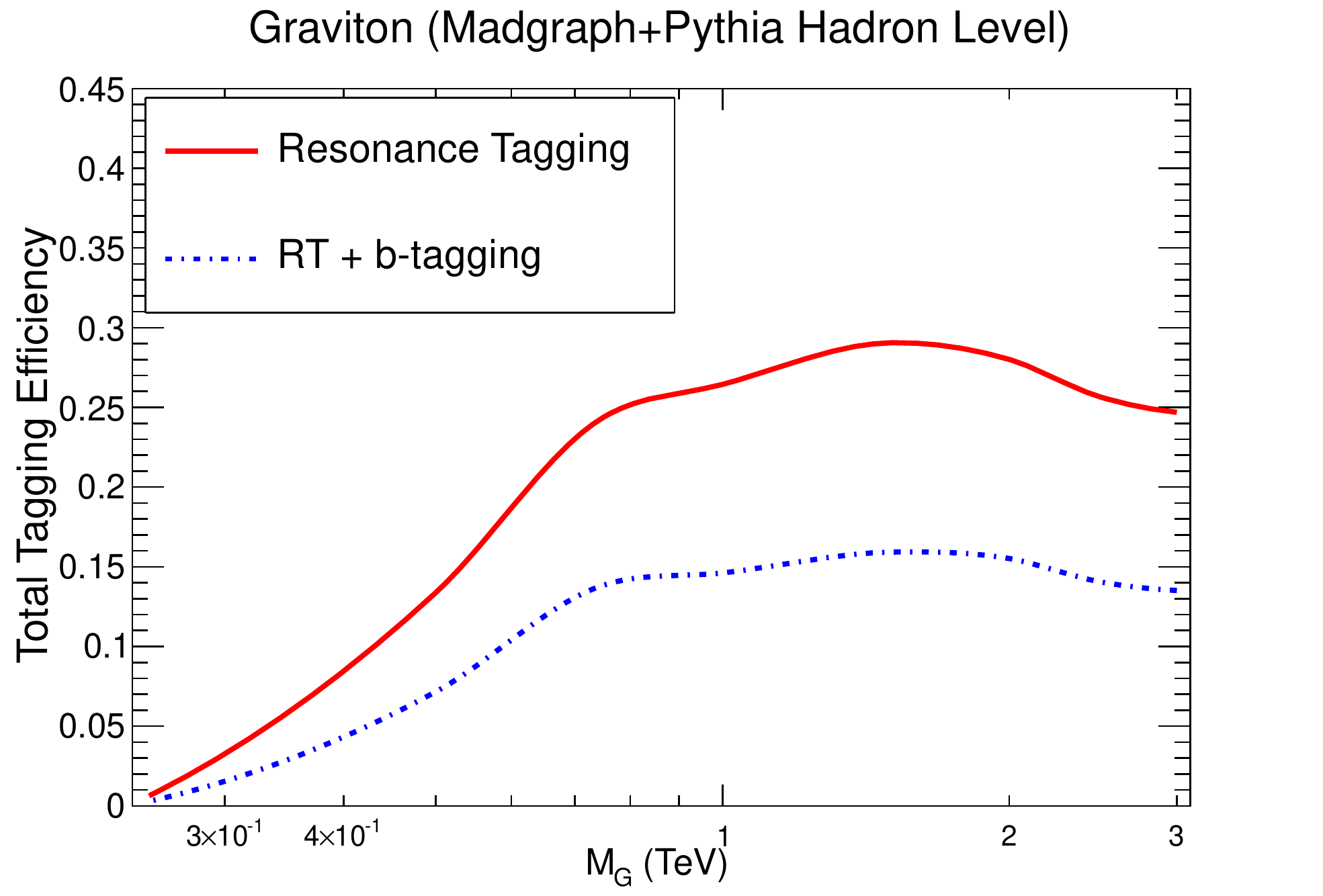}
\caption{\small The $b$-tagging efficiencies for graviton mediated Higgs pair production at the LHC 8 TeV. We show the efficiency of the resonance tagging
(same as in the upper right plot of Fig.~\ref{fig:efficiency-madgraph})
together with that including the $b$-tagging. }
\label{fig:btageff}
\end{figure}

Results for the $b$-tagging efficiencies
for graviton mediated Higgs pair production at the LHC 8 TeV 
as a function of $r_M$
are shown in Fig.~\ref{fig:btageff}. 
As we can see,
for the relevant mass range we have approximately a 15\% constant
signal efficiency after taking into account the resonance tagging
algorithm and the $b$-tagging. 
%

\subsection{QCD multijet background simulation}

The dominant Standard Model
 background to multijet final states that 
leads to event topologies similar to the signal is QCD jet
production. We have therefore produced a large sample of
QCD multijets with   {\tt Pythia8}~\cite{pythia},
starting from dijet configurations and
 with the shower radiation taking care
of generating the higher-order jet topologies.
These events include a subset with two and also four $B$-hadrons in
the final state.
The resulting hadron level events are then processed through the 
same analysis chain as the signal events.
There are several ways in which QCD radiation can mimic the
conditions for resonance tagging: for example, fake mass drops
can be generated from
a sufficiently symmetric splitting of a quark or gluon.
Note that while {\tt Pythia8} is known
to underestimate the amount of QCD multi-jet topologies by a factor up to two with respect to experimental data~\cite{CMS:2012yf}, for the accuracy
requested for this feasibility study we consider this
 precision  to be sufficient.
 Similar results have been obtained with the {\tt Alpgen} parton level
 event generator~\cite{alpgen} matched to {\tt Pythia8} using the MLM
 matching~\cite{Mangano:2001xp}.\footnote{As for the case of the
   signal, for the background too we have only considered leading
   order predictions. Next-to-leading order corrections are known for
   the two of the main backgrounds, namely 4-jet
   production~\cite{Bern:2011ep,Badger:2012pg} and two jets produced
   in association with two heavy quarks~\cite{Bevilacqua:2010ve}.
   Note however that our background involves multiple scales, which
   may limit the predictivity of NLO calculations.
 }

\begin{figure}[t]
\centering
\includegraphics[scale=0.37]{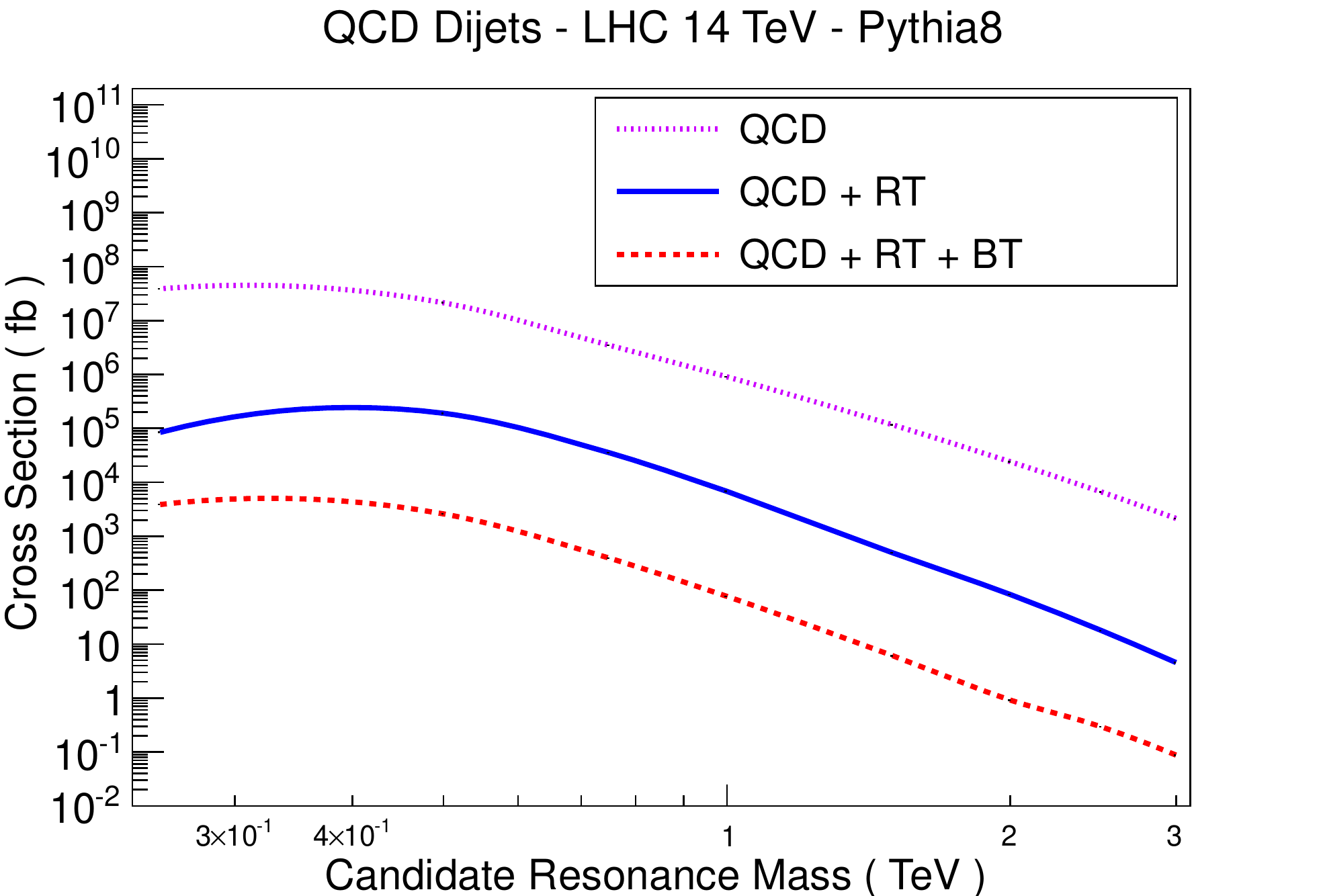}
\includegraphics[scale=0.37]{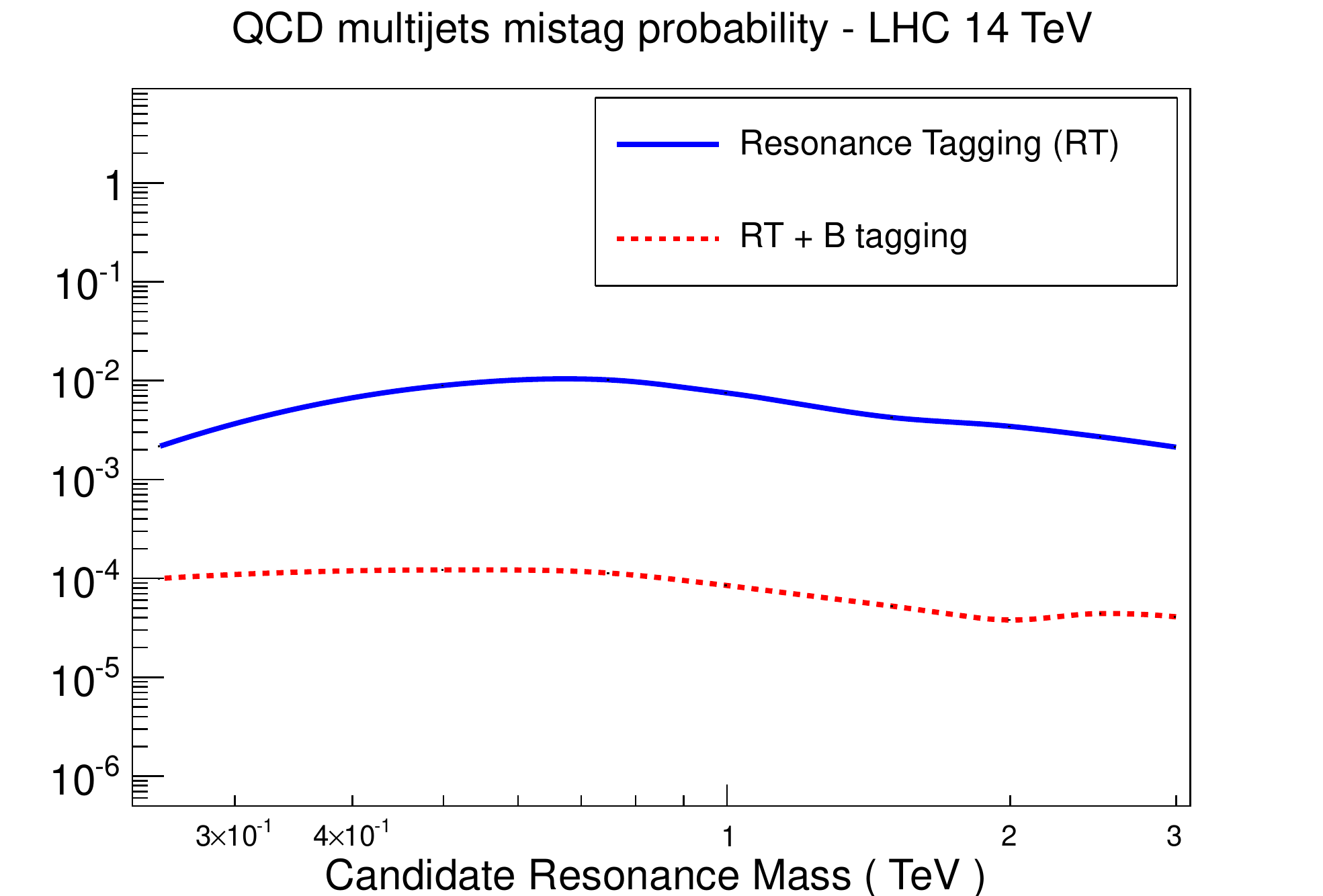}
\caption{\small Left
plot: the QCD dijet cross section before and after the events are
processed through the resonance-tagging algorithm, for
LHC 14 TeV.
Right plot: the mistag probability of
QCD dijet events with the resonance tagging (RT) algorithm,
defined as the ratio of mistags over the QCD cross section,
without and with $b$-tagging,
as a function of the mass point
$M$.  The mistag probability of QCD dijets at LHC 8 TeV
is very similar.}
\label{fig:mistagprob}
\end{figure}

\begin{figure}[t]
\centering
\includegraphics[scale=0.37]{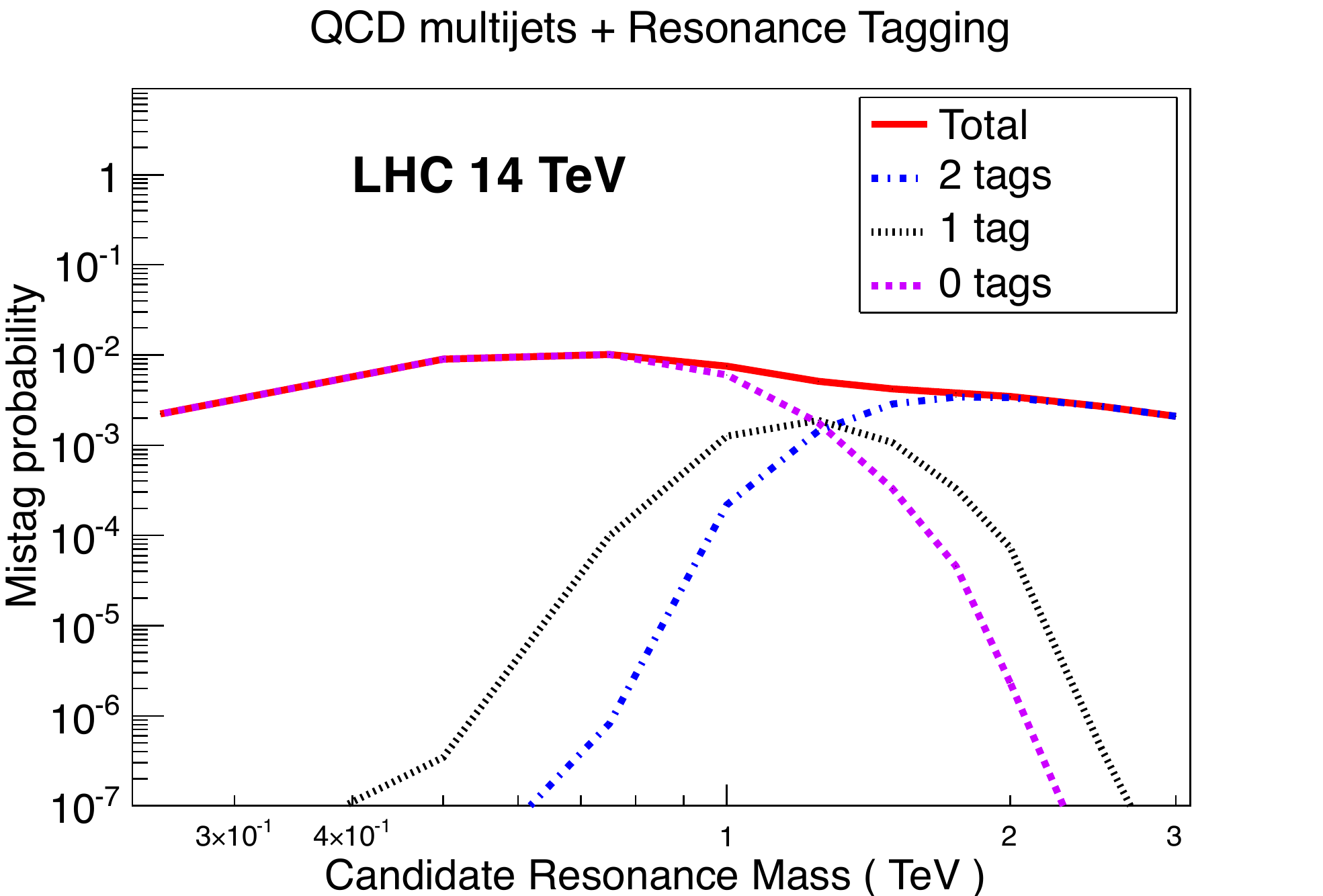}
\includegraphics[scale=0.37]{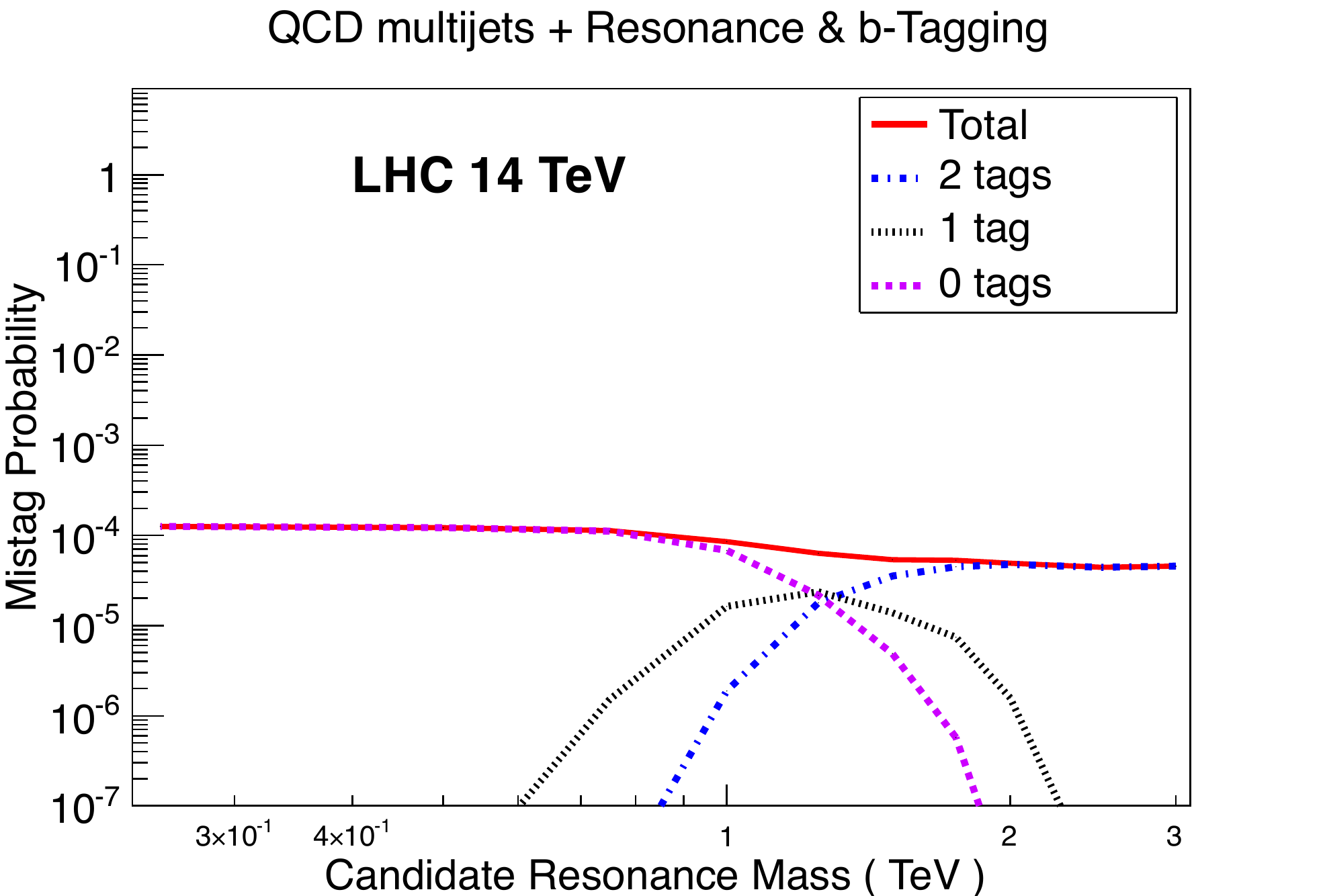}
\caption{\small Decomposition of the mistag probabilities according to
  the number of boosted-object tags in the event, without (left) and
  with (right) $b$-tagging.}
\label{fig:mistagprob-decomposition}
\end{figure}

In Fig.~\ref{fig:mistagprob} (left) we show
the QCD dijet cross section obtained from the {\tt Pythia8}
multijet sample at LHC 14 TeV. 
The dijet cross section is defined, for each
 mass point $M$, as the number
of QCD  events that survive the
 basic selection cuts Eq.~(\ref{eq:cuts_final}) and lead to
 an invariant mass
within the mass resolution window around $M$ given by
$\lc M(1-f_m),M(1+f_m)\rc$. 
In addition, we demand that
the two leading jets are separated in rapidity by less than
$\Delta y_{\rm max}$. 
Note that the dijet cross sections
flattens at small masses because there the selection cuts
 Eq.~(\ref{eq:cuts_final}) have a sizable effect.

In order to achieve an efficient QCD multijet
event generation, for any candidate resonance mass 
$M$  we have generated dijet events with a generation cut
of $ p_T \ge M/5$, and no generation cut in rapidity.
To motivate this choice, let us recall that
the kinematics of massless jet pair production determine that the mass of the
dijet will be given in terms of the $p_T$ of the jets and their
rapidity separation in the laboratory frame $\Delta y$ by $M = 2 p_T \cosh \lp \Delta y /2\rp$. Therefore,
to properly cover all phase space the generation cut for
QCD dijets should be at least
\be
p_T^{\rm min} \sim \frac{M}{2}\frac{1}{\cosh \lp \Delta y_{\rm max}/2\rp}
\ee
for any candidate resonance mass $M$. 
For the four jet configuration,
it is reasonable to require a minimum $p_T$ value of half
of that of above. Since we are using a rapidity cut of
 $\Delta y_{\rm max}=1.3$, 
we find that the {\tt Pythia8} minimum $p_T$ in generation should be
 $p_T^{\rm min}\sim M/5$.
We have explicitly verified that the QCD dijet cross-section is
not modified if looser generation cuts are adopted.

We also show in Fig.~\ref{fig:mistagprob} (right) 
the background rejection factors,
defined as
 the fraction of the QCD dijet events which
are mistagged as arising from a heavy resonance,
both with and without $b$-tagging.
Note that 
 the background rejection probability is approximately scale invariant:
similar mistag probabilities are obtained for all values of
the mass.
It is clear that the QCD background cross sections is  reduced
by a combination of the resonance tagging and $b$--tagging
by several orders of magnitude.
For example, with $b$--tagging 
the mistag probability is about $10^{-4}$, constant to
very good approximation in all the relevant mass
range. 
This improvement is due to the requirement that each Higgs candidate
should be associated with two 
identified b-hadrons, a topology that is less frequent
in QCD multijets.
The decomposition of the mistag probabilities according to the number
of boosted object tags in shown in
Fig.~\ref{fig:mistagprob-decomposition}, and is qualitatively similar
to what was seen for the signal in Fig.~\ref{fig:efficiency-parton}.

In summary, our study of the QCD background rejection
  confirm the consistency  of the resonance tagging algorithm, since
it makes possible to simultaneously explore  the low
mass and high mass region,  achieving similar signal efficiencies
and background rejection factors in all the mass range.

\subsection{Model independent exclusion limits}

We will now combine the results of the signal efficiencies and the multijet 
background
rejection of the resonance tagging algorithm to derive model
independent bounds on BSM scenarios with enhanced Higgs pair
production in the $4b$ final state.
This information is enough to derive
the values of the cross section times branching fraction
$\sigma\lp pp \to X \rp {\rm BR}\lp X \to HH\rp$ that can be excluded
at the 95\% confidence level from a measurement of the QCD $b$-tagged
multijet cross sections, as a function of the
mass of this hypothetical resonance.
In the following, to compute the number of 
signal and background events, we will assume a total
integrated luminosity of $\mathcal{L}=25$ fb$^{-1}$ at 8 TeV and
of 500 fb$^{-1}$ at 14 TeV.

For each candidate resonance mass, $M$, we compute the number of
 background events in a mass window of width $f_m=$15\% around
$M$.
 The local $p$--value
for each mass point $M$ based on the expected number
of signal and background events, $N_s$ and $N_b$ respectively,
in the mass window considered, is given by
\be
\label{eq:pvalue}
p = \frac{1}{2}\lp 1-{\rm Erf}\lc \frac{N_{s}}{\sqrt{2N_{b}}}\rc\rp \, ,
\ee
where ${\rm Erf}$ is the error function, and one
assumes that the number of background events in each
mass bin is a Poisson distribution with mean $N_b$.\footnote{Eq.~(\ref{eq:pvalue}) is only valid where both $N_s$ and $N_b$
are much larger than one, in the opposite case one has to
use the corresponding discrete Poisson formula for the $p$-value,
\be
p\lp M \rp =1-\frac{\Gamma\lp N_s+N_b,N_b\rp}{\Gamma\lp N_s+N_b\rp} \, ,
\ee
which involves the incomplete Gamma function.
}
 Then
requiring the condition $p=0.05$ determines
 the number of signal events $N_{s}$ that would allow an
exclusion of the background-only hypothesis at the 95\% confidence level,
namely
\be
\label{eq:ns}
N_s=\sqrt{2N_b} \cdot {\rm Erf}^{-1}\lp 1-2\cdot 0.05\rp \ .
\ee
Using Eq.~(\ref{eq:ns}) to determine the value of $N_s$ in a given mass window,
we can obtain the model independent bound on the combination 
$\sigma\lp pp \to X \rp {\rm BR}\lp X \to HH\rp$ by
correcting the number of events for the signal tagging efficiency,
with and without $b$-tagging, the Higgs to $b\bar{b}$ branching fraction
and the assumed total integrated luminosity $\mathcal{L}$,
\be
\lc \sigma\lp pp \to X \rp {\rm BR}\lp X \to HH\rp \rc_{\rm excl.~95\%~CL}(M) = 
\frac{N_s}{\lp BR(H\to b\bar{b})\rp^2 {\rm SignalEff}(M) \mathcal{L}} \, ,
\ee
where the signal efficiency ${\rm SignalEff}(M)$ is derived
from the {\tt MadGraph} radion and graviton samples, 
see Fig.~\ref{fig:btageff}.  
We have used ${\rm BR}\lp H\to b\bar{b}\rp=0.577$
from the Higgs Cross Section Working Group 
recommendations~\cite{Dittmaier:2012vm}.

\begin{figure}[t]
\centering
\includegraphics[scale=0.37]{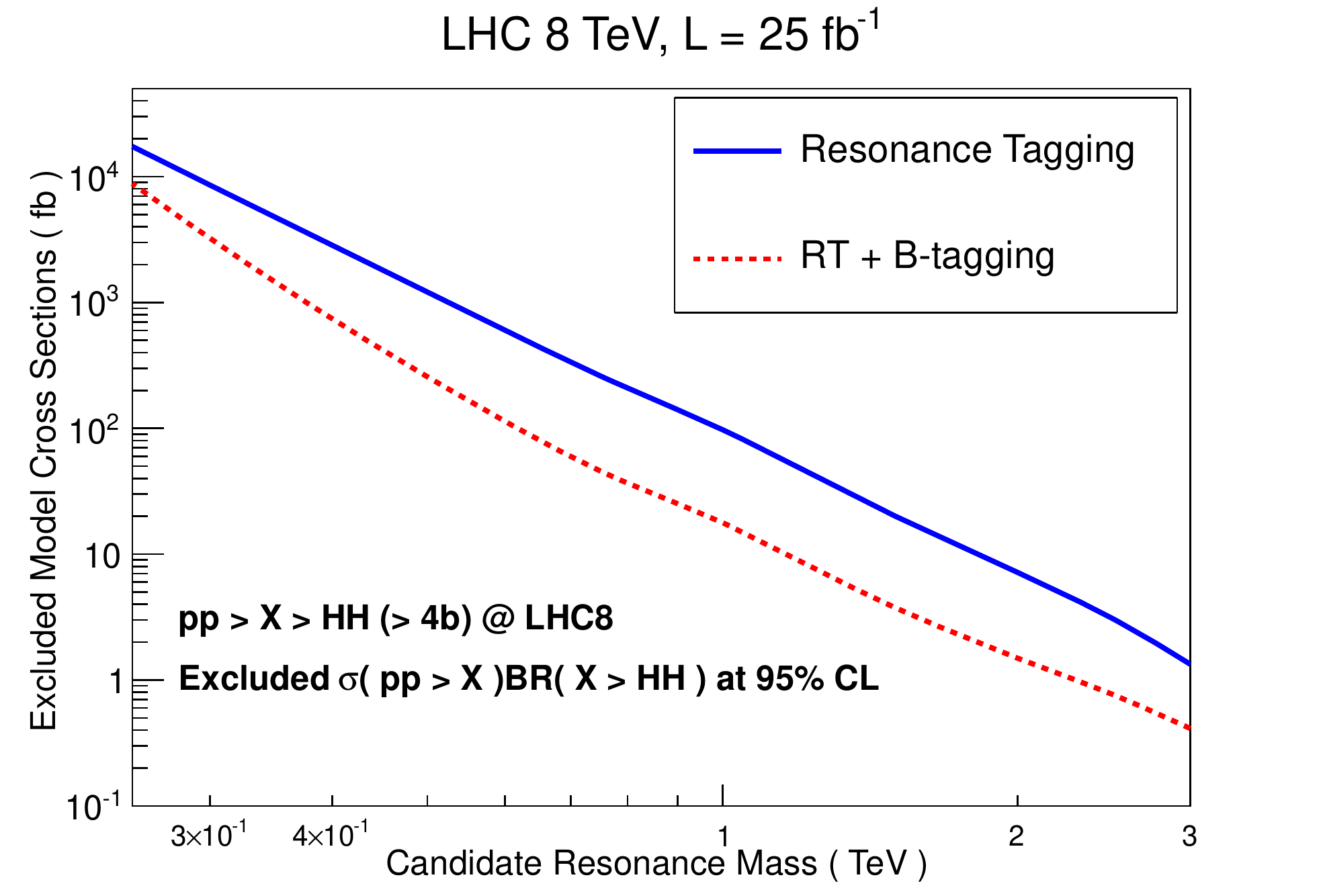}
\includegraphics[scale=0.37]{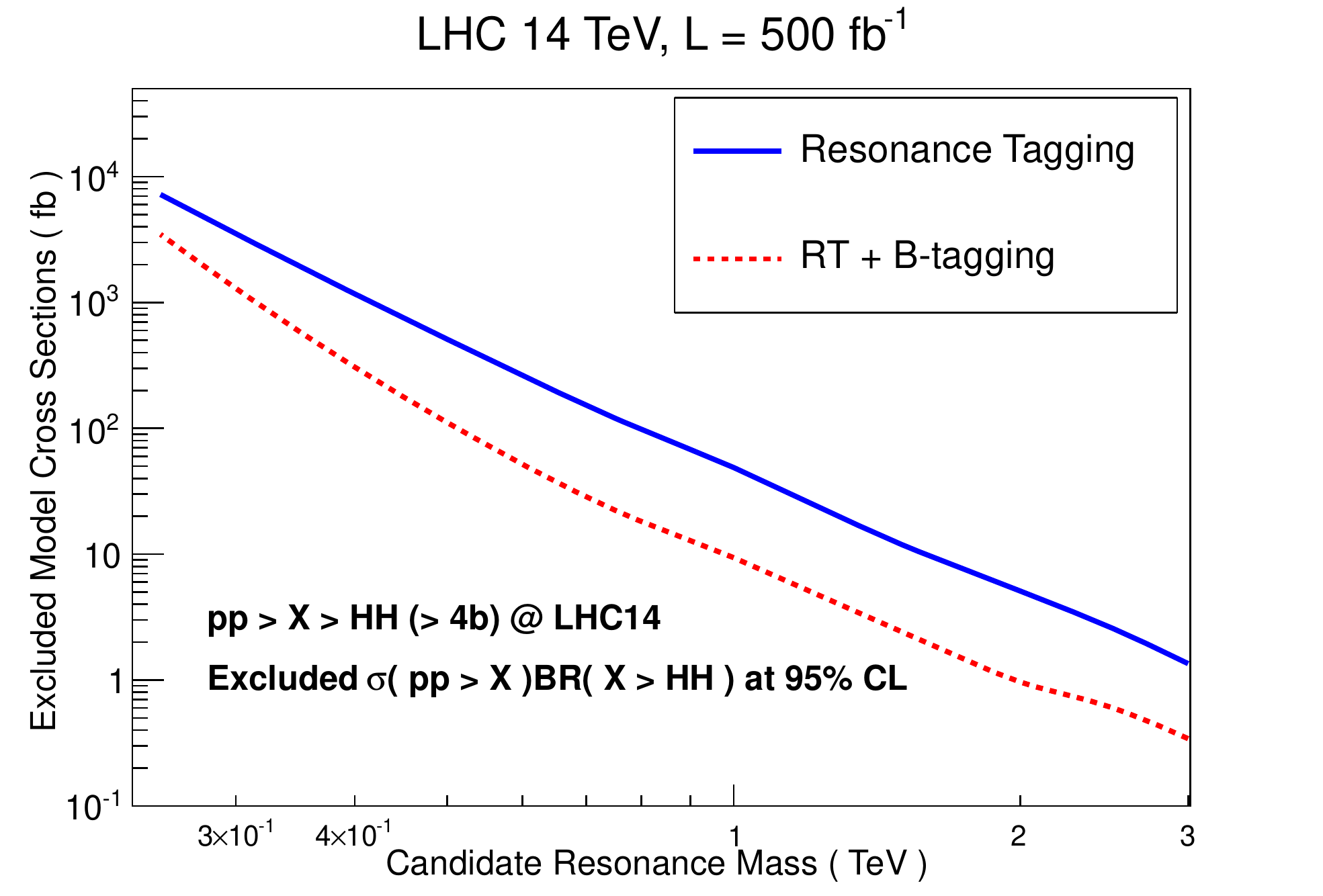}
\caption{\small  The model independent exclusion ranges
in $\sigma\lp pp \to X \rp {\rm BR}\lp X \to HH\rp$
at the 95\% confidence level for the
production of a heavy resonance $X$ which then decays into
a  Higgs boson pair, which decay subsequently
into a bottom-antibottom pair. The left plot
show the exclusion ranges at 8 TeV with $\mathcal{L}=25$ fb$^{-1}$
while the right plot corresponds to 
 14 TeV with $\mathcal{L}=500$ fb$^{-1}$.  }
\label{fig:model-independent-exclusion}
\end{figure}

The 95\% excluded model-independent cross sections times
branching fractions are shown in 
Fig.~\ref{fig:model-independent-exclusion}.
We see that
that we are sensitive to cross sections as small
as 200 fb (50 fb) at $M\sim 500$ GeV at LHC 8 TeV (14 TeV),
while at higher masses, $M\sim 2$ TeV, the $4b$ final
state is sensitive to cross sections as small as
1 fb at both energies.
Note that the increase in luminosity when going
from 8 to 14 TeV is partially canceled by
the corresponding increase of the high mass QCD dijet
cross sections.
On the other hand, signal cross sections in relevant models
also increase  when going from 8 to 14 TeV,
so all in all we obtain a substantial improvement in
exclusion reach when increasing the center of mass energy.

These results confirm that the $4b$ final state
can be relevant for many new physics scenarios that lead to
 enhanced cross sections for resonant Higgs pair production,
and the search strategy that we propose makes it possible to efficiently
explore a wide range of resonance masses within a common analysis.

\subsection{Graviton and radion searches in the $2H\to 4b$ channel}

Now we consider the specific benchmark scenarios
for radion and graviton production introduced in Sect.~\ref{sec:models}.
First of all, we summarize in Table~\ref{tab:bench} 
the model parameters that
we adopt here. 
The mass scales and
branching fractions to Higgs boson pairs are kept fixed,
and only the couplings of the radion and graviton to
gluons will be varied.
 For the graviton, we consider two
different values of the coupling, $c_g=1$ (G-Brane) which corresponds
to the RS1 model and $c_g=1/35$ (G-Bulk) as in the bulk models.

For the radion we study the nominal coupling $\kappa^{\phi}_g$
(R-Bulk),
as well as the case in which this coupling is enhanced
by a factor of ten by some unspecified mechanism, such as 
when the radion arises as a composite bound state (R-Comp). 
We use
the same mass scale in the two cases, $\Lambda_{\phi}=\Lambda_{G}=2$
TeV: although the two scales are related, we prefer to explore
independently the radion and graviton scenarios.
Let us recall that the cross section scale as $1/\Lambda_{\phi,G}^2$,
so any different choice of the mass scale $\Lambda_{\phi,G}$ will
lead to a trivial rescaling of the cross section.

%
\begin{table}[t]
\centering
\small
\begin{tabular}{c|c|c|c}
\hline
\multicolumn{4}{c}{radion Production} \\
\hline
Scenario  & $|\kappa^{\phi}_g|$  & $\Lambda_{\phi}$  & ${\rm BR}(\phi\to 2H)$  \\
\hline
\hline
radion Bulk (R-Bulk)  &  $|- \alpha_s b_3/8\pi -1/4 kL| \sim 0.04$ & 2 TeV & 1/4 \\
radion Composite (R-Comp)  &  $0.4$ & 2 TeV & 1/4 \\
 \hline
\hline
\multicolumn{4}{c}{graviton Production} \\
\hline
Scenario  & $c_g$  & $\Lambda_{G}$  & ${\rm BR}(G\to 2H)$  \\
\hline
\hline
graviton RS1 (G-Brane)  &  1 & 2 TeV & 1/4 \\
graviton Bulk (G-Bulk)  &  $1/kL=1/35$ & 2 TeV & 1/4 \\
 \hline
\end{tabular}
\caption{\small Parameters of the benchmark scenarios for
radion and graviton production. 
For the radion we consider both the nominal value of
$\kappa^{\phi}_g$  
(denoted by R-Bulk),
and a coupling ten times larger that could arise
for example in composite dual scenarios (denoted by R-Comp).
For the graviton we consider
two different values of the gluon-gluon-graviton coupling,
$c_g=1$ as in RS1 (denoted by G-Brane settings) and $c_g=0.02$ as in bulk models
(denoted by G-Bulk). 
 \label{tab:bench}
}
\end{table}

\begin{figure}[t]
\centering
\includegraphics[scale=0.37]{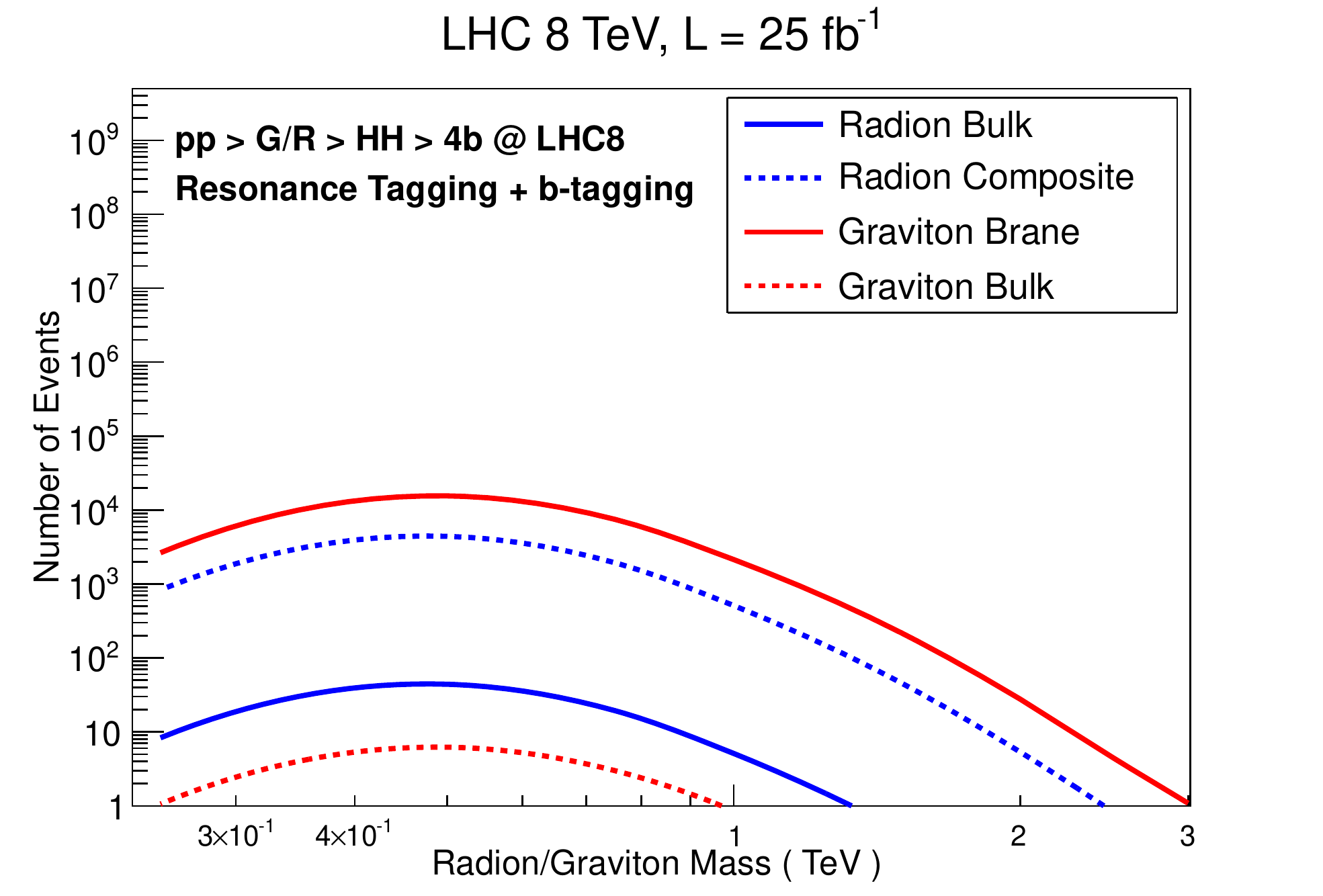}
\includegraphics[scale=0.37]{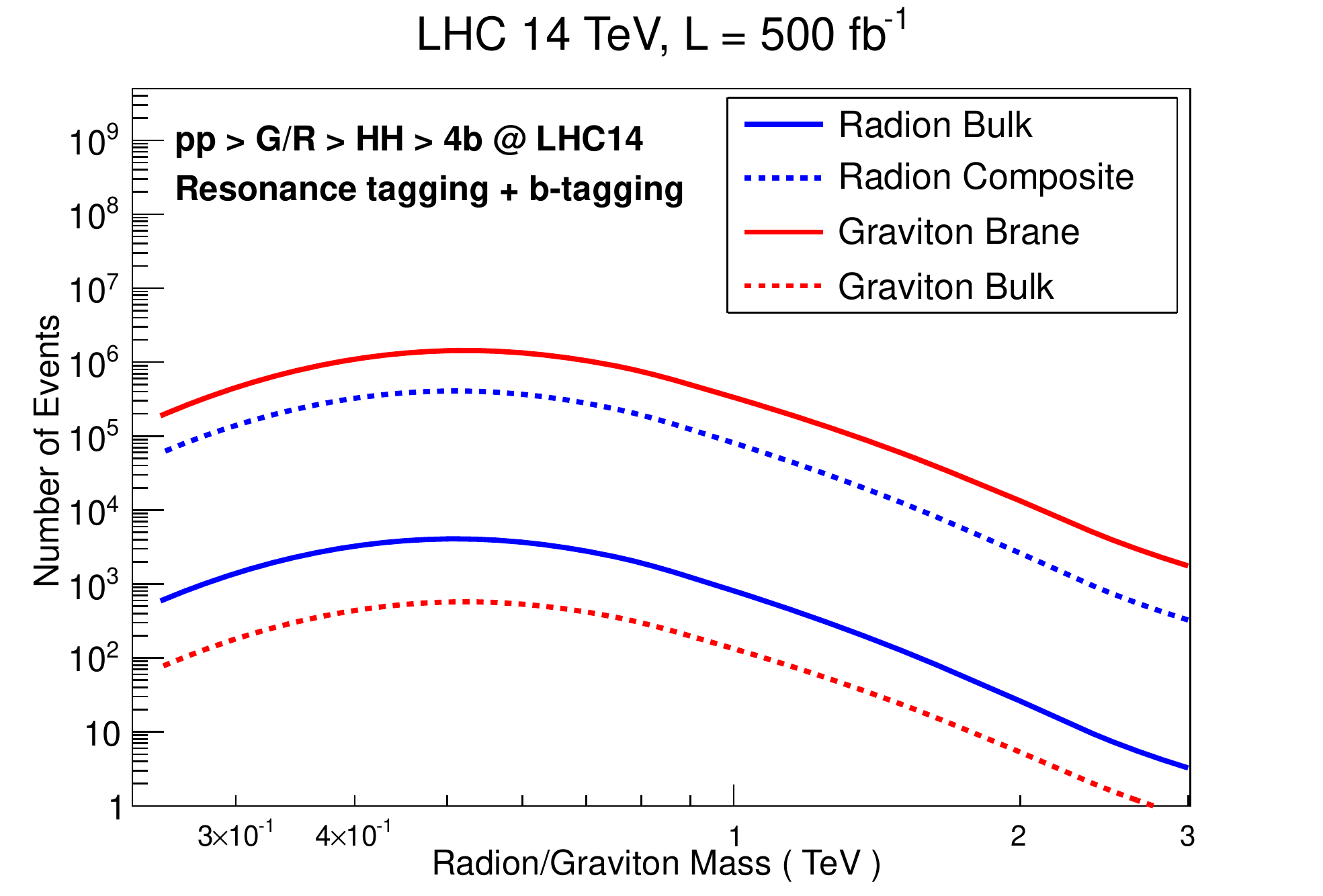}
\caption{\small The expected number of events 
in the benchmark scenarios after both resonance
tagging and $b$--tagging, at the
LHC 8 (left plot) and 14 TeV (right plot). The integrated
luminosities are $\mathcal{L}=30$ fb$^{-1}$ at 8 TeV
and $\mathcal{L}=500$ fb$^{-1}$ at 14 TeV.  }
\label{fig:nevents}
\end{figure}

\begin{figure}[t]
\centering
\includegraphics[scale=0.37]{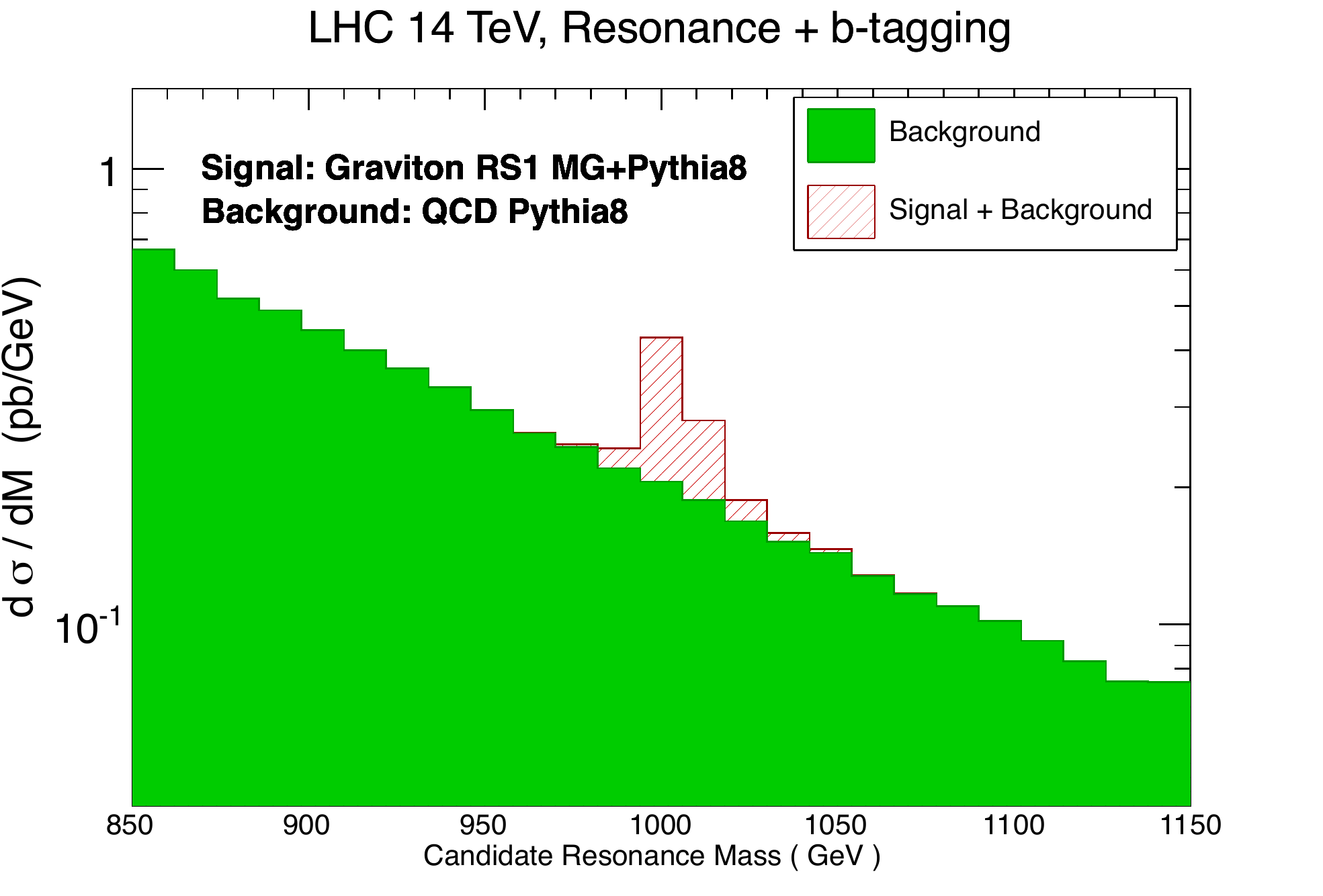}
\caption{\small Illustration of a signal mass peak superposed on the
  QCD background, for the case of an RS1 graviton with a mass of
  $1$~TeV, in 14~TeV pp collisions.}
\label{fig:signal-peak}
\end{figure}

First of all we evaluate the expected number of events for these
four benchmark points at LHC 8 and 14 TeV, using the
results of Sect.~\ref{sec:models}.
 We take into account the
branching fraction of the Higgs bosons into $b\bar{b}$ pairs.
As before, we assume total integrated luminosities of $\mathcal{L}=25$ 
fb$^{-1}$ at 8 TeV and
of 500 fb$^{-1}$ at 14 TeV. 
The model cross sections for 
the benchmark scenarios can be easily obtained from the 
results of Sect.~\ref{sec:models}, in
particular from Fig.~\ref{xsec}.
The number of expected events is shown in Fig.~\ref{fig:nevents}, 
after accounting for the
selection efficiencies from resonance and $b$-tagging analysis. 
At 8 TeV we expect just a handful of events at low masses
for the R-Bulk and G-Bulk points, and about one thousand events
(few tens of events) and low (high) masses for the 
R-Comp and G-Brane points. 
At LHC 14 TeV on the other hand
we have a large enough number of events for all masses
and all benchmark points, thanks to both the increased
resonance production cross sections and the higher
integrated luminosity.
An illustration of the type of signal that one might observe is given
in Fig.~\ref{fig:signal-peak} for the case of an RS1 graviton with a
mass of $1$~TeV in 14~TeV pp collisions.
Note that at this mass, the signal involves 0, 1 and 2-tag categories
combined. 

\begin{figure}[t]
\centering
\includegraphics[scale=0.37]{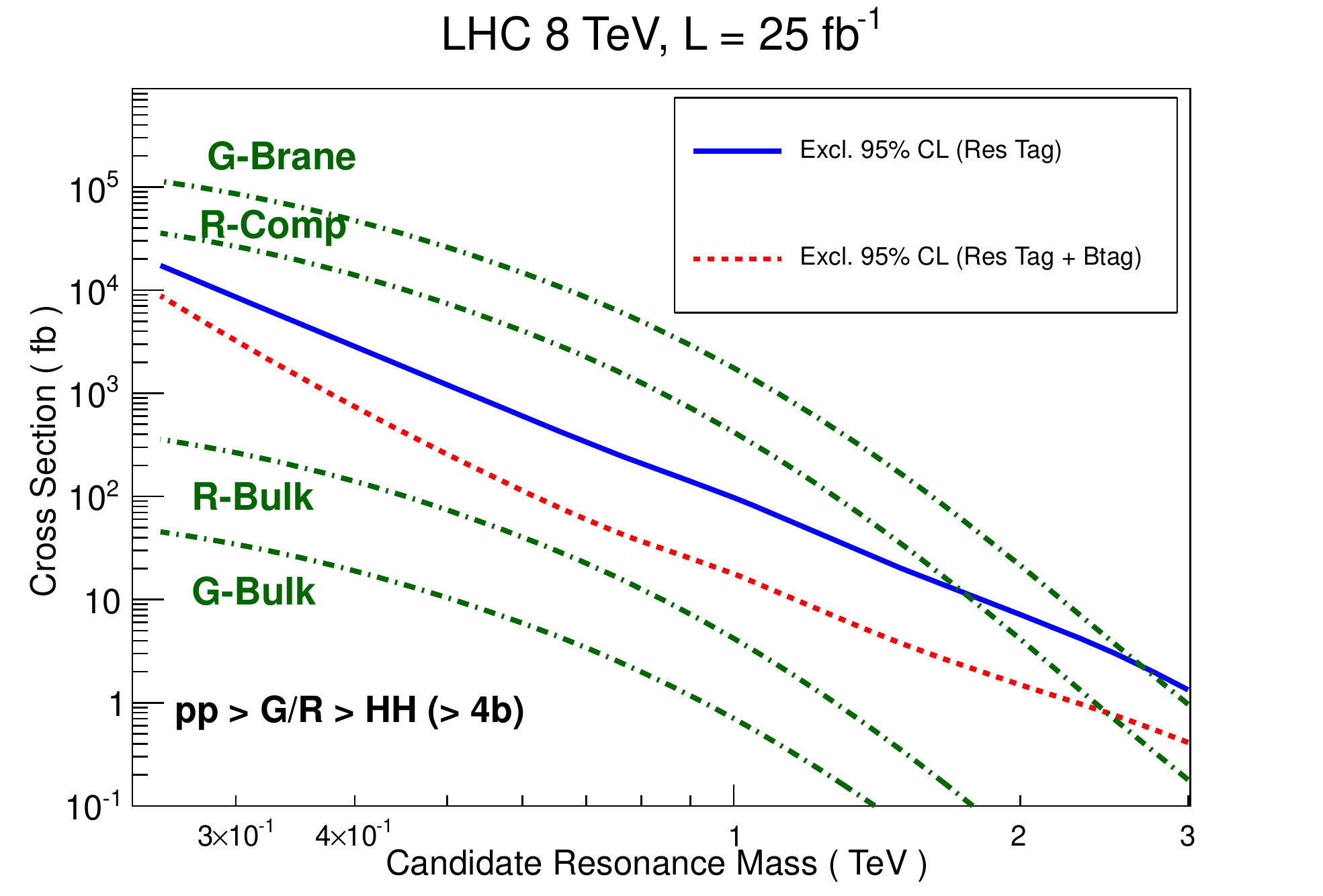}
\includegraphics[scale=0.37]{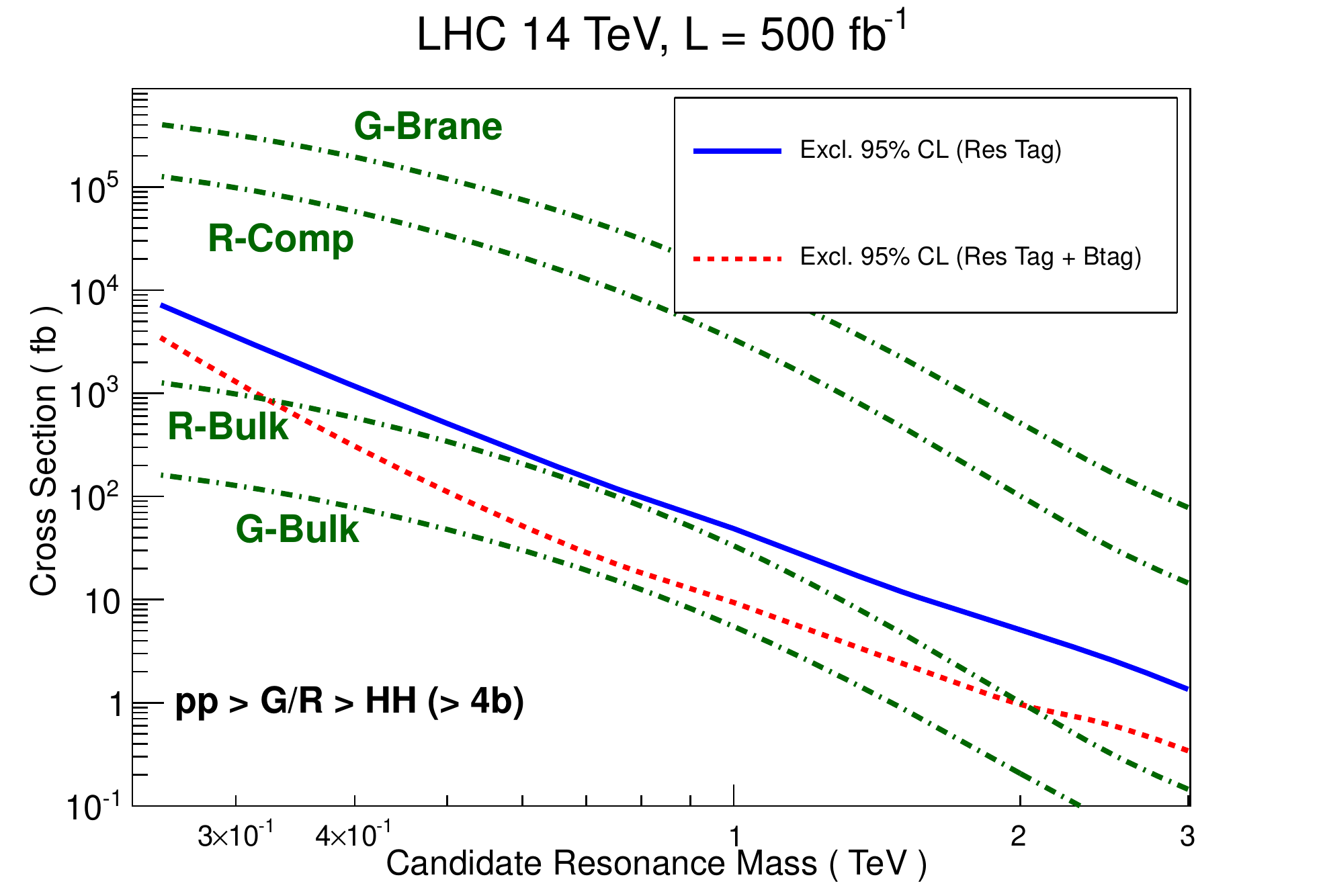}
\caption{\small The 95\% confidence level exclusion ranges
at 8 TeV (left plot) and 14 TeV (right plot) compared
to the specific cross sections of the four
different model scenarios of Table~\ref{tab:bench}.  }
\label{fig:model-dep}
\end{figure}

Now in
Fig.~\ref{fig:model-dep} we show the same excluded
cross sections at the 95\% confidence level 
of Fig.~\ref{fig:model-independent-exclusion} but
this time adding the specific model cross sections of the four
 scenarios of Table~\ref{tab:bench}.
The improvement in exclusion power when going
from 8 to 14 TeV is clear.\footnote{Note that we only
consider statistical errors in determining the exclusion 
limits.
In the small and intermediate mass regions, the exclusions
are based on a large number of events, corresponding
to small signal over background ratios.
This can be understood from Eq.~(\ref{eq:ns}), which tells us that at
the exclusion limit,
$N_{s}/N_{b}\sim 1/\sqrt{N_b}$.
}
 At LHC 8 TeV we can explore a large part of the
parameter range of the graviton models up to 2 TeV, 
but the default radion scenario seems to be out of reach, unless
its cross section is enhanced by some mechanism, for instance
as in the composite duals discussed in Sect.~\ref{sec:models}.
At the LHC 14 TeV on the hand we are sensitive
to R-Bulk scenario with $\kappa_g^{\phi}=1$,
for most of the mass range up to $M_{\phi}=2$ TeV.
Likewise, we could exclude a Bulk graviton 
up to masses of 2.5 TeV.
Therefore, after the energy increase to 14 TeV most
of the parameter space of the radion and
massive KK graviton models will become
accessible in the $4b$ final state.

Using these results, it is also
possible to determine the 95\% confidence level exclusion
ranges for some of the parameters of the benchmark scenarios.
We can keep all
the parameters as in Table~\ref{tab:bench} and determine the
exclusion ranges for the couplings of the gluons
to the massive KK graviton $c_g$ and to the radion $\kappa^{\phi}_g$
and scan the allowed values for $\Lambda_{\phi}$. 

\begin{figure}[t]
\centering
\includegraphics[scale=0.37]{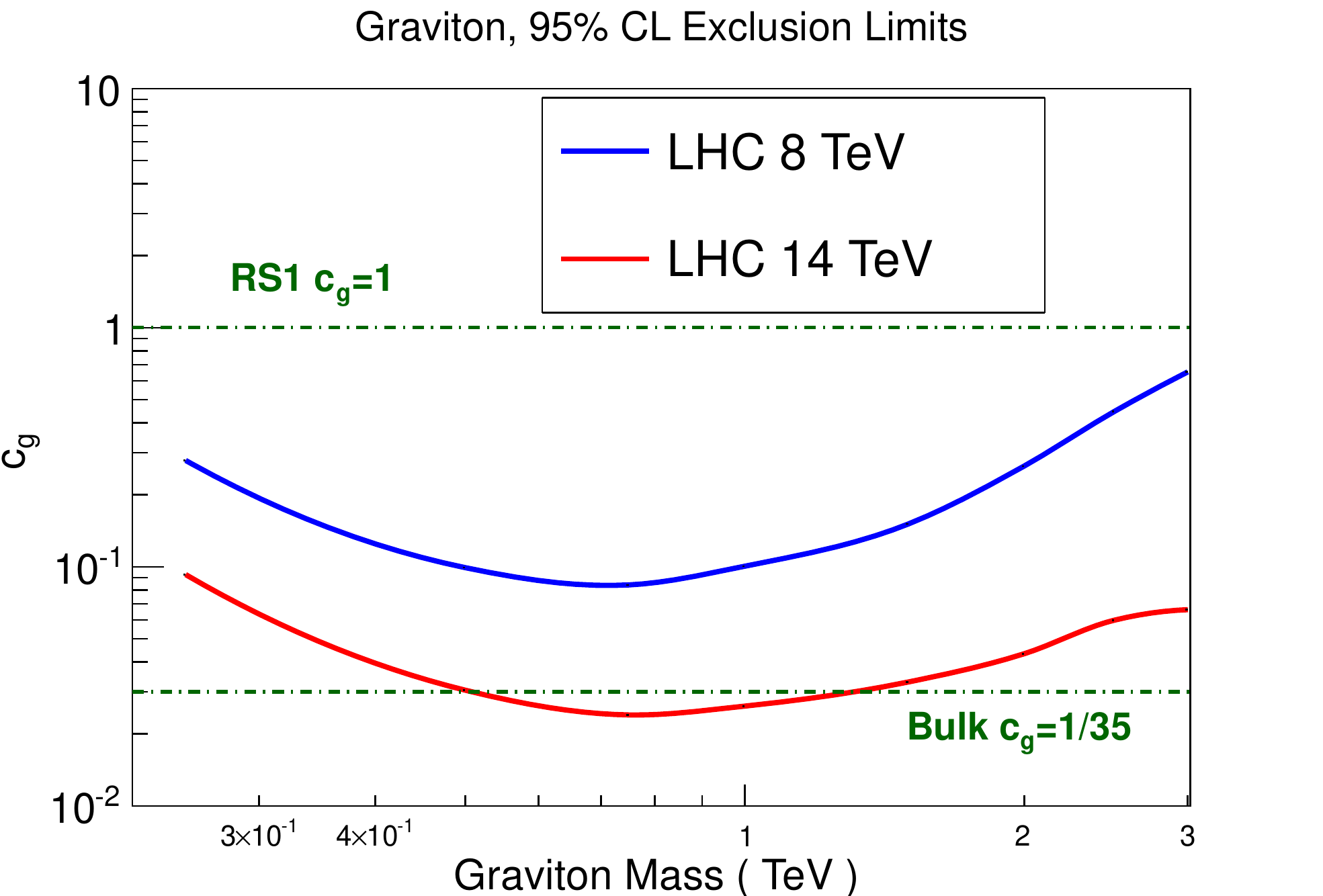}
\includegraphics[scale=0.37]{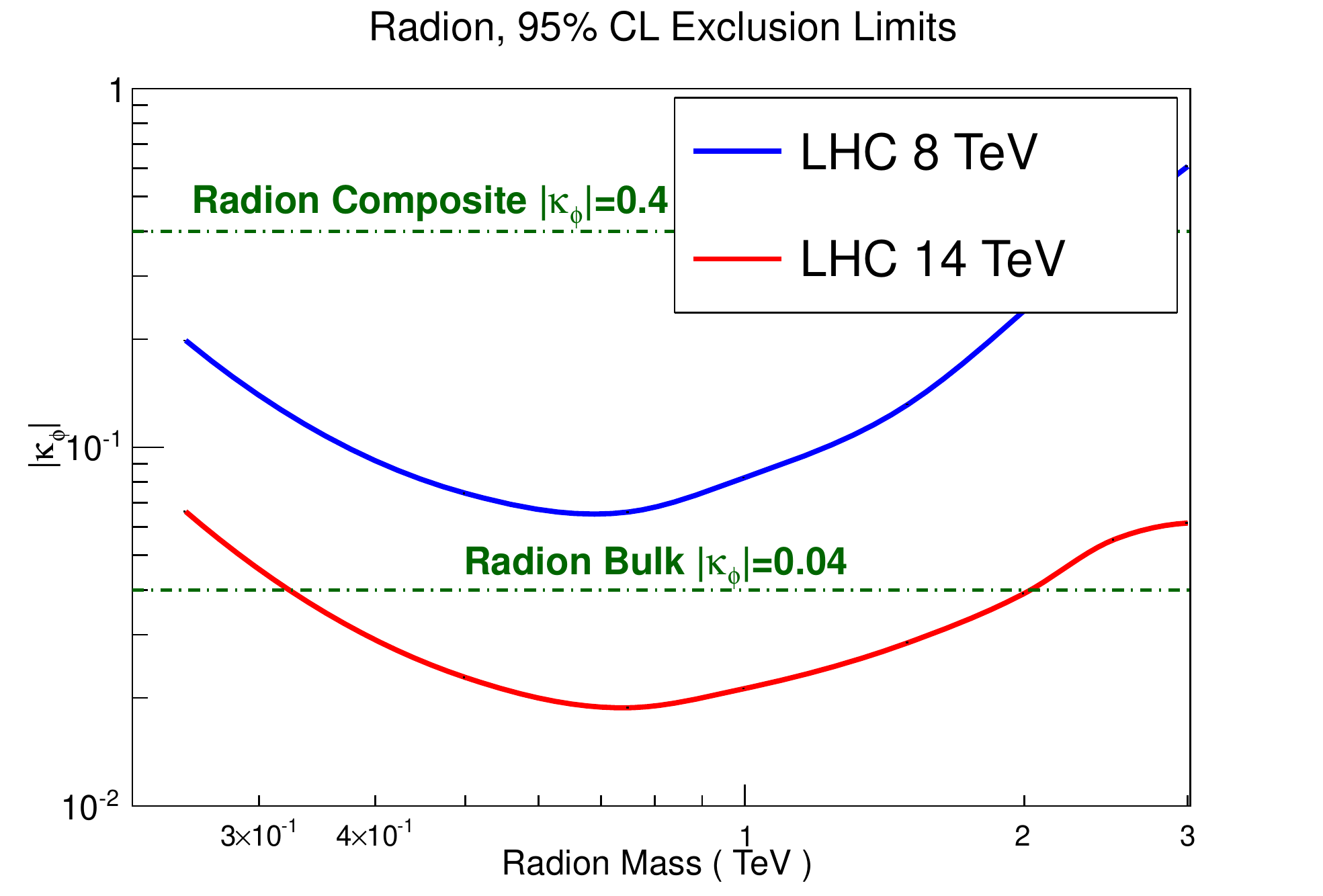}
\caption{\small Left plot: the values of the massive KK graviton-gluon
coupling $c_g$ that can be excluded at the 95\% CL as a function
of the graviton mass for 8 and 14 TeV.
For 14 TeV we provide the results for both $b$-tagging scenarios.
Right plot: same for the radion-gluon coupling $\kappa^{\phi}_g$. }
\label{fig:excluded}
\end{figure}

We show the results in Fig.~\ref{fig:excluded}. In the case of the
graviton coupling, we see that at 14 TeV the $2H\to 4b$ final state
can access essentially all the relevant range, from the RS1 value
$c_g=1$ down to the bulk value of $c_g=1/35$.
 For the case of the radion 
coupling $\kappa_{g}^{\phi}$, 
we see that at 8 TeV we are sensitive to values
down to  $\kappa_{g}^{\phi} \simeq 0.06$ around 750 GeV,
while at 14 TeV the LHC can exclude a bulk
radion (with the default value for the coupling) 
for masses between 300 GeV and 2 TeV at least.

In summary, we have shown in this section that the $2H\to 4b$ final state
offers a promising channel to probe enhanced Higgs pair production
at the LHC, despite the overwhelming QCD multijet background.
The combination of
jet substructure techniques and $b$-tagging makes it possible to probe
a wide region in the parameter space of various benchmark
models. 
Therefore, we advocate that the experiments explore this new
channel in order to complement existing searches of new heavy
resonances in other, more traditional, channels.


\section{Conclusions and outlook}
\label{sec:conclusions}

In this paper we have presented a new strategy for heavy-resonance
searches in multijet final states, which attempts to unify
in a single approach the techniques used in the boosted 
and resolved regimes. 
By classifying events as a function
of the number of mass-drop tags, we can smoothly interpolate
between the boosted regime, where jet substructure techniques
can be used, and the resolved regime, where the final state
particles appear as well separated jets.
In particular, we have considered the process $X \to YY \to 4z$
in which the resonances $Y$ are pair produced from the decay of
a heavier resonance $X$ and then decay into a pair of QCD partons, then
observed as jets.
 We have shown that our
strategy leads to approximately scale-invariant signal selection
efficiencies and background rejection rates.

As a benchmark scenario, we have considered Higgs pair production in
extra dimension models, where the Higgs pair is produced from the decay of a heavy graviton or radion, and then decays into four $b$ quarks. 
Note however that the kinematical structure of the final state of the benchmark model holds
for other scenarios, such as composite models, with more 
freedom on the couplings and therefore in the cross sections strengths.
By comparing with
the QCD multijet background, we have derived the model independent
95\% confidence level exclusion ranges for the cross sections
for $\sigma \lp  pp \to X \to HH\rp$ where the Higgs bosons
decay into a $b\bar{b}$ pair, and showed that a substantial
region of the parameter space of these models can be successfully explored
in this final state with the tagging strategy that has been proposed.

In the particular case of  graviton and radion production,
it would be especially interesting to study the feasibility of
radion/graviton searches in the $b\bar{b}\gamma\gamma$ decay
channel~\cite{Ball:2007zza,Baur:2003gp}. 
This is a cleaner channel than the all hadronically
decaying case, since the two high $p_T$ photons
substantially reduce the QCD background. 
In this final state, by varying
$r_M$ one also moves from the boosted regime (with one single fat jet
in the final state) to the resolved limit, with two well separated
jets in the final state. 
Another interesting final state to search
for enhanced Higgs pair production
would be $b\bar{b}ZZ$.

The approach advocated in this paper could also be applied to other 
relevant problems, for example top quark pair production, again
providing a smooth coverage across the 
transition between the resolved regime,
relevant for SM measurements, and the boosted regime, where
substructure techniques~\cite{Plehn:2011tq,Plehn:2009rk} are used to enhance the potential of
new physics searches.

\bigskip
\bigskip
\begin{center}
\rule{5cm}{.1pt}
\end{center}
\bigskip
\bigskip

{\bf\noindent  Acknowledgments \\}
We acknowledge Xin Fang for collaboration in early stages of this project.
We warmly thank Pierluigi Catastini, Dinko Ferencek and
Alexander Schmidt for their
important inputs on the $b$-tagging capabilities of ATLAS and CMS,
Maurizio Pierini and Andreas Hinzmann for useful discussions. 
J.~R. is supported by a Marie Curie 
Intra--European Fellowship of the European Community's 7th Framework 
Programme under contract number PIEF-GA-2010-272515.
G.~P.~S gratefully acknowledges support from the French Agence
Nationale de la Recherche, grant ANR-09-BLAN-0060, and from the EU ITN
grant LHCPhenoNet, PITN-GA- 2010-264564.
R.~R. is partially supported by a CNPq research grant and by a Fapesp Tematico grant 2011/11973-4. R.~R. thanks the CERN Theory Group, where this work was initiated, for the hospitality. 


\end{document}